\documentclass[floats,floatfix,showpacs,amssymb,prd,twocolumn,superscriptaddress,nofootinbib]{revtex4-1}

\usepackage{graphicx,epsf, epsfig, amssymb}
\usepackage{bm}
\usepackage{longtable,tabularx}
\usepackage{color}
\usepackage[breaklinks,bookmarksopen=true]{hyperref}
\usepackage{amsfonts,amsmath,amssymb,mathrsfs,gensymb}
\usepackage{natbib}
\usepackage{prd_macros}
\usepackage{array}
\usepackage{multirow}
\usepackage{rotating,array}
\usepackage{graphicx}
\usepackage[normalem]{ulem}
\usepackage[caption=false]{subfig} 
\usepackage{ulem}
\usepackage{dcolumn}
\usepackage{braket}
\usepackage{appendix}
\usepackage{comment}
\usepackage{siunitx}

\usepackage{float} 
\usepackage{enumitem}

\DeclareRobustCommand{\rchi}{{\mathpalette\irchi\relax}}
\newcommand{\irchi}[2]{\raisebox{\depth}{$#1\chi$}} 

\usepackage [english]{babel}
\usepackage [autostyle, english = american]{csquotes}
\MakeOuterQuote{"}

\begin{document}

\setlength{\parskip}{0.2pt}

\title{Pathways for producing binary black holes \\ with large misaligned spins in the isolated formation channel}
\author{Nathan Steinle}
\email{nathan.steinle@utdallas.edu}
\affiliation{Department of Physics, The University of Texas at Dallas, Richardson, TX 75080, USA}
\author{Michael Kesden}
\email{kesden@utdallas.edu}
\affiliation{Department of Physics, The University of Texas at Dallas, Richardson, TX 75080, USA}
\date{\today}

\begin{abstract}
Binary black holes (BBHs) can form from the collapsed cores of isolated high-mass binary stars. The masses and spins of these BBHs are determined by the complicated interplay of phenomena such as tides, winds, accretion, common-envelope evolution (CEE), supernova natal kicks, and stellar core-envelope coupling. The gravitational waves emitted during the mergers of BBHs depend on their masses and spins and can thus constrain these phenomena. We present a simplified model of binary stellar evolution and identify regions of the parameter space that produce BBHs with large spins misaligned with their orbital angular momentum. In Scenario A (B) of our model, stable mass transfer (SMT) occurs after Roche-lobe overflow (RLOF) of the more (less) massive star, while CEE follows RLOF of the less (more) massive star. Each scenario is further divided into Pathways 1 and 2 depending on whether the core of the more massive star collapses before or after RLOF of the less massive star, respectively. If the stellar cores are coupled weakly to their envelopes, highly spinning BBHs can be produced if natal spins greater than 10\% of the breakup value are inherited from the stellar progenitors. BBHs can alternatively acquire large spins by tidal synchronization during the Wolf-Rayet stage in Scenario A or by accretion onto the initially more massive star during SMT in Scenario B. BBH spins can be highly misaligned if the kicks are comparable to the orbital velocity, which is more easily achieved in Pathway A1 where the kick of the more massive star precedes CEE.
\end{abstract}

\maketitle

\section{Introduction} 
\label{sec:Intro}

The LIGO/Virgo Collaboration reported ten stellar-mass binary black-hole (BBH) mergers and one binary neutron-star merger during its O1 and O2 observing runs \cite{LIGO2019catalog}. Hundreds of additional BBH mergers are expected in the next decade as the LIGO/Virgo detectors improve in sensitivity and as new gravitational-wave (GW) detectors come online \cite{Punturo2010,Abbott2017,Baibhav2019,Kuns2020}.

Two possible channels in which BBHs form are the dynamical channel, where individual BHs form binaries through dynamical interactions with a cluster of stars, and the isolated channel, in which BBHs evolve from isolated binary stars. Both channels can explain the origin of the current LIGO/Virgo sources, and future observations may identify the fractional contribution from each channel \cite{Breivik2016,Nishizawa2016,Nishizawa2017,Zevin2017,Vitale2017,Bouffanais2019,LIGO2019astro,Zevin2020b,Wong2020}. Measurements of BBH spin orientations help to discriminate between the possible formation channels \cite{Mandel2010,Gerosa2013,Rodriguez2016,Talbot2017,Farr2017,Stevenson2017,Gerosa2018,Farr2018,Talbot2018,Wysocki2018,LIGO2019astro,Zevin2020b,Miller2020,Callister2020}: in the dynamical channel the spin directions are isotropic, whereas in the isolated channel the spin directions are determined by astrophysical processes like tides, accretion, and supernova natal kicks.

Most of the massive stars that evolve into BHs are found in binary or triple systems \cite{Kobulnicky2007,Sana2012,Moe2017}. Rather than using existing comprehensive models of binary stellar evolution, we parameterize zero-age main sequence (ZAMS) binary stars by their initial masses, spins, metallicity, and binary separation, then evolve the systems as the stars develop through the various stages of nuclear evolution and interact with each other. This work focuses on how stellar and binary processes determine BBH spins in the isolated channel. 

The most significant uncertainties in predicting the masses and spins of the BBHs formed from binary stars arise from the transport of angular momentum within each star (i.e. the strength of core-envelope coupling and of stellar winds), from the transport of mass and angular momentum throughout the binary (i.e. the relevance of mass transfer and of tides), and from the gravitational collapse of each star into a compact object (i.e. the supernova and kick mechanisms). We explore how these processes affect BBH spins and identify key regions of the parameter space from which candidate precessing BBH systems arise.

\subsection{Isolated channel of BBH formation}

Stars generally increase in radius as they age. If the binary separation is sufficiently small, a star will fill its Roche lobe (RL) - the region around the star in which material is gravitationally bound to it. This Roche lobe overflow (RLOF) initiates a mass-transfer event. As more massive stars evolve more rapidly, the initially more massive star in a binary will undergo RLOF first, followed by the initially less massive star.

The stability of the mass transfer, which depends on the response of the donor's RL and stellar radius to changes in the mass-transfer rate, determines the type of mass-transfer event. If the mass transfer is stable, the companion (a star or a BH) can be spun-up by accreting a fraction $f_{\rm a}$ of the envelope of the donor star. If the mass transfer is unstable, a common envelope engulfs the binary and energy and angular momentum are transferred from the binary orbit to the envelope, expelling it from the system. This common-envelope evolution (CEE) shrinks the orbital separation by a factor of $\gtrsim 100$ and may prematurely merge the binary. The physics of CEE is uncertain \cite{Ivanova2013,Postnov2014,Schroder2020}, yet it is crucial for the isolated channel since only compact binaries with small enough separations can merge within the age of the Universe and thus emit observable GWs.

After a donor star loses its envelope, its core is exposed and emerges as a Wolf-Rayet (WR) star which experiences mass loss through winds \cite{WRprogenitors1990,Tutukov2003,WRprogenitors2018}. The initial spin of the WR star depends crucially on the strength of the spin coupling between the core and envelope of its progenitor: the more efficiently angular momentum is extracted from the core by the envelope, the smaller the initial spin of the WR star. This process is uncertain \cite{Meynet2005,Maeder2012}, but it is important in modeling the spin of BH progenitors \cite{Meynet2013}. 

For small separations, winds and tides compete to determine the final spin magnitude of the WR star, and tides may align the stellar spin with the orbital angular momentum. At the end of the WR lifetime, stellar collapse occurs and may result in a supernova (SN).
For convenience, we refer to these stellar collapses as SN regardless of whether they are accompanied by a luminous transient.
Matter and energy may be ejected asymmetrically in this supernova, causing the compact remnant to receive a natal kick \cite{Kalogera2000}. We assume that these natal kicks are isotropic in direction with magnitudes given by a Maxwellian distribution of velocity dispersion $\sigma$. Natal kicks directed out of the initial orbital plane will tilt the orbital angular momentum, and kicks of sufficient strength will unbind the binary. Binaries that survive the kick may have significant spin-orbit misalignment depending on how the kick velocity compares to the orbital velocity prior to the supernova. If the dimensionless spin magnitude of the collapsing stellar core exceeds unity, it must lose angular momentum (potentially accompanied by significant mass loss) to satisfy the Kerr spin limit for black holes and avoid becoming naked singularities \cite{Penrose1969}.

Compact-object formation depends on the mass and metallicity of the progenitor \cite{Heger2003,OConnor2017}. We consider initial masses between 30~M$_\odot$ and 100~M$_\odot$ and assume that all stars in this mass range collapse directly into black holes at the end of their WR lifetimes (no fallback accretion or pair-instability supernovae).
The resulting BBHs inspiral due to GW emission for $\lesssim$~Gyr until coalescence. The BBH masses $m_1$ and $m_2$ and dimensionless spin magnitudes $\rchi_1$ and $\rchi_2$ are constant, while the misalignment angles $\theta_1$ and $\theta_2$ between the BBH spins and the orbital angular momentum evolve on the precession timescale. Nonetheless, the effective aligned spin parameter \cite{Damour2001}
\begin{align} \label{E:ChiEff}
\rchi_{\rm eff} \equiv \frac{\rchi_1 \cos\theta_1 + q\rchi_2 \cos\theta_2}{1 + q}~,
\end{align}
with mass ratio $q \equiv m_2/m_1 \leq 1$, is conserved through the inspiral \cite{Racine2008}. This implies that some information describing the initial BBH spin orientation, which can be used to constrain aspects of the BBH formation described above, is preserved until the BBHs emit GWs at detectable frequencies near merger \cite{Kesden2015,Gerosa2015}.

Eight of the ten BBH mergers in the O1 and O2 LIGO/Virgo catalog \cite{LIGO2019catalog} have posteriors within the 90\% credible interval  $-0.1 \leq \rchi_{\rm eff} \leq 0.1$, although this conclusion depends on the choice of priors \cite{Vitale2017b,Zevin2020}.
This may be explained by BH spins that are either: (1) both small, (2) both high and directed in the orbital plane, or (3) nearly equal in magnitude and opposite in direction. The other two mergers, GW170729 and GW151226, are consistent with $0.11 \leq \rchi_{\rm eff} \leq 0.58$ and $0.06 \leq \rchi_{\rm eff} \leq 0.38$, respectively. Consistently modelling the evolution of the BBH progenitor masses, spin magnitudes, and spin directions is important for understanding BBH formation.

BBH spin precession modulates the GW amplitude and frequency and has recently been detected at marginal significance in LIGO/Virgo events \cite{GW190412,GW190521}. Exploring how this precessional modulation can be produced in the isolated formation channel is the primary motivation for this work. If BHs form with large spins, spin precession is generic for a BBH formed in the dynamical channel where isotropic spin orientations are expected. Various aspects of stellar-binary evolution conspire to suppress spin precession in the isolated channel, but there are still regions in the stellar-binary parameter space that produce precessing BBH systems. We attempt to identify these regions, although we do not predict the distributions of initial parameters as would be provided by genuine population synthesis. Calculating the observability of BBH precession is an open question \cite{OShaughnessy2005,Fairhurst2019}.

\subsection{Executive Summary}
\label{subsec:ExecSum}

\begin{figure*}[!t]
\centering
\includegraphics[width=0.9\textwidth]{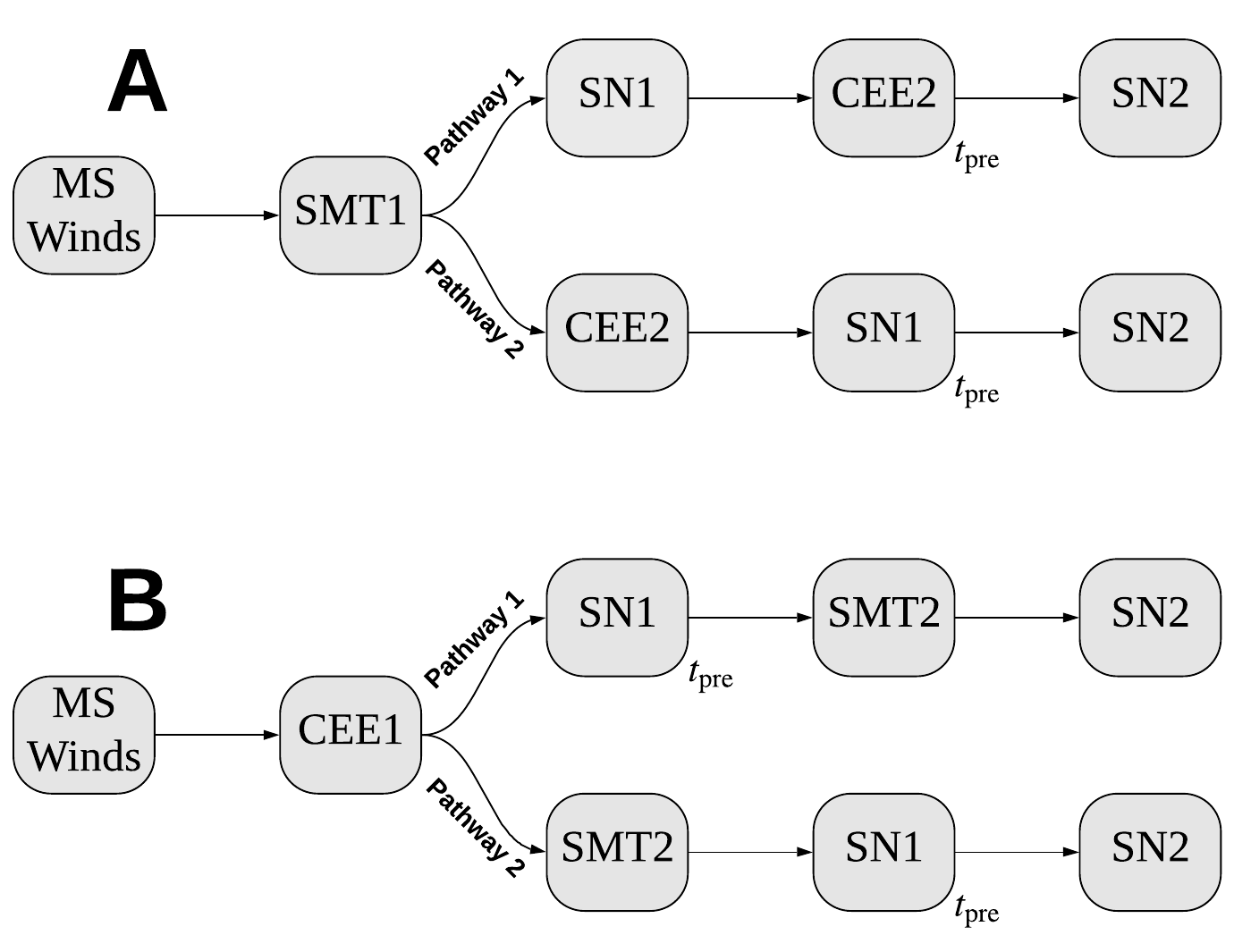}%
\caption{
Diagram of the four evolutionary pathways that we explore. $\it{Top}$: Scenario A, where the primary RLOF leads to stable mass transfer (SMT1), and the secondary RLOF leads to common-envelope evolution (CEE2). $\it{Bottom}$: Scenario B, where the primary RLOF leads to common-envelope evolution (CEE1), and the secondary RLOF leads to stable mass transfer (SMT2). In Pathway 1 (Pathway 2) of each scenario, the primary supernova SN1 occurs before (after) secondary RLOF. We also allow the misaligned spins produced from SN1 to precess prior to the secondary SN (SN2) if the precession timescale following both SN1 and CEE, given by Eq.~(\ref{E:TpreSN}) and indicated by $t_{\rm pre}$ in the diagram, is shorter than the time until SN2.
} \label{F:Diagram}%
\end{figure*}

\begin{table*}[t]
\begin{center}
\def\arraystretch{1.2}
\setlength\tabcolsep{0.5em} 
\begin{tabular}{|c|c|c|c|c||c|c|c|c|}
 \hline
 Pathway & A1 & A2 & B1 & B2 & A1 & A2 & B1 & B2 \\
 \hline
 Coupling & weak & weak & weak & weak & strong & strong & strong & strong \\
 $m_{2,\rm ZAMS}$ (M$_{\odot}$) & 50 & 67 & 50 & 60 & 50 & 67 & 50 & 60 \\
 $a_{\rm ZAMS}$ ($10^3$ R$_{\odot}$) & 6 & 6 & 12 & 12 & 6 & $1.5$ & 12 & 12 \\
 $\sigma$ (km/s) & 30 & 200 & 200 & 200 & 30 & 200 & 200 & 200 \\
 $f_{\rm B}$ & 0.01 & 0.01 & 0.01 & 0.01 & --- & --- & --- & --- \\
 \hline
 \hline
 $\overline{a}_{\rm BBH}$ (R$_{\odot}$) & 19 & 20 & 20 & 16 & 19 & 11 & 20 & 16 \\
 $\overline{m}_{1}$ (M$_{\odot}$) & 23 & 23 & 29 & 20 & 23 & 21 & 29 & 20 \\
 $\overline{m}_{2}$ (M$_{\odot}$) & 19 & 25 & 16 & 19 & 19 & 23 & 16 & 19 \\
 $\overline{\rchi}_{1}$ & 0.15 & 0.15 & 0.7 & 1 & 0.006 & 1 & 0.6 & 1 \\
 $\overline{\rchi}_{2}$ & 0.35 & 0.16 & 0.17 & 0.16 & 0.2 & 1 & 0.004 & 0.005 \\
 $\overline{\cos\theta_{1}}$ & 0.4 & 0.6 & 0.7 & 0.7 & 0.4 & 0.9 & 0.7 & 0.7 \\
 $\overline{\cos\theta_{2}}$ & 0.6 & 0.6 & 0.7 & 0.7 & 0.6 & 0.9 & 0.7 & 0.7 \\
 $f_{\rm bound,SN1}$ & 0.4 & 0.7 & 0.7 & 0.7 & 0.4 & 0.9 & 0.7 & 0.7 \\
 $f_{\rm bound,SN2}$ & 1 & 0.7 & 0.7 & 0.7 & 1 & 0.9 & 0.7 & 0.7 \\
 $\overline{\rchi}_{\rm eff}$ & 0.1 & 0.1 & 0.35 & 0.4 & 0.1 & 0.9 & 0.3 & 0.35 \\
 $\overline{t}_{\rm merge}$ (Gyr) & 3 & 2 & 3 & 2 & 3 & 0.5 & 3 & 2 \\
 $f_{\rm merge}$ & 0.75 & 0.6 & 0.5 & 0.6 & 0.75 & 0.9 & 0.5 & 0.6 \\
 \hline
\end{tabular}
\caption{Typical examples of BBHs with large, misaligned spins produced in the evolutionary pathways depicted in Fig.~\ref{F:Diagram}.  All the binaries listed in this table have an initial metallicity $Z = 0.0002$, a primary ZAMS mass $m_{1,\rm ZAMS} = 70$~M$_{\odot}$, and an accreted fraction $f_{\rm a} = 0.2$ during stable mass transfer.
The first column lists our assumptions, initial parameters and final parameters (evaluated after SN2) of the binaries: core-envelope coupling (strong or weak), secondary ZAMS mass $m_{2,\rm ZAMS}$ (which determines the pathway of evolution), ZAMS separation $a_{\rm ZAMS}$, one-dimensional velocity dispersion $\sigma$, initial fraction of WR breakup spin $f_B$, average BBH semi-major axis $\overline{a}_{\rm BBH}$, average primary (secondary) BH mass $m_1$ ($m_2$), average primary (secondary) BH dimensionless spin $\rchi_1$ ($\rchi_2$), average cosine of primary (secondary) BH misalignment $\overline{\cos\theta_{1}}$ ($\overline{\cos\theta_{2}}$), fraction of binaries that remain bound after each SN $f_{\rm bound}$, average BBH aligned effective spin $\overline{\rchi}_{\rm eff}$, average time until BBH merger $t_{\rm merge}$ and the fraction $f_{\rm merge}$ of binaries that merge within the age of the Universe. The next eight columns show the results of our model.}
\label{T:table1}
\end{center}
\end{table*}

Previous studies of BBHs formed in the isolated channel considered how stellar processes such as core-envelope coupling, stellar winds, and core collapse, and binary processes such as tidal interactions, mass transfer and accretion affected the distribution of BBH masses and spins.  If the BBH spins are initially aligned, the main astrophysical source of spin-orbit misalignment is the recoil kick possibly received by a star undergoing gravitational collapse. Tides or strongly torqued accretion disks may realign the spin with the orbital angular momentum. Typically, the BBH spin orientation is calculated in binary stellar-evolution models either from arbitrary prior distributions \cite{Postnov2018,Sedda2018}, by completely ignoring misalignments \cite{Hotokezaka2017,Qin2018,Fuller2019}, or by post-processing the effects of SN kicks \cite{BK2008,Gerosa2018,Wong2019,Belczynski2020}. Some models address the uncertainties of BH accretion but ignore tidal realignment of the WR spin \cite{BKspins2008,Fragos2010}. 

We use our own model that incorporates the various effects described above to identify regions of parameter space that lead to precessing BBHs. We assume either maximal stellar core-envelope spin coupling, resulting in low natal WR spins, or minimal coupling in which the newly born WR star inherits an initial spin that we parameterize by the fraction $f_{\rm B}$ of its breakup value. We assume that RLOF immediately initiates a mass-transfer event either through common-envelope evolution (CEE), which drastically shrinks the binary separation, or in stable mass transfer (SMT) in which a fraction $f_{\rm a}$ of the donor's envelope is accreted.  If the dimensionless spin of the collapsing WR star exceeds the Kerr spin limit ($\rchi = 1$), we assume that mass loss reduces the BH spin to this limiting value. We consider the two extreme possibilities of negligible mass loss or isotropic mass loss from the stellar surface.

Our prescription for the tidal evolution of the BH progenitor - the WR star - consistently evolves the spin magnitude and misalignment as mass is lost due to winds (see Eqs. (\ref{E:TWspinODE}) and (\ref{E:TWalignODE})). BBH spin-orbit misalignments are determined by supernova (SN) natal kicks (parameterized by $\sigma$). Misalignments from SN kicks are imperative for producing precessing systems in our model. 

We define the primary (secondary) as the initially more (less) massive star. Fig.~\ref{F:Diagram} depicts the two scenarios of binary evolution that we explore defined by whether CEE, required to produce BBHs with sufficiently small separations that merge within the age of the Universe, occurs following the RLOF of the primary or secondary star. Scenario A, in which RLOF of the primary (secondary) leads to SMT (CEE), is expected to dominate for small natal kicks. Scenario B, in which RLOF of the primary (secondary) leads to CEE (SMT), dominates for large natal kicks where CEE is needed to prevent the binaries from becoming unbound following the first kick \cite{Gerosa2018}. These kicks are needed to produce misaligned BBHs in the isolated formation model given our assumption that the initial spins are aligned with the orbital angular momentum $\bf{L}$ \cite{Breivik2017}.

As shown in Fig.~\ref{F:Diagram}, Scenarios A and B are each split into two unique pathways of binary stellar evolution depending on whether the supernova of the primary (SN1) occurs before (Pathway 1) or after (Pathway 2) RLOF of the secondary. Pathway 1 occurs for binaries with ZAMS mass ratio below a transition value $q_{\rm trans}$, where we define the ZAMS mass ratio to be less than unity. This work is the first to systematically explore how these different scenarios and pathways of stellar-binary evolution affect the properties of the BBHs they produce.

The fraction of binaries that remain bound after a SN kick monotonically decreases with $\sigma$ (see Fig.~\ref{F:TiltsFbound}), so we expect Pathways A2, B1, and B2 in which CEE precedes the first natal kick will dominate at large values of $\sigma$. In Pathway A1, a smaller value of $\sigma$ is required for binaries to survive the SN kick that occurs at pre-CEE separations. We stress that the boundary $q_{\rm trans}$ between Pathways 1 and 2 is distinct from the criterion for mass-ratio reversal (MRR), where the primary star evolves into the less massive BH. The occurrence of MRR in Scenario A depends on the fraction $f_{\rm a}$ of the donor's envelope that is accreted during SMT. In Pathway B2, it additionally depends on the amount of mass loss that accompanies the angular-momentum loss needed during core collapse of the primary to preserve the Kerr spin limit on the resulting BH.

Initial separations $a_{\rm ZAMS}$ that allow a binary to survive until BBH formation depend on the interplay of stellar evolution and mass transfer. This implies the existence of a model-dependent "Goldilocks zone" for the production of highly precessing BBHs; properties of typical BBHs produced in this region of parameter space for each pathway are listed in Table~\ref{T:table1}. We chose a low metallicity ($Z = 0.0002$) when preparing this table to reduce the effects of stellar winds and emphasize those of binary evolution.

The early onset of CEE in Scenario B creates the possibility of the secondary filling its Roche lobe before leaving the main sequence and thus destroying the binary \cite{Ivanova2004,BK2007}. This necessitates larger $a_{\rm ZAMS}$ (i.e. $\gtrsim 10{,}000$~R$_{\odot}$) than in Scenario A, as seen in the $a_{\rm ZAMS}$ row of Table~\ref{T:table1}. These wider initial separations imply that only accretion during stable mass transfer can spin up the primary in Scenario B, whereas tidal synchronization may spin up the WR star in Scenario A.

In Scenario A, binaries with initial separations that are too small (i.e. $\lesssim 3{,}000$~R$_{\odot}$) have negligible misalignments due to tidal alignment (contrast the strong and weak A2 columns of Table~\ref{T:table1}). Tides can temporarily preserve or even increase spin misalignment for the large spins that can be attained with minimal core-envelope coupling or from accretion onto a WR star (see Fig.~\ref{F:TidesSep}), but we do not explore this possibility since it only occurs in a finely-tuned portion of the parameter space.

In Pathway A1, significant spin misalignments ($\cos\theta_i \lesssim 0.7$) are possible with modest kicks ($\sigma \gtrsim 30$~km/s). If we assume maximal core-envelope coupling, natal BH spins are very small, i.e. $\rchi \sim 0.001$, unless the initial separation is small enough, i.e. $a_{\rm ZAMS} \lesssim 3{,}000$~R$_{\odot}$, for tidal synchronization to operate during the WR stage (contrast the strong-coupling A1 and A2 columns of Table~\ref{T:table1}). If we assume minimal coupling with $f_{\rm B} \gtrsim 0.01$, the BBHs can have significant spin magnitudes ($\rchi \gtrsim 0.1$) and significant misalignments as shown in the weak coupling columns of Table~\ref{T:table1}. Also, the spin of the secondary BH in Pathway A1 is on average larger than the spin of the primary BH since SN1 produces a spread in the semimajor axis which are then shrunk considerably by circularization in RLOF and by CEE of the secondary prior to the tides acting on the secondary WR star. 

In Pathway A2, since CEE of the secondary occurs before the primary SN kick, a larger value of $\sigma$ ($\gtrsim 150$~km/s) is needed than in A1 to obtain significant misalignments. The spin magnitudes remain small when maximal core-envelope coupling is assumed. Tidal synchronization can produce large spins at small initial separations, but tidal alignment also causes the misalignments to vanish. For a fairly narrow range of binary separations, we can obtain large spins through tidal synchronization without having enough time for complete tidal alignment as shown in the strong coupling column for A2 of Table~\ref{T:table1}. 

If we instead assume minimal coupling, an initial WR spin of $f_{\rm B} \sim 0.01$, and modest kicks, the BBH has both significant spin magnitudes ($\rchi \sim 0.1$) and misalignments for a broad range of initial binary separations $a_{\rm ZAMS}$ as shown in the weak coupling column for A2 of Table~\ref{T:table1}.  We find that $f_{\rm B} \gtrsim 0.1$ produces highly spinning BHs for all masses and metallicities, 
making Pathway A2 a likely source of precessing BBHs.

In Scenario B, significant spin misalignments ($\cos\theta \lesssim 0.7$) also require large kicks ($\sigma \gtrsim 150$~km/s). If we assume maximal core-envelope spin coupling, avoiding RLOF on the MS implies that the secondary spin remains small. Accretion by the primary BH in Pathway B1 or by the primary WR star in Pathway B2 during SMT from the secondary results in a high primary BH spin ($\rchi \gtrsim 0.5$). The amount of the secondary's envelope that is accreted is uncertain.  For Pathway B1, if $f_{\rm a} = 0.2$ as in Table~\ref{T:table1}, then $\rchi \sim 0.6$, and if $f_{\rm a} = 0.5$, then $\rchi \sim 0.8$.  For Pathway B2, $\rchi = 1$ for $f_{\rm a} > 0$ because of the larger specific angular momentum of the gas that is accreted at the stellar surface of the WR star.  If we assume minimal coupling with $f_{\rm B} \gtrsim 0.01$, then both BHs have significant spins.

This paper focuses on the spin magnitudes and misalignments produced in our four pathways of isolated binary stellar evolution. In future work \cite{SteinleFuture1}, we shall use the code \texttt{PRECESSION} \cite{Gerosa2016} to evolve the BBHs that we identify in this work as expected precessing candidates down to the small binary separations at which they produce detectable GWs in order to quantify their precessional properties.

\subsection{Outline of the Paper}

This paper is organized as follows. Sec.~\ref{sec:BinEvoModel} explains our model of binary stellar evolution. Sec.~\ref{sec:Res} presents our results pertaining to the BBH masses and spins, the transition mass ratio $q_{\rm trans}$ between Pathways 1 and 2, and the possibility of mass-ratio reversal. Sec.~\ref{sec:Disc} summarizes these results and discusses their implications. Additional details of our model are described in the Appendices.

\section{Binary-Evolution Prescriptions} 
\label{sec:BinEvoModel}

Studies that evolve binary stars into compact binaries typically adopt one of two strategies. In the first approach, large statistical samples of binary populations are synthesized using fits to simplified stellar-evolution models \cite{BK2008,Belczynski2016,Breivik2017,Belczynski2020,Stevenson2017b,Postnov2018,Wysocki2018,Gerosa2018,Giacobbo2018,Sedda2018,Bogomazov2018,Kruckow2018,Vigna2018,Breivik2020}. In the second approach, numerical models of stellar structure are used to perform more accurate and computationally expensive stellar evolution on a smaller number of systems while binary processes are simplified 
\cite{Hotokezaka2017,Hotokezaka2017Implications,Qin2018,Fuller2019}. Given current computational limitations, both approaches have challenges predicting the complicated dependence of BBH spins on the astrophysical evolution of the BBH progenitors. A recent hybrid approach found BBH spin magnitudes that are consistent with each strategy, but they did not study BBH spin misalignments in depth \cite{Bavera2020}. Rather than utilizing these existing comprehensive models, we simulate binary stellar evolution with simplified stellar-evolution formulae and with binary processes pertinent to the evolution of the spin magnitudes and directions.

\citeauthor{Gerosa2018}~\cite{Gerosa2018} showed that the detection rates of GW events are dominated by binaries that experience a single episode of common-envelope evolution (CEE) (see their Fig.~3). We therefore focus on such formation scenarios, rather than those with zero or two episodes of CEE (Channels II and III of \citeauthor{Neijssel2020}~\cite{Neijssel2020}). When this CEE occurs after the supernova (SN) of the primary, the kick velocity dispersion $\sigma$ must be small to avoid unbinding the binary. Larger values of $\sigma$ are allowed when CEE occurs before the SN of the primary, due to the small post-CEE binary separation.

Inspired by these results, we define two scenarios of binary stellar evolution according to how mass transfer proceeds from Roche-lobe overflow (RLOF). In Scenario A, the initially more massive star (primary) fills its Roche lobe first and loses its envelope by stable mass transfer to the initially less massive star (secondary). The secondary RLOF then leads to CEE. In Scenario B, the roles are switched so that the primary CEE occurs first and then the secondary stably transfers mass to the primary.

For both scenarios, when the ZAMS mass ratio ($q_{\rm ZAMS} \leq 1$) is sufficiently small, the secondary evolves slowly enough that it experiences RLOF after the primary SN at the end of its Wolf-Rayet (WR) lifetime. We call this possibility Pathway 1, and Pathway 2 is when RLOF of the secondary precedes the primary SN. The sequence of events in these four different evolutionary Pathways (A1, A2, B1, B2) are depicted in Fig.~\ref{F:Diagram}. The analysis of \citeauthor{Gerosa2018}~\cite{Gerosa2018} suggests that Pathway A1 dominates the event rate for low $\sigma$, while some combination of the remaining three, all of which have CEE prior to the primary SN, dominate for high $\sigma$.

Our initial stellar binaries are parameterized by the ZAMS mass $m_{\rm ZAMS}$ of each star, the metallicity $Z$, the initial binary separation $a$, the fraction $f_{\rm a}$ of the donor envelope accreted during SMT, the SN kick strength $\sigma$, and the initial spin magnitude of each star defined as a fraction $f_{\rm B}$ of the breakup spin at which the star's centrifugal acceleration exceeds its self-gravity.  For maximal core-envelope coupling, $f_{\rm B}$ is chosen at ZAMS and for minimal coupling, $f_{\rm B}$ is chosen for the natal WR star.  For simplicity, we assume both members of the binary have the same $f_{\rm B}$ and $Z$. The primary (secondary) is indexed by the subscript 1 (2), so the ZAMS mass ratio is $q_{\rm ZAMS} \equiv m_{2, {\rm ZAMS}}/m_{1, {\rm ZAMS}} \leq 1$. Depending on the value of $f_{\rm a}$, stable mass transfer can cause mass-ratio reversal (MRR) in which the primary evolves into a BH less massive than the secondary. We simulate single stellar evolution by directly implementing the formulae of \citeauthor{Hurley2000}~\cite{Hurley2000} and incorporate binary interactions.

\subsection{Initial Stellar Spin} 
\label{subsec:StarSpin}

The initial dimensionless spin $\rchi_{\rm 0} = f_{\rm B} \rchi_{\rm B}$ of each star in our binary is parameterized with the fraction $f_{\rm B}$ of the dimensionless breakup spin,
\begin{align} \label{E:Break}
\rchi_{\rm B}  = \frac{c |\bf{S}_{\rm B}|}{G m^2} = \frac{c r_{\rm g}^2 R^2 \Omega_{B}}{G m} = r_{\rm g}^2 \left( \frac{c^2R}{Gm} \right)^{1/2},
\end{align}
where $m$ is the mass of the star, $R$ is the stellar radius, $r_{\rm g}$ is the radius of gyration, and $\Omega_{\rm B}$ is the breakup angular frequency. A numerical fit \cite{Hurley2000} to observations of the initial mean-equatorial velocity of high-mass main-sequence stars \cite{Lang1992} suggests that, with ZAMS radius given by Tout~{\it et al.}~\cite{Tout1996}, $f_{\rm B}$ ranges from 0.004 at high mass ($m_{\rm ZAMS} = 100$~M$_{\odot}$) and metallicity ($Z = 0.02$) to 0.1 at low mass ($m_{\rm ZAMS} = 10$~M$_{\odot}$) and metallicity ($Z = 0.0002$). There is no simple formula for the initial spin of WR stars \cite{Crowther2007}. We treat $f_{\rm B}$ as a free parameter to explore the affects of uncertainty in the initial spin magnitudes on the final BBH spins.

Another uncertainty is the extent to which the rotation of the stellar core is coupled to that of the stellar envelope \cite{Sweet1950,Zahn1992,Maeder1998,Maeder2012,Meynet2013,Georgy2013}. Detailed stellar-evolution codes have been used to explore the effects of core-envelope spin coupling on the spin magnitudes of compact binaries \cite{Postnov2016,Belczynski2020,Qin2018,Fuller2019}, but the results are model dependent and the strength of the coupling likely depends on the stage of nuclear evolution \cite{Tayar2019}. Since this issue remains elusive, we simply consider two extreme possibilities for core-envelope spin coupling: $maximal$ coupling in which the entire star is a rigid rotator so that the initial spin of each star, $\rchi_{\rm 0}$, is chosen at ZAMS, and $minimal$ coupling in which the stellar core rotates independently from its envelope so that the initial spin of each star, $\rchi_{\rm 0}$, is chosen at zero-age Wolf-Rayet (ZAWR). The effect of core-envelope spin coupling on stellar misalignment evolution is very uncertain, so we assume that the core and envelope share the same misalignment at all times.

\begin{figure}[t!]
  \centering
  \includegraphics[width=0.49\textwidth]{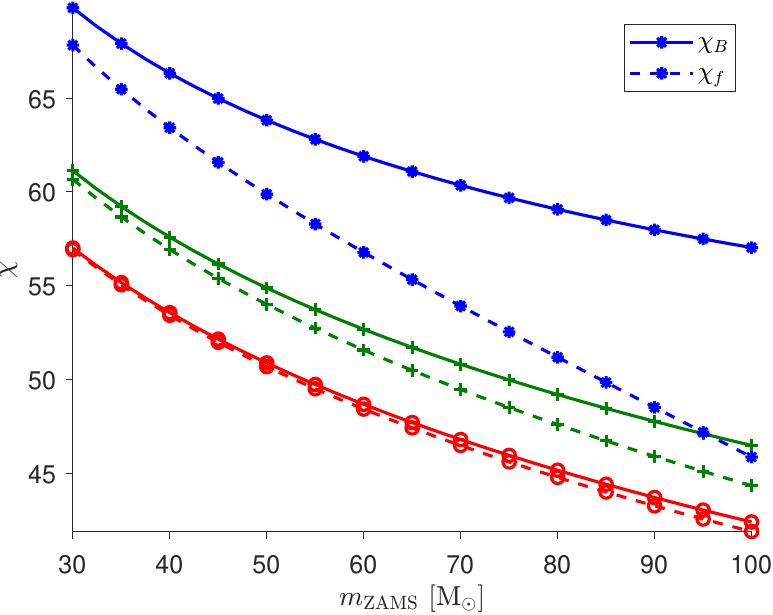}
   \caption{The dimensionless spin $\rchi_{\rm B}$ as a function of zero-age main-sequence stellar mass $m_{\rm ZAMS}$.  The solid (dashed) lines correspond to spins before (after) mass loss due to winds over the MS lifetime. Red-circle, green-cross and blue-star curves correspond to metallicities $Z = 0.0002$, $0.002$, and $0.02$ respectively.
   } \label{F:ZAMSspin}
\end{figure}

Fig.~\ref{F:ZAMSspin} shows the dimensionless spin $\rchi$ as a function of ZAMS mass $m_{\rm ZAMS}$ at the beginning and end of the MS for stars that begin the ZAMS spinning at breakup. The ZAMS stellar radius $R$ increases less than linearly as a function of $m_{\rm ZAMS}$, implying that $\rchi_{\rm B}(m_{\rm ZAMS})$ monotonically decreases (see Eq.~(\ref{E:Break})). As the metallicity $Z$ increases, $\rchi_{\rm B}$ increases since $R$ increases. Winds are more effective at higher metallicity, implying a greater gap between the solid and dashed curves (which denote the beginning and end of the MS) for high values of $Z$.

\begin{figure}[t!]
  \centering
      \includegraphics[width=0.49\textwidth]{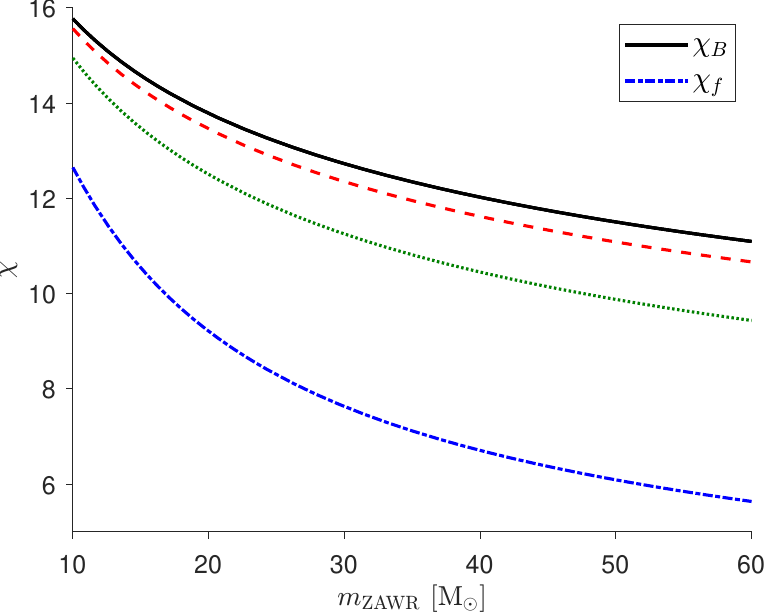}
   \caption{The dimensionless spin $\rchi$ of Wolf-Rayet (WR) stars as a function of their zero-age stellar mass $m_{\rm ZAWR}$. The solid black curve corresponds to the beginning of the WR stage for stars spinning at breakup, while the dashed red, dotted green and dot-dashed blue curves correspond to spins after mass loss due to winds over the WR lifetime (Eq.~(\ref{E:WRlifetime})) for metallicities $Z = 0.0002$, $0.002$, and $0.02$ respectively.} \label{F:WRSpin}
\end{figure}

Fig.~\ref{F:WRSpin} displays the dimensionless spin $\rchi$ of Wolf-Rayet (WR) stars. WR stars that begin the WR stage spinning at breakup have a dimensionless spin $\rchi_{\rm B,WR} \approx 15.8\left(m_{\rm WR}/10~{\rm M}_{\odot}\right)^{-0.2}$ as a function of their mass $m_{\rm WR}$ shown by the solid black curve. High-mass WR stars have much smaller radii than MS stars of comparable mass and correspondingly smaller moments of inertia and breakup spin angular momenta. We assume that all WR stars have zero-age radii $R_{\rm WR}$ given by Eq.~(\ref{E:WRradius}) independent of metallicity. As on the MS, the stronger winds at higher $Z$ lead to smaller spins at the end of the WR stage.

The initial spins of WR stars depend on the choice of core-envelope coupling during their evolution. In the case of maximal coupling, the initial spin that the WR star inherits from losing its envelope in RLOF via Eq.~(\ref{E:IsoSpin}) is small. For minimal coupling, MS stars can be differentially rotating and therefore the initial WR spins depend on the value of $f_{\rm B}$ chosen at the start of the WR stage.

Although much progress has been achieved in explaining the formation of binary stars and multiple systems, many uncertainties remain \cite{Tohline2002,Goodwin2007,Duchene2013}. Recent studies of stellar clusters suggest that the initial misalignments between binary star spins and their orbital angular momentum are very small \cite{Corsaro2017}, but turbulent and dynamical processes may produce misaligned spins \cite{Offner2016}. If stellar spins in isolated binaries are initially isotropic, this isotropy is likely retained by the final BBHs unless binary interactions like tides or accretion realign the spins \cite{Postnov2018}, but we do not explore this here. Since we are interested in whether natal kicks can generate significant misalignments, we make the conservative assumption that the initial stellar spins are aligned with the orbital angular momentum.

\subsection{Mass Transfer}
\label{subsec:MT}

A high-mass star generally expands as it ages and may fill its Roche lobe if the binary separation is sufficiently small. Roche lobe overflow (RLOF) causes a star to transfer mass and angular momentum to its companion or out of the system entirely. In our model, we assume RLOF causes a star to lose its entire envelope. The physics of binary mass transfer has long been studied, but it is still poorly understood \cite{Podsiadlowski2012,Vanbeveren2018}. A fully consistent implementation of RLOF and mass transfer is difficult to obtain \cite{Campos2018}.

The Roche lobe (RL) radius of an object in a binary is the distance from the center of the object within which material remains gravitationally bound to it. For a star with mass $m_1$ that is synchronized in a circular binary, the RL radius is approximated \cite{Eggleton1983} by,
\begin{align} \label{E:RL}
R_{\rm RL} = \frac{0.49}{0.6 + q^{2/3}\ln(1 + q^{-1/3})} a_i
\end{align}
\noindent
where $q = m_2/m_1$ and $a_i$ is the binary separation at the start of RLOF.

Our ZAMS binaries are circular, so RLOF and the core collapse of the primary (SN1) each occur in a circular orbit. In Pathway 1 (see Fig.~\ref{F:Diagram}), the secondary experiences RLOF after the kick in SN1 has generated orbital eccentricity. However, Eq.~(\ref{E:RL}) is only valid for circular orbits. We approximate the RL radius on eccentric orbits by replacing the binary separation $a_i$ in Eq.~(\ref{E:RL}) with $a_i(1 - e_i^2)$, the binary separation of the circular orbit with the same orbital angular momentum. This approximation is unnecessary in Pathway 2 where RLOF always occurs before either of the core-collapse events. 

The outcome of RLOF depends on the stability of the mass transfer, which is determined by how the stellar radius and Roche lobe change relative to the mass transfer rate \cite{Soberman1997}. Population-synthesis codes typically implement mass transfer using a formalism \cite{Hurley2002,Postnov2014} where time-dependent mass-radius exponents describe the response of the stellar radius and RL to mass loss. Since we model mass transfer as a discrete event rather than evolving the stellar masses and radii dynamically, we must adopt a different approach. We find the time at which RLOF occurs (i.e. $R_\ast(t_{\rm RLOF}) = R_{\rm RL}$) and then calculate the stellar core mass at that time using the formulae of \citeauthor{Hurley2000}~\cite{Hurley2000}. This procedure provides the core mass and the binary separation after a mass transfer event and is explained in greater detail in Appendix~\ref{app:RLOF}. We assume that after a donor star loses its envelope, its core emerges as a WR star.

In unstable mass transfer, the mass-transfer rate is large enough that the donor star's RL radius shrinks faster than its stellar radius. Common-envelope evolution (CEE) ensues when the companion's accretion rate cannot keep pace with the mass transfer rate. Energy and momentum are transferred to the envelope by the viscous friction of the binary's motion through the envelope on the very short dynamical timescale of the donor. This transfer either ejects the envelope or merges the binary \cite{Ivanova2013}, and the orbital separation shrinks by several orders of magnitude even in systems that survive.

The result of CEE depends on the stage of nuclear evolution of the donor star. For instance, donors on the MS and Hertzsprung gap (HG) are argued to prematurely merge the binary during CEE since they lack a steep core-envelope density gradient \cite{Ivanova2004,BK2007}. Several authors have explored this with population synthesis, e.g. \cite{Dominik2012,Pavlovskii2017,Mapelli2018,Bavera2020b}, but the physics of CEE remains highly uncertain \cite{Ivanova2013,Postnov2014,Schroder2020}. To more broadly explore the parameter space of BBH spins and misalignments, we allow binaries to survive CEE if the donor fills its Roche lobe on the HG \cite{deMink2015} or later stellar stages. The ratio of the binary separation $a_f$ after CEE to its initial value $a_i$ follows from conservation of the sum of the stellar envelope's binding energy and the binary's orbital energy \cite{Heuvel1976,Paczynski1976,Tutukov1979,Webbink1984},
\begin{align}\label{E:CE}
\frac{a_f}{a_i} = \frac{m_{f, \rm D}}{m_{i, \rm D}}\left( 1 + \frac{2a_i}{\lambda R_{\rm RL}}\frac{m_{i, \rm D} - m_{f, \rm D}}{m_{\rm A}}
\right)^{-1}.
\end{align}
Here $m_{i, \rm D}$ and $m_{f, \rm D}$ are the initial and final masses of the donor star, $m_{\rm A}$ is the mass of the companion which remains fixed during CEE \cite{King1999,MacLeod2015,Chamandy2018}, and $\lambda$ is a dimensionless parameter of order unity which depends on the structure and mass of the donor star. We use a numerical fit for $\lambda(R_{\rm RL})$ first presented in Eq.~(A31) of \citeauthor{Gerosa2013}~\cite{Gerosa2013} based on the results of Dominik {\it et al.}~\cite{Dominik2012}. We assume that CEE is too brief for significant accretion or change in the spin directions to occur.

In stable mass transfer (SMT), the mass transfer rate is low enough that the donor's RL radius does not shrink uncontrollably relative to the stellar radius. We assume that some fraction $f_{\rm a}$ of the envelope is accreted by the companion. The value of $f_{\rm a}$ is uncertain and can range from fully conservative ($f_{\rm a} = 1$) to fully non-conservative ($f_{\rm a} = 0$) \cite{Meurs1989}. Following previous work \cite{Belczynski2020,Stevenson2017}, we consider the two values: $f_{\rm a} = 0.2$ and $f_{\rm a} = 0.5$. The effects of this choice on whether systems experience mass-ratio reversal are explored in Sec.~\ref{subsec:MRR}. SMT generally occurs over the donor's thermal timescale and changes the binary separation due to isotropic re-emission of the donor's envelope \cite{Postnov2014} according to,
\begin{equation} \label{E:SMT}
\frac{a_f}{a_i} = \frac{m_{i, \rm D} + m_{i, \rm A}}{m_{f, \rm D} + m_{f, \rm A}} \left( \frac{m_{i, \rm D}}{m_{f, \rm D}} \right)^2 e^{-2(m_{i, \rm D} - m_{f, \rm D})/m_{i, \rm A}}~.
\end{equation}

RLOF and the subsequent mass transfer determines the range of initial binary separations allowed in the BH formation scenarios we consider. If the binary stars are too widely separated initially, the RLOF criterion given by Eq.~(\ref{E:RL}) will not be satisfied implying the binary separation will never be reduced by CEE given by Eq.~(\ref{E:CE}) to a value small enough for the BBHs to merge within the age of the Universe. However, if the initial separation is too small, one or both of the stars may fill their Roche lobes during the MS or WR stages leading to the destruction of the binary. RLOF is discussed in greater detail in Appendix~\ref{app:RLOF}.

As a donor star's envelope is lost during mass transfer, it will carry away angular momentum and reduce the spin of the emergent stellar core by an amount that depends on the strength of core-envelope coupling. To avoid introducing further uncertainties (see Sec.~\ref{subsec:StarSpin}),
we consider two limiting cases. For maximal core-envelope coupling, we assume the envelope is shed isotropically in a spherical shell so that conservation of angular momentum yields a zero-age Wolf-Rayet (ZAWR) dimensionless spin,
\begin{equation} \label{E:IsoSpin}
\rchi_f = \rchi_i \Big(\frac{m_f}{m_i} \Big)^{2/3r_{\rm g}^2 - 2}~.
\end{equation}
This equation was derived assuming that the stellar radius and the radius of gyration $r_{\rm g}$ remain constant at their WR values during mass transfer, analogous to wind-driven mass loss.  This is a conservative estimate of the angular momentum loss, since the larger stellar radii prior to RLOF imply that a given amount of mass will carry away even more angular momentum. For $r_{\rm g}^2 = 0.075$ \cite{Kushnir2016}, these ZAWR spins are small ($\rchi_f \sim 0.01$) for metallicities $Z = 0.02$ and 0.002, and even smaller ($\rchi_f \sim 0.001$) for lower metallicity $Z = 0.0002$ because of the smaller radii and thus smaller dimensionless spins $\rchi_i$ of these stars at RLOF as shown in Fig.~\ref{F:ZAMSspin}. In the case of minimal core-envelope coupling, the core's spin is unaffected by the loss of its envelope and is freely specified after mass transfer by its fraction $f_{\rm B}$ of the WR breakup spin given by Eq.~(\ref{E:Break}).

\subsection{Tides and Winds} 
\label{subsec:TidesWinds}

\subsubsection{Tides}

Tidal forces in a close, detached binary system drive the exchange of kinetic energy and angular momentum between the rotation of the components and their orbital motion. The binary approaches an equilibrium state of minimum kinetic energy where the orbit is circular, and the spins are synchronized and aligned with the orbital angular momentum. The strength of the tidal interaction is determined largely by the orbital separation and by the efficiency of the mechanisms that dissipate kinetic energy \cite{Zahn2008}. Various dissipation mechanisms misalign the tidal bulge relative to the line connecting the two stars' centers, which produces a torque as angular momentum is exchanged between the orbit and the rotation of the star. Two dissipation mechanisms are typically considered: (1) viscous/turbulent dissipation (convective damping) acting on the equilibrium tide, and (2) radiative damping acting on the dynamical tide \cite{Zahn1975}.

We only consider tides acting on Wolf-Rayet (WR) stars after CEE has occurred, since the tidal torque is negligible at the very large pre-CEE separations and since stable mass transfer alone shrinks the separation insufficiently \cite{Bavera2020b}. Tidal evolution depends on the orbital eccentricity of the binary, but in our model tides only operate after circularization due to RLOF (Pathway 1) or after the SN of the primary has produced modest eccentricity ($e < 0.4$) (Pathway 2). For these reasons, we assume circular orbits in our tidal evolution. 

WR stars have radiative envelopes and convective cores implying that radiative damping of the dynamical tide is the dominant mechanism of tidal dissipation \cite{Zahn1977}. However, we assume for simplicity that tidal dissipation is given by that of an equilibrium tide with a constant time lag in the weak-friction approximation, which leads to the following synchronization, alignment, and circularization timescales \cite{Hut1981} for tides acting on a star of mass $m$ due to its companion of mass $m_{\rm c}$ in a circular binary:
\begin{subequations} \label{E:timescales}
\begin{align}
t_{\rm sync} &= \left| \frac{\Omega_{\rm orb} - \Omega}{\dot{\Omega}} \right| = \frac{Tr_{\rm g}^2}{3kq_{\rm t}^2} \left( \frac{a}{R} \right)^6 \label{E:tsync1} \\
t_{\rm align} &= \left| \frac{\theta}{\dot{\theta}} \right| = \left|
\frac{\Omega_{\rm orb}}{\Omega} - \frac{1}{2} (1 - \eta) \right|^{-1} t_{\rm sync} \label{E:talign1} \\
t_{\rm circ} &= \left| \frac{e}{\dot{e}} \right| = \frac{q_{\rm t}}{9r_{\rm g}^2(1 + q_{\rm t})} \left|1 - \frac{11}{18} \frac{\Omega}{\Omega_{{\rm orb}}}\right|^{-1} \left(\frac{a}{R}\right)^2
t_{\rm sync}
\end{align}
\end{subequations}
where $k$ is the dimensionless apsidal motion constant proportional to the quadrupole moment raised on the WR star \cite{Lecar1976}, $q_{\rm t} \equiv m_{\rm c}/m$ is the ratio of the companion and WR masses, $R$ is the WR radius [Eq.~(\ref{E:WRradius})], $\Omega$ is the rotational frequency of the WR star, $\Omega_{\rm orb} = (GM/a^3)^{1/2}$ is the orbital frequency,
\begin{align}\label{E:eta}
\eta &= r_{\rm g}^2 \left(\frac{m + m_{\rm c}}{m_{\rm  c}}\right)
\left(\frac{R}{a}\right)^2 \left(\frac{\Omega}{\Omega_{\rm orb}}\right) \notag \\
&= \frac{f_{\rm B,WR} r_{\rm g}^2}{q_{\rm t}} \left[ \frac{R(1 + q_{\rm t})}{a} \right]^{1/2}
\end{align}
is the ratio of the rotational angular momentum of the WR star to the orbital angular momentum, $f_{\rm B,WR} = \Omega/\Omega_{\rm B}$ is the ratio of the rotational frequency of the WR star to its value at breakup, and
\begin{equation} \label{E:tidalT}
T = \left| \frac{\Omega - \Omega_{\rm orb}}{\phi_{\rm lag}\Omega_{\rm B}^2} \right|
\end{equation}
is the time on which significant tidal evolution changes the orbit.  In Eq.~(\ref{E:tidalT}), $\phi_{\rm lag}$ is the angle between the tidal bulge raised on the WR star and the separation vector of the binary. For the radiative damping of the dynamical tide relevant to WR stars,
\begin{align} \label{E:kT}
\frac{k}{T} &= \num{1.9782e4}~{\rm yr}^{-1} (1 + q_{\rm t})^{5/6} E_2  \notag \\
& \quad \times \left( \frac{m}{\rm M_\odot} \right)^{1/2} \left( \frac{R}{\rm R_\odot} \right) \left( \frac{a}{\rm R_\odot} \right)^{-5/2}
\end{align}
after accounting for the typo in Eq.~(42) of \cite{Hurley2002}.

The value of the dimensionless tidal torque constant $E_2$, first introduced by Zahn (1975) \cite{Zahn1975}, is uncertain because of its dependence on the stellar core radius \cite{Siess2013,Kushnir2016,Qin2018}. For simplicity, we follow the prescription presented in \cite{Hurley2002} where
\begin{align} \label{E:E2}
E_2 = \num{1.592 e-9} \left( \frac{m}{\rm M_\odot} \right)^{2.84}~.
\end{align}

\begin{figure}[t!]
\centering
  \includegraphics[width=0.48\textwidth]{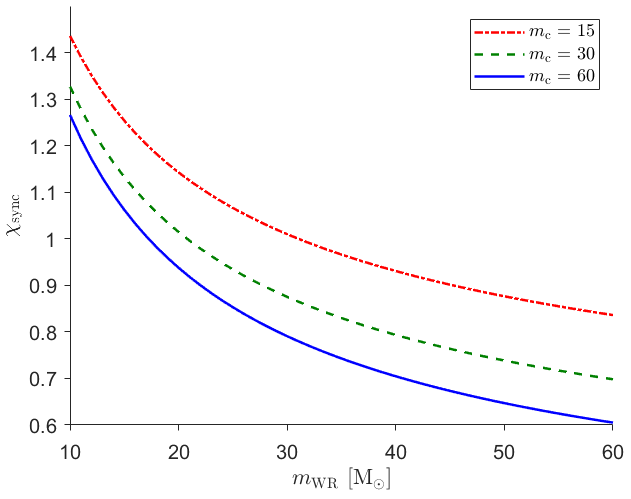}
   \caption{The dimensionless spin as a function of mass of a WR star at the widest separation that tidal synchronization can occur.  The style of the lines correspond to different values of $m_{\rm c}$: The dot-dashed red line, dashed green line, and solid blue line correspond to companion masses $m_{\rm c} = 15,~30$, and $60$~M$_{\odot}$, respectively. These spins are well below the WR breakup spin shown in Fig.~\ref{F:WRSpin}.}\label{F:Chisync}
\end{figure}

Substituting $r_{\rm g}^2 = 0.075$ \cite{Kushnir2016}, Eqs.~(\ref{E:kT}) and (\ref{E:E2}), and the large-mass limit of the WR mass-radius relation (in which Eq.~(\ref{E:WRradius}) implies $R \propto m^{0.6}$) into Eq.~(\ref{E:tsync1}) yields,
\begin{align} \label{E:Tsync}
t_{\rm sync} = \frac{17.8~{\rm Myr}}{(1 + q_{\rm t})^{5/6}q_{\rm t}^2} \left(\frac{a}{\rm R_{\odot}}\right)^{17/2}\left(\frac{m}{\rm M_{\odot}}\right)^{-7.54} 
\end{align}
where the exponent of the WR mass arises from $m^{-[7(0.6) + 0.5 + 2.84]} = m^{-7.54}$. The rotation of the WR will synchronize ($\Omega = \Omega_{\rm orb}$) if its lifetime $t_{\rm WR}$ is longer than this synchronization timescale, which occurs for binary separations less than
\begin{align} \label{E:async}
\frac{a_{\rm sync}}{\rm R_{\odot}} &= 4.77\left[\frac{t_{\rm WR}}{0.3~{\rm Myr}}(1 + q_{\rm t})^{5/6}q_{\rm t}^2\left(\frac{m}{10~\rm M_{\odot}}\right)^{7.54}\right]^{2/17}
\end{align}
where $0.3$~Myr is a typical WR lifetime given by Eq.~(\ref{E:WRlifetime}). The dimensionless WR spin $\rchi_{\rm sync}$ at the synchronization separation $a_{\rm sync}$ as a function of WR mass $m$ for three different companion masses $m_{\rm c}$ is shown in Fig.~\ref{F:Chisync}. If the binary separation is small enough for tidal synchronization to occur ($a \leq a_{\rm sync}$), Fig.~\ref{F:Chisync} suggests that tides are capable of producing a highly spinning BH progenitor.

\begin{figure}[!t]
\centering
  \includegraphics[width=0.48\textwidth]{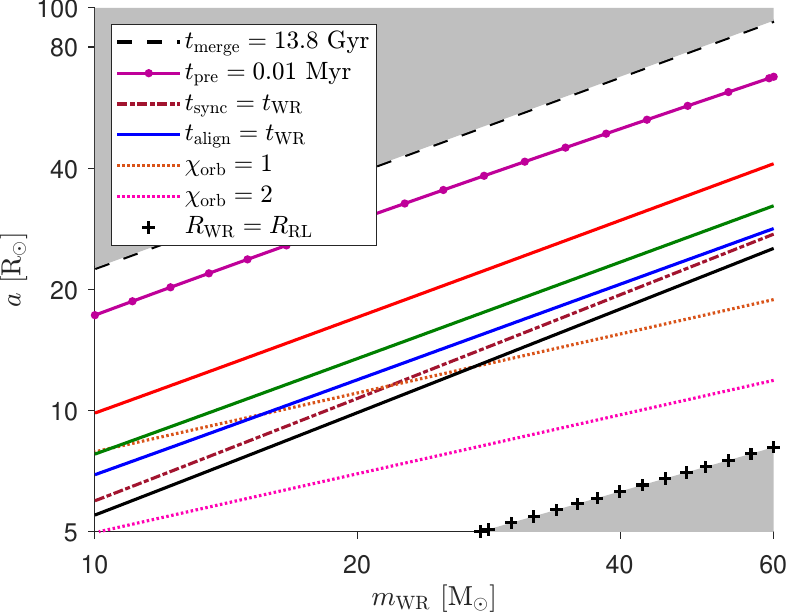}
   \caption{Separations at which various processes operate as a function of the mass of a WR star in a binary with a companion that has a 10\% larger mass.  The dashed black, dotted purple, and dash-dotted brown lines are the upper bounds on binaries that: (1) merge through GW emission in less than the age of the Universe, (2) have spins that precess with a period of 0.01 Myr comparable to the time between SN, and (3) tidally synchronize the WR spin.  The solid lines are upper bounds on tidal alignment for WR stars with dimensionless spins of 0.01, 0.1, and 0.3 (red, green, and blue) and synchronized with the orbit (black). The dotted brown and pink lines show WR stars with synchronized dimensionless spins of 1 and 2. The black crosses show the lower bound on binaries that avoid destroying the WR star through RLOF. The shaded gray regions above the dashed line and below the crosses do not produce observable GW sources.} \label{F:tidaltimes}
\end{figure}

We show the approximate synchronization separation $a_{\rm sync}$ given by Eq.~(\ref{E:async}) as a function of WR mass with the dot-dashed brown line in Fig.~\ref{F:tidaltimes}. The solid lines in this figure show the binary separations below which the WR spin is aligned with the orbital angular momentum ($t_{\rm align} < t_{\rm WR}$) for various WR spin magnitudes.  As indicated by Eq.~(\ref{E:talign1}), larger initial spins take longer to align and therefore can only be aligned within the WR lifetime at smaller separations. Almost all WR spins acquired through tidal synchronization (below the dot-dashed brown line) are also aligned (below the solid black line). Tides may briefly increase the misalignments of large spins (see Fig.~\ref{F:TidesSep} in the Appendix), but we neglect that possibility here as it only occurs in a narrow portion of parameter space.

To determine the relevance of tidal synchronization and alignment to GW sources, we must identify the allowed range of initial binary separations. The upper bound is determined by the requirement that the BBHs merge within the lifetime of the Universe.
The time to merger for circular orbits, which is an upper bound on the time to merger for eccentric orbits \cite{Peters1964}, is
\begin{align} \label{E:merge}
t_{\rm merge} &= \frac{5(1+q_t)^2}{256q_t} \frac{GM}{c^3} \left( \frac{a}{GM/c^2} \right)^4 \notag \\
&\approx 602~{\rm Myr}\frac{(1+q_t)^2}{4q_t} \left( \frac{M}{M_\odot} \right)^{-3} \left( \frac{a}{R_\odot} \right)^4
\end{align}
where $M$ is the total mass. Setting $t_{\rm merge}$ equal to 13.8~Gyr \cite{Planck2016} and solving for the binary separation $a$ yields the dashed black line in Fig.~(\ref{F:tidaltimes}).  The lower bound on the allowed range of initial separations is determined by the requirement that the WR star avoids RLOF until it can collapse into a BH. Setting Eq.~(\ref{E:RL}) for the RL radius equal to Eq.~(\ref{E:WRradius}) for the WR radius and solving for the separation $a$ yields the black crosses in Fig.~(\ref{F:tidaltimes}).

The upper bounds for tidal synchronization and alignment fall within this allowed range. The solid colored lines show that WR stars which acquire spins through other means (such as accretion or a weakly coupled core) can avoid tidal alignment. The dotted orange and pink lines are contours of constant dimensionless spin $\rchi = 1$ and $\rchi = 2$ for synchronized systems ($\Omega = \Omega_{\rm orb}$).  The portions of these lines below the dot-dashed brown synchronization line show that tides can produce large BH spin, which can subsequently become misaligned by the recoil in the SN of the secondary.

We also estimate the upper bound on the binary separation of systems that experience differential spin precession between the primary and secondary SN. Setting the differential precession time $t_{\rm pre}$ given by Eq.~(\ref{E:TpreSN}) equal to an estimate of the time between the SN ($\approx 0.01$~Myr) and solving for the binary separation $a$ yields the dotted purple line in Fig.~\ref{F:tidaltimes}. Differential precession is important because it allows the misalignment angles $\theta_1$ and $\theta_2$ to differ even in the absence of tidal alignment.

\subsubsection{Winds}

Winds from a hot, luminous star's surface are best explained by the line-driven model \cite{Lamers1999} in which the wind mass-loss rate is an increasing function of stellar mass and metallicity \cite{Abbott1982,Shimada1994,Vink2001}. Although winds occur throughout the star's life, for simplicity we only consider them during the main-sequence (MS) and WR stages. Winds are important for stellar spin evolution, because the fractional change in the angular momentum of a WR star due to a wind launched from its surface is a factor $r_g^{-2} \simeq 13$ larger than the fractional change in its mass. We ignore the effects of magnetic fields which would increase the specific angular momentum of the wind. 

For a MS star of mass $m$, we use the the $Z$-modified Neiuwenhuijen \& de Jager mass-loss rate
\begin{align} \label{E:NJ}
\dot{M}_{\rm NJ} &= -\num{9.6e-15}~{\rm M_\odot}/{\rm yr} \left(\frac{Z}{\rm Z_\odot} \right)^{1/2} \left( \frac{R}{\rm R_\odot} \right)^{0.81} \notag \\ 
& \quad \times \left( \frac{L}{\rm L_\odot} \right)^{1.24} \left( \frac{m}{\rm M_\odot} \right)^{0.16},
\end{align}
\noindent
where $R$ is the MS radius and $L$ is the MS luminosity \cite{Hurley2000}.
To compute the amount of mass lost over the MS lifetime, we integrate Eq.~(\ref{E:NJ}) while accounting for the time-dependent stellar radius and luminosity. We ignore winds during the brief HG and CHeB stages since the large radii during these stages imply that Eq.~(\ref{E:NJ}) overestimates the mass loss \cite{Smith2014,Renzo2017}.

The rate at which WR stars lose mass due to winds is crucial in determining whether a neutron star or black hole forms for a given mass and metallicity \cite{Pols2002}. We calculate the final WR mass and spin using the recent mass-loss rate of Vink \cite{Vink2017},
\begin{align} \label{E:V}
\dot{M}_{\rm V} = -\num{e-13.3}~{\rm M_\odot}/{\rm yr} \left(\frac{Z}{\rm Z_\odot} \right)^{0.61} \left( \frac{L}{\rm L_\odot} \right)^{1.36},
\end{align}
with a mass-dependent WR luminosity \cite{Hurley2000}.

We assume wind mass loss is isotropic \cite{Hotokezaka2017} implying that the star is spun down according to Eq.~(\ref{E:IsoSpin}). The radius of gyration generally depends on time, but for simplicity we assume for MS stars $r_{\rm g}^2 = 0.2$ \cite{Motz1952,Claret1989} and for WR stars $r_{\rm g}^2 = 0.075$ \cite{Kushnir2016}. We do not consider accretion by the companion of mass lost due to winds.

\subsubsection{Wolf-Rayet Spin Evolution}

Tides and winds compete to determine the final WR spin. Tides dominate at small separations due to the strong separation dependence of the tidal synchronization timescale given by Eq.~(\ref{E:tsync1}), while winds dominate at large separations. Assuming constant binary separation (since $t_{\rm sync} \ll t_{\rm circ}$) and radius, the change in the dimensionless spin and the spin-orbit misalignment angle per unit change in mass is
\begin{subequations}
\begin{align} 
\frac{d\rchi}{dm} &= \left( \frac{d\rchi}{dm} \right)_{\rm tid,L} + \left( \frac{d\rchi}{dm} \right)_{\rm w} \notag \\
\label{E:TWspinODE}
&= \frac{\rchi_{\rm orb} - \rchi}{\dot{m} t_{\rm sync}}  + \left( \frac{2}{3 r_{\rm g}^{2}} - 2 \right) \frac{\rchi}{m}, \\
\label{E:TWalignODE}
\left( \frac{d\theta}{dm} \right)_{\rm tid,L} &= -\frac{\theta}{\dot{m} t_{\rm sync}} \left[ \frac{\rchi_{\rm orb}}{\rchi} - \frac{1}{2} \left( 1 - \eta \right) \right],
\end{align}
\end{subequations}
where
\begin{equation} \label{E:chi_orb}
\rchi_{\rm orb} = r_{\rm g}^2 \left[ \left( \frac{m + m_c}{m} \right) \left( \frac{R}{Gm/c^2} \right) \left( \frac{R}{a} \right)^3 \right]^{1/2}
\end{equation}
is the dimensionless spin of the WR star at synchronization.  The first term in Eq.~(\ref{E:TWspinODE}) and Eq.~(\ref{E:TWalignODE}) are the linearized tidal-evolution equations given by Eqs.~(11) and (13), respectively, of Hut (1981) \cite{Hut1981}. The second term in Eq.~(\ref{E:TWspinODE}) is derived assuming isotropic winds in the co-rotating frame.

In the limit of large $m$, $L \propto m^{1.25}$ \cite{Hurley2000} implying according to Eq.~(\ref{E:V}) that $\dot{m} = \dot{M}_{\rm V} \propto m^{1.7}$ during the WR stage.  This allows us to integrate the coupled Eqs.~(\ref{E:TWspinODE}) and (\ref{E:TWalignODE}) to determine the final WR spin prior to core collapse. The solutions to these equations are discussed in Appendix~\ref{app:Tides}.

\subsection{Natal Kicks} 
\label{subsec:Kicks}

Observations of pulsar motion suggest that neutron stars receive natal kicks (or recoils) due to asymmetric energy/mass ejection during their formation \cite{Hobbs2005}. Natal kicks are the most likely source of spin-orbit misalignment in close compact binaries in the isolated channel \cite{Fryer1999}. Currently, two types of kick mechanisms are favored in supernova (SN) explosions: neutrino-driven kicks caused by asymmetric neutrino emission, and hydrodynamical kicks associated with asymmetric mass ejection. The mechanism that drives the natal kick is uncertain and it is unclear whether BHs receive kicks \cite{Martin2010,Janka2012,Repetto2012,Repetto2015,Mandel2016}. Future observations may constrain the prevalence of the two kick mechanisms for BHs \cite{Janka2013,Taylor2018,Wong2019}. Hydrodynamical kicks may be suppressed by the fallback of ejected material \cite{Fryer2012}, which is often modeled in population synthesis via a fallback parameter. We do not include this parameter since it would be degenerate in our simplified model with the kick velocity dispersion, and instead we assume that BHs are formed through direct collapse.

We assume that the kick velocity $\mathbf{v}_{\rm  k}$ is randomly drawn from an isotropic Gaussian distribution with one-dimensional velocity dispersion $\sigma$. The value of $\sigma$ may be as low as 15~km/s for double neutron star binaries undergoing electron capture-supernova, and as large as 265~km/s for iron core-collapse supernova \cite{Mapelli2018}. This uncertainty justifies treating $\sigma$ as a free parameter. For an initially circular orbit, the binary will be unbound for 
\begin{equation} \label{E:ukUB}
u_{\rm k} \geq (2\beta - \sin^2\theta_{\rm k})^{1/2} - \cos\theta_{\rm k}~,
\end{equation}
where $u_{\rm  k}  = v_{\rm  k}/v_{\rm orb}$ is the kick magnitude $v_{\rm  k}$ normalized by the orbital velocity $v_{\rm orb} \equiv (GM/a)^{1/2}$ before the kick, $\theta_{\rm  k}$ is the angle between the kick and the velocity of the collapsing star, and $\beta$ is the ratio of the final and initial total masses \cite{Kalogera2000,Hurley2002,Gerosa2013}.

\begin{figure*}[!t]
\centering
\includegraphics[width=1.0\textwidth]{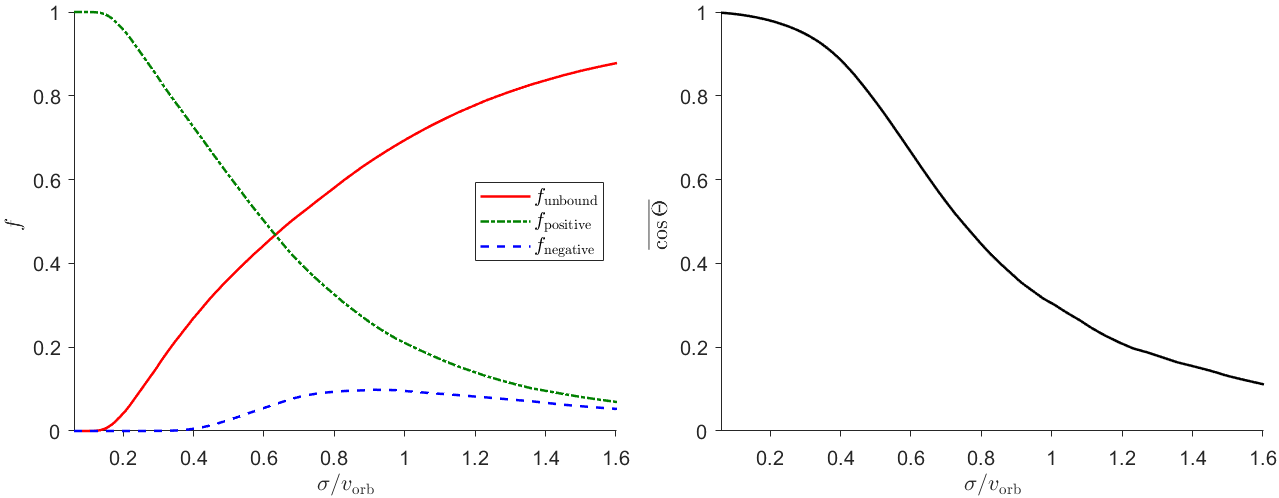}
\caption{{\it Left panel:} The solid red line shows the fraction of binaries that are unbound by a SN as a function of the ratio $\sigma/v_{\rm orb}$ of the kick dispersion to the orbital speed.  The dot-dashed green (dashed blue) line shows the fractions of binaries that remain bound but change the direction of their orbital angular momentum by a tilt angle $\Theta < (>)~90^\circ$.  The exploding WR star has an initial mass of $35~{\rm M}_{\odot}$ and an equal-mass companion; it loses 10\% of its mass during the SN.  {\it Right panel:} The average value of $\cos\Theta$ as a function of the ratio $\sigma/v_{\rm orb}$.} \label{F:TiltsFbound}%
\end{figure*}

Assuming initial alignment, the angle between the spins $\mathbf{S}_i$ and the orbital angular momentum $\mathbf{L}$ following the primary natal kick (for systems that remain bound) is given by
\begin{equation} \label{E:SN1}
\cos\Theta = \frac{1 + u_{\rm  k} \cos\theta_{\rm  k}}{[(1 + u_{\rm  k}  \cos\theta_{\rm  k})^2
+ (u_{\rm  k}  \sin\theta_{\rm  k} \cos\phi_{\rm  k})^2]^{1/2}}
\end{equation}
where $\phi_{\rm  k}$ is the azimuthal angle of the kick in the plane perpendicular to the stellar velocity, where $\phi_{\rm  k} = 0$ corresponds to alignment with the initial orbital angular momentum.  Fig.~\ref{F:TiltsFbound} shows the fraction of binaries that remain bound and the average value of $\cos\Theta$ as functions of $\sigma/v_{\rm orb}$.

Prior to the second SN, the binary components may have different misalignments due to tidal realignment of the secondary WR star
or differential spin precession. After the second SN, the angles $\theta_i$ between each BH spin and the orbital angular momentum are
\begin{equation} \label{E:SN2}
\cos\theta_i = \frac{(1 + u_{\rm  k} \cos\theta_{\rm  k})\cos\gamma_{i}
- u_{\rm  k}  \sin\theta_{\rm  k} \cos\phi_{\rm  k} \sin\gamma_{i} \sin\varpi_i}{[(1 + u_{\rm  k} \cos\theta_{\rm  k})^2 +  
(u_{\rm  k}  \sin\theta_{\rm  k}\cos\phi_{\rm  k})^2)]^{1/2}}
\end{equation}
where $\gamma_{i}$ is the angle between $\mathbf{S}_i$ and $\mathbf{L}$ prior to the second SN and $\varpi_i$ is the angle between the projection of $\mathbf{S}_i$ into the orbital plane and the binary-separation vector. Tidal alignment can cause $\gamma_i < \Theta$, while differential spin precession described below can cause $\varpi_1 \neq \varpi_2$.
Eq.~(\ref{E:SN2}) reduces to Eq.~(\ref{E:SN1}) in the limit that $\gamma_{i} = 0$. We use these equations to translate the kick distribution into the distributions of the BBH misalignment angles $\theta_i$ and we also compute the corresponding distributions of the BBH semi-major axis and eccentricity.

The natal kick from the SN of the primary tilts the direction of $\mathbf{L}$, but it leaves $\mathbf{S}_1$ and $\mathbf{S}_2$ aligned with each other and thus $\varpi_1 = \varpi_2$.  However, post-Newtonian (PN) spin precession can occur on a fairly short timescale even at the $\approx 10~{\rm R}_\odot$ binary separations following CEE.  Spin-orbit coupling causes the spins of unequal-mass stars to precess at different rates about $\mathbf{L}$, implying that $\varpi_1 - \varpi_2$ is randomized on a timescale
\begin{align} \label{E:TpreSN}
t_{\rm pre} &\equiv \frac{2\pi}{|\Omega_1 - \Omega_2|} \notag \\
&= 99.5~{\rm yrs} \frac{1 + q_t}{1 - q_t}\left(\frac{a}{\rm R_{\odot}}\right)^{5/2}\left(\frac{M}{\rm M_{\odot}}\right)^{-3/2},
\end{align}
where $\Omega_i$ are the precession frequencies at lowest PN order \cite{Kidder1995}.

We compare this timescale to the time $\Delta t_{\rm SN}$ between CEE and the SN of the secondary (Pathway A1) or between the two SN (Pathways A2, B1, B2) as shown in Fig.~\ref{F:Diagram}.  We set $\varpi_1 = \varpi_2$ if $\Delta t_{\rm SN} < t_{\rm pre}$ or draw the angles $\varpi_i$ independently from a flat distribution between 0 and $2\pi$ if $\Delta t_{\rm SN} > t_{\rm pre}$.

Throughout this work, we assume that 10\% of the pre-SN WR mass is lost due to neutrino emission \cite{Lattimer1989,Timmes1996,Belczynski2020}, but the amount of this mass loss remains uncertain \cite{BK2016,Janka2017}. We also assume that these neutrinos do not carry away any angular momentum. Furthermore, we enforce an upper bound $\rchi \leq 1$ on the dimensionless BH spin consistent with cosmic censorship \cite{Penrose1969}. Due to the enormous uncertainties in the understanding of black-hole formation \cite{OConnor2011}, we explore two possible extremes for the mass loss that accompanies this loss of angular momentum: (1) either mass is ejected isotropically, in which case $m_f/m_i = \chi_i^{-1/6.89}$ consistent with wind mass loss in Eq.~(\ref{E:IsoSpin}) with $r_{\rm g}^2 = 0.075$, or (2) there is negligible mass lost consistent with angular-momentum transport outwards through an extended accretion disk about the newly formed BH.

The newly formed BBHs inspiral for a time $t_{\rm merge}$ given by Eq.~(\ref{E:merge}), reduced for non-zero eccentricity, before coalescing \cite{Peters1964}. We use this time to compute the fraction $f_{\rm merge}$ of binaries that merge within the age of the Universe in Table~\ref{T:table1}. The spin misalignments given by Eq.(\ref{E:SN2}) will continue to precess until the BBHs merge. Once GW observatories are sensitive enough to detect the effects of this spin precession \cite{Fairhurst2019}, we will be able to constrain the many aspects of stellar-binary evolution described in this section that affect BBH spin distributions.

\section{Results} 
\label{sec:Res}

\begin{figure*}[!t] 
  \centering
  \includegraphics[width=1.0\linewidth]{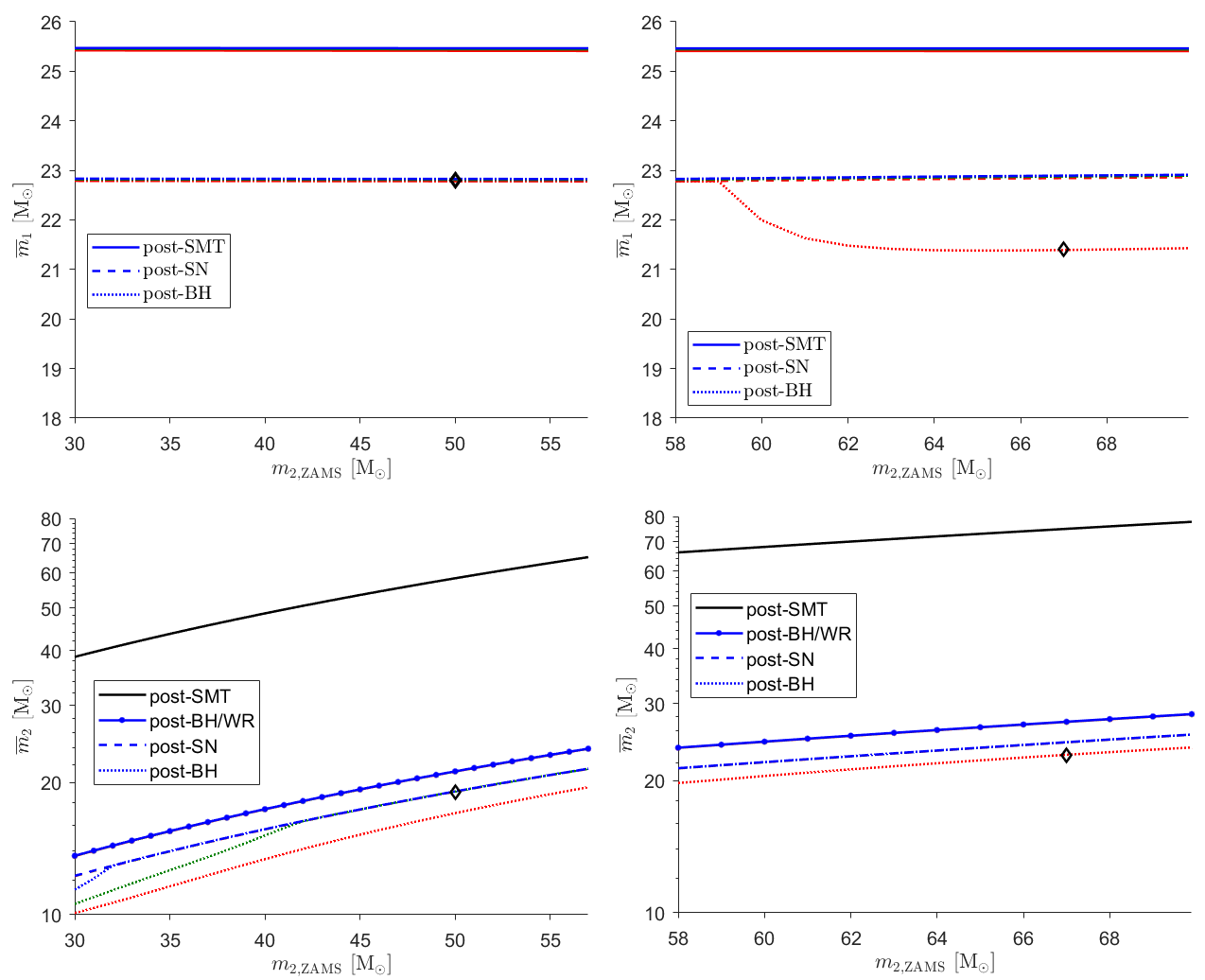}
   \caption{The average mass $\overline{m}_1$ ($\overline{m}_2$) of the primary (secondary) at different stages of their evolution as a function of the secondary ZAMS mass $m_{2,\rm ZAMS}$ in Scenario A. The initial parameters are $m_{1, \rm ZAMS} = 70$~M$_{\odot},~Z = 0.0002,~f_{\rm B} = 0.1,~\sigma = 30$~km/s$,~f_{\rm a} = 0.2$, and there is maximal core-envelope coupling.  Each color corresponds to a different ZAMS separation: red, green, and blue are $a_{\rm ZAMS} = 1{,}500$~R$_\odot,~3{,}000$~R$_\odot$ and $6{,}000$~R$_\odot$. The left (right) panels are Pathway A1 (A2).  The solid black line (post-SMT) is after the primary loses its envelope, the dashed line (post-SN) is after each star loses 10\% of its final WR mass to neutrino emission, and the dotted line is after any additional mass and angular-momentum loss from the surface of each WR star to preserve the Kerr spin limit. For the secondary, additional solid lines marked with dots (post-BH/WR) shows the final WR mass. The black diamonds denote the binaries listed in Table~\ref{T:table1}.
   }\label{F:ScenAMasses}
\end{figure*}

\begin{figure*}[!t] 
  \centering
  \includegraphics[width=1.0\linewidth]{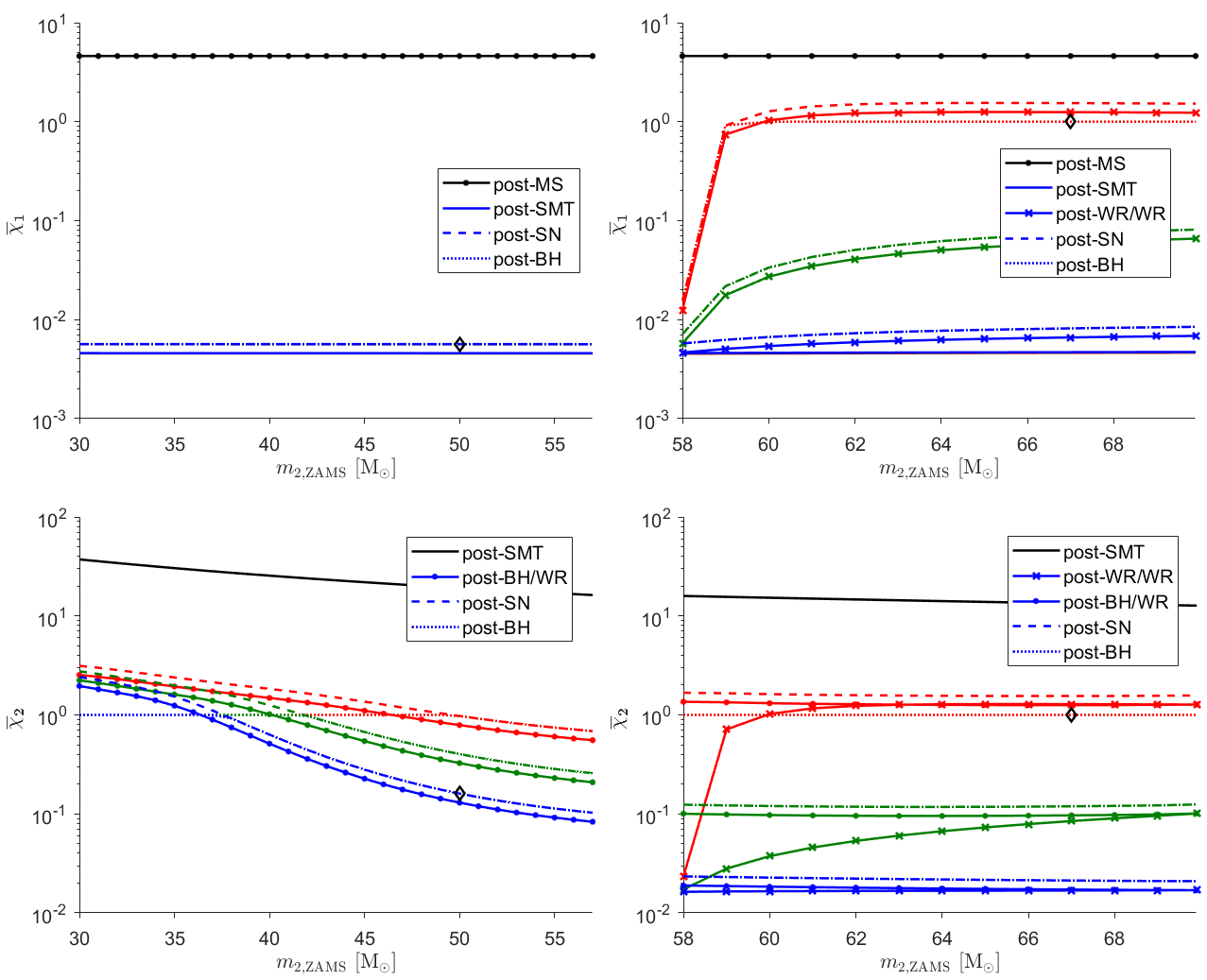}
   \caption{The average dimensionless spin $\overline{\rchi}_1$ ($\overline{\rchi}_2$) of the primary (secondary) at different stages of their evolution as a function of the secondary ZAMS mass $m_{2,\rm ZAMS}$ in Scenario A with maximal core-envelope coupling and the same initial parameters and separations (with same corresponding colors) as in Fig.~\ref{F:ScenAMasses}. The left (right) panels are Pathway A1 (A2). The solid black lines marked by dots (post-MS) show the primary spin at the end of the MS, the solid lines (post-SMT) show the spins after SMT following primary RLOF, the dashed lines (post-SN) show the spins after neutrino emission, and the dotted lines (post-BH) show the spins with the imposition of the Kerr spin limit. The solid lines marked by X's (post-WR/WR) show the spins at the end of the binary WR stage in Pathway A2, while the solid colored lines marked by dots (post-BH/WR) show the secondary spin at the end of its WR stage.}\label{F:ScenAChi}
\end{figure*}

\begin{figure*}[!t] 
  \centering
  \includegraphics[width=1.0\linewidth]{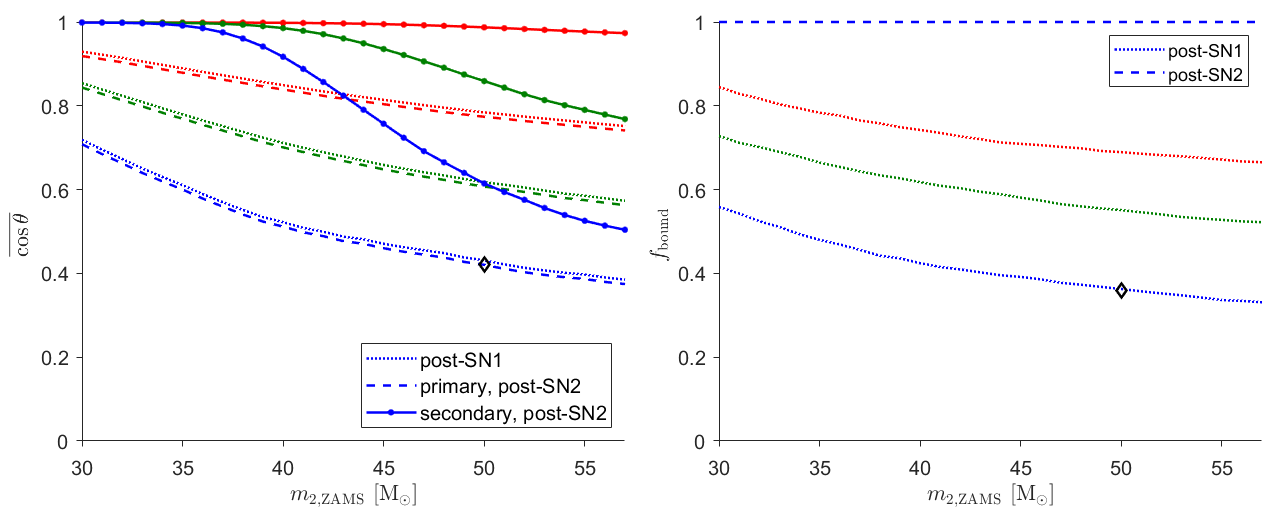}
   \caption{The average of the cosines of the misalignment angles $\theta_i$ (left panel) and the fraction of binaries that remain bound (right panel) after the two SN as a function of secondary ZAMS mass in Pathway A1 for the same initial parameters and separations as in Fig.~\ref{F:ScenAMasses}. The dotted lines show the values after the primary SN (post-SN1). In the left panel, the dashed (primary, post-SN2) and solid lines marked with dots (secondary, post-SN2) show the misalignment angles after the secondary SN which differ depending on the separation due to tidal alignment of the secondary WR. In the right panel, the dashed line (post-SN2) shows the fraction of binaries bound after the primary SN that remain bound after the secondary SN.}\label{F:ScenATiltsFbound}
\end{figure*}

Our model of stellar-binary evolution depends upon seven initial parameters: the masses $m_{i, \rm ZAMS}$, separation $a_{\rm ZAMS}$, metallicity $Z$, fraction of breakup spin $f_{\rm B}$, kick-velocity dispersion $\sigma$, and accreted fraction $f_{\rm a}$. Various physical motivations, as summarized in Sec.~\ref{subsec:ExecSum}, constrain the values of these parameters. Low $Z$ systems best illustrate the effects of binary processes on the component spins since high $Z$ produces strong winds which suppress such effects. 

In both Scenario A and B, the final BH spin is independent of the ZAMS breakup spin fraction $f_{\rm B}$ in the case of maximal core-envelope coupling. However, for minimal coupling, the final BH spin is dependent upon the chosen value of $f_{\rm B}$ for the WR progenitor. Small values ($f_{\rm B} \lesssim 0.001$) produce results that are essentially identical to the maximal coupling case, intermediate values ($f_{\rm B} \sim 0.01$) produce significantly spinning BHs, and high values ($f_{\rm B} \gtrsim 0.1$) produce highly spinning BHs for all masses and separations. In the results below, we assume maximal core-envelope coupling for Scenario A and we assume minimal coupling with $f_{\rm B} = 0.01$ for Scenario B to assess whether tides or accretion during SMT, respectively, can produce highly spinning BHs.

We present results assuming mass is lost isotropically from the stellar surface of the WR star at core collapse if the star has a spin in excess of the Kerr limit. The other extreme, of negligible mass loss during core collapse, resulted in higher BBH masses only when small ZAMS separations led to strong tides following CEE in Scenario A or when stellar accretion during SMT increased the primary's spin significantly in Pathway B2. The higher mass of the primary in this case can reduce the probability of a mass-ratio reversal (MRR) which has important consequences for spin precession \cite{Gerosa2013}.

The value of $\sigma$ determines the likelihood that a binary remains bound after a supernova (SN) kick \cite{Kalogera2000,Gerosa2018}. In Pathway A1, the primary SN occurs before CEE, implying binaries only have a high probability of remaining bound for $\sigma \lesssim 110$~km/s. In the other pathways, we consider $\sigma \gtrsim 150$~km/s, since for smaller $\sigma$ the resulting misalignments are insignificant ($\overline{\cos\theta} \gtrsim 0.9$). Overlining here and in the rest of the paper refers to averaging over our Gaussian distributions of natal kicks.

\subsection{Binary Black-Hole Masses and Spins}

\subsubsection{Scenario A}
\label{subsec:ScenA}

To illustrate Scenario A, we choose initial parameters $m_{1, \rm ZAMS} = 70$~M$_{\odot},~Z = 0.0002,~f_{\rm B} = 0.1,~\sigma = 30$~km/s$,~f_{\rm a} = 0.2$ with $m_{2, \rm ZAMS} \in (30, 69.9)$~M$_{\odot}$ and $a_{\rm ZAMS} \in \{1{,}500,~3{,}000,~6{,}000\}$~R$_{\odot}$. We assume maximal core-envelope coupling and isotropic mass loss for final spins above the Kerr limit. For these initial parameters, secondary RLOF occurs after the primary SN (Pathway 1) in binaries with ZAMS mass ratios below the transition $q_{\rm trans} \approx 0.83$, i.e. $m_{\rm 2, ZAMS}/{\rm M}_\odot \approx 0.83 \times 70 = 58$. The figures in this section have $m_{\rm 2, ZAMS}$ on the horizontal axis and are divided into left and right panels at this transition value (except Fig.~\ref{F:ScenATiltsFbound}), with the legends for each panel specifying the orders of events in Pathways 1 and 2 as shown in Fig.~\ref{F:Diagram}. This value of $q_{\rm trans}$ implies that the majority of binaries evolve in Pathway A1 ($q_{\rm ZAMS} < q_{\rm trans}$) for a uniform distribution of ZAMS mass ratios.

Fig.~\ref{F:ScenAMasses} shows the mass evolution of the primary and secondary in Scenario A. The primary mass is independent of ZAMS secondary mass $m_{2, \rm ZAMS}$ in Pathway A1 as shown in the top left panel. The primary loses mass through winds on the MS, then loses its envelope following RLOF during its core helium-burning (CHeB) stage (Fig.~\ref{F:RLOF} shows the stage at which RLOF occurs as a function of $a_{\rm ZAMS}$ and $Z$). After losing a small amount of mass from WR winds, it loses 10$\%$ of its mass to neutrino emission during core collapse to form a BH of mass $m_1 \simeq 23$~M$_{\odot}$. The secondary evolves through a similar sequence of events in Pathway A1 as shown in the bottom left panel, except that it accretes rather than loses mass through SMT following RLOF of the primary.  For small values of $m_{\rm 2, ZAMS}$ and $a_{\rm ZAMS}$, tides can spin up the secondary above the Kerr limit during the WR stage requiring additional mass loss during core collapse shown by the dotted curves. The mass evolution in Pathway A2 is similar as shown in the right panels, except that the primary is still in the WR stage at the time of CEE following secondary RLOF, implying that it too can experience tidal synchronization above the Kerr limit and subsequent mass loss during core collapse for $a_{\rm ZAMS} = 1{,}500$~R$_\odot$ shown by the dotted red curve.

\begin{figure*}[!t] 
  \centering
  \includegraphics[width=1.0\linewidth]{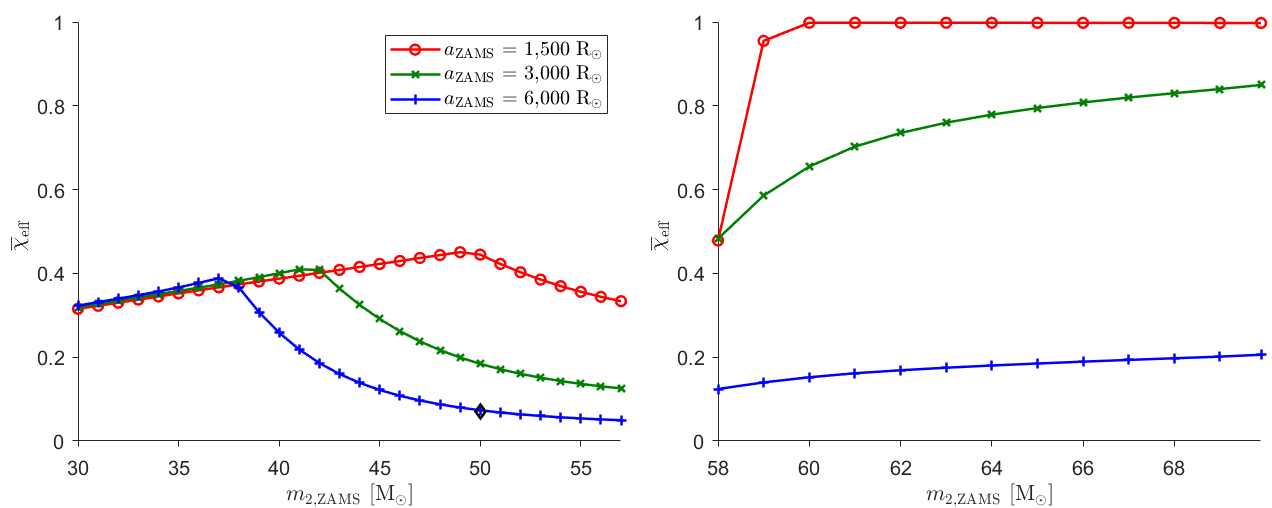}
   \caption{The average BBH aligned effective spin $\rchi_{\rm eff}$ as a function of the secondary ZAMS mass $m_{2, \rm ZAMS}$ in Pathway A1 (left panel) and Pathway A2 (right panel) with maximal core-envelope coupling and the same initial parameters as Fig.~\ref{F:ScenAMasses}.
   }\label{F:ScenAChiEff}
\end{figure*}

The top panels of Fig.~\ref{F:ScenAChi} show the spin evolution of the primary in Scenario A. In Pathway A1 (top left panel), the SN of the primary occurs before CEE implying that tides (which depend on the secondary mass and separation) cannot influence its spin. Maximal core-envelope coupling causes the primary to lose most of its spin during its RLOF ($\lesssim 0.01$).  Neutrino emission during core collapse causes a slight increase in the dimensionless spin because we assume that these neutrinos carry away mass but not angular momentum. In Pathway A2 (top right panel), CEE following secondary RLOF occurs before the primary SN, implying that the primary can be spun up by tides during the binary WR stage.  At the transition between Pathways A1 and A2 ($m_{\rm 2, ZAMS} \simeq 58$~M$_\odot$), the secondary RLOF and primary SN occur simultaneously, implying that the binary WR stage has zero duration and therefore tides do not have time to influence the primary spin.  As $m_{\rm 2, ZAMS}$ increases, the delay between secondary RLOF and primary SN also increases and the primary spin asymptotically approaches its synchronized value $\rchi_{\rm sync}$ shown in Fig.~\ref{F:Chisync}. Wider initial separations $a_{\rm ZAMS}$ imply wider separations during the binary WR stage and thus smaller synchronized spins.

The bottom panels of Fig.~\ref{F:ScenAChi} show the spin evolution of the secondary in Scenario A. SMT following primary RLOF spins the secondary up to large dimensionless spins $\rchi \gtrsim 10$, but maximal core-envelope coupling causes the secondary to lose most of this spin angular momentum during its own RLOF. In Pathway A1 (bottom left panel), the primary SN occurs before secondary RLOF so the primary is a BH during the entire WR stage of the secondary. As $m_{\rm 2, ZAMS}$ increases, the WR mass of the secondary increases as well and its synchronized dimensionless spin $\rchi_{\rm sync}$ decreases as shown in Fig.~\ref{F:Chisync}. In Pathway A2 (bottom right panel), secondary RLOF occurs before the primary SN allowing the existence of a binary WR stage. As for the primary spin shown in the top right panel, the increasing duration of this binary WR stage as $m_{\rm 2, ZAMS}$ increases causes the secondary spin to asymptote to its synchronized value. By the end of the secondary WR lifetime shown by the solid curves marked with dots, there is always enough time for the secondary spin to become tidally synchronized, with smaller spins for wider separations.

The left panel of Fig.~\ref{F:ScenATiltsFbound} shows the average values of the cosines of the misalignment angles $\theta_i$ between the spins and the orbital angular momentum due to SN kicks in Pathway A1. Wider initial separations $a_{\rm ZAMS}$ lead to larger misalignments after the primary SN, as does larger $m_{2,\rm ZAMS}$ because of the exponential dependence of the separation $a_f$ following SMT on the mass $m_{i, {\rm A}}$ of the accreting secondary in Eq.~(\ref{E:SMT}). Since CEE occurs before the secondary SN and since we use a small value of $\sigma$ here, the secondary kick has a small effect on the misalignments as indicated by the proximity of the dashed and dotted lines. The primary collapses into a BH before CEE in Pathway A1, so tides cannot realign its spin with the orbital angular momentum prior to the secondary SN. However, tides can fully realign the secondary spin ($t_{\rm sync} \to 0, \overline{\cos\theta} \to 1$) for small initial separations ($a_{\rm ZAMS} = 1{,}500~{\rm R}_\odot$ shown by the solid red line marked with dots), implying that generally $\theta_1 \neq \theta_2$ once the BBH has formed. The increasing residual misalignment as $m_{2, \rm ZAMS} \to q_{\rm trans}m_{1, \rm ZAMS} \simeq 58~{\rm M}_\odot$ for the wider initial separations results from the wider separations (and thus longer alignment times $t_{\rm align}$) of these binaries following SMT as described above. Although CEE shrinks the binary separation by several orders of magnitude, in our model the separation remains an increasing function of $m_{2, \rm ZAMS}$.

The right panel of Fig.~\ref{F:ScenATiltsFbound} shows the fraction of binaries that remain bound after each SN. The smaller separations following CEE between the two SN leads to a bound fraction near unity after the secondary SN (post-SN2) for all three ZAMS separations. We do not show the results for Pathway A2, as there are only small spin misalignments and few unbound binaries for $\sigma = 30$~km/s when CEE occurs before the primary natal kick. For separations $a_{\rm ZAMS} \lesssim 1{,}500~{\rm R}_\odot$, binaries can also be destroyed by RLOF of the secondary during the WR stage.

Fig.~\ref{F:ScenAChiEff} displays the aligned effective spin $\rchi_{\rm eff}$, Eq.~(\ref{E:ChiEff}), of the final BBHs as a function of the secondary ZAMS mass $m_{2, \rm ZAMS}$. At small $m_{2, \rm ZAMS}$, the slightly larger $\rchi_{\rm eff}$ for the green and blue lines than the red line originates from the dependence on the BBH mass ratio: binaries at larger initial separations experienced less mass-loss in BH formation (see Fig.~\ref{F:ScenAMasses}) and hence have larger BBH mass ratio. As $m_{2, \rm ZAMS}$ increases in Pathway A1 (left panel), $\rchi_{\rm eff}$ first increases as the weight of the contribution from the more highly spinning secondary increases, then decreases once the secondary spin is no longer fully aligned as seen in the left panel of Fig.~\ref{F:ScenATiltsFbound}. When $m_{2, \rm ZAMS}$ increases above $q_{\rm trans}m_{1, \rm ZAMS}$, the binary evolution transitions to Pathway A2 (right panel) in which CEE occurs before the primary core collapse. In this case, the small value of the natal kick velocity dispersion $\sigma = 30$~km/s yields negligible misalignments, so $\rchi_{\rm eff}$ depends strongly on the spin magnitudes. Tides can spin up the primary and secondary during the binary WR stage for $a_{\rm ZAMS} \lesssim 3{,}000~{\rm R}_\odot$ as shown by the red and green lines marked by X's in the upper right panel of Fig.~\ref{F:ScenAChi}. This leads to the larger values of $\rchi_{\rm eff}$ for the corresponding values of $a_{\rm ZAMS}$ in the right panel of Fig.~\ref{F:ScenAChiEff} provided that $m_{2, \rm ZAMS}$ is sufficiently above the threshold that the binary WR stage lasts longer than the synchronization time $t_{\rm sync}$.

\begin{figure*}[!t] 
  \centering
  \includegraphics[width=1.0\linewidth]{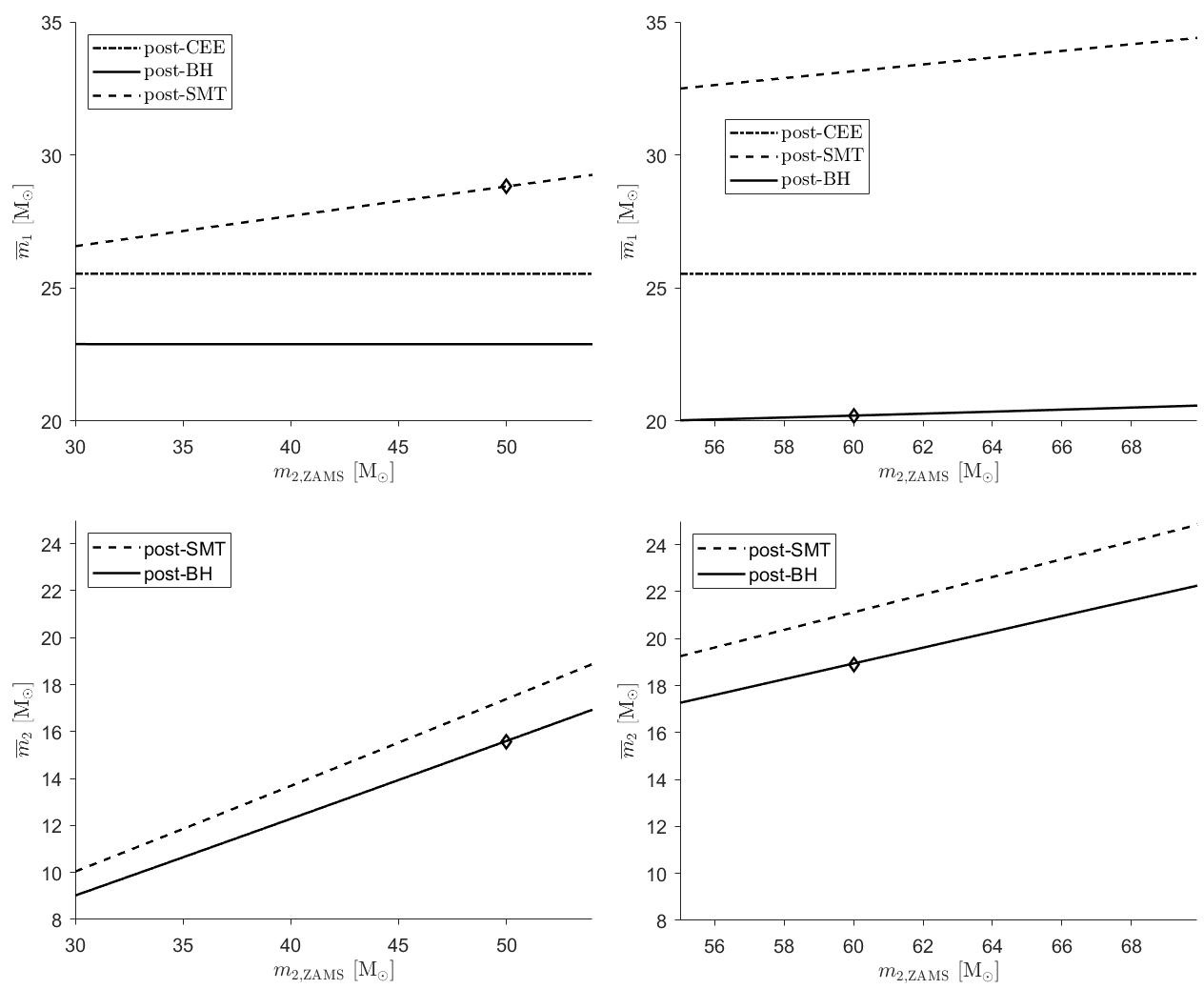}
   \caption{
   The average mass $\overline{m}_1$ ($\overline{m}_2$) of the primary (secondary) at different stages of their evolution as a function of the secondary ZAMS mass $m_{2,\rm ZAMS}$ in Pathways B1 (left panels) and B2 (right panels).  The initial parameters are $m_{1, \rm ZAMS} = 70$~M$_{\odot},~Z = 0.0002,~f_{\rm B} = 0.01,~\sigma = 200$~km/s$,~f_{\rm a} = 0.2$, and there is minimal core-envelope coupling. The dot-dashed lines (post-CEE) show the primary mass after CEE following its RLOF, the dashed lines (post-SMT) show each mass after SMT following secondary RLOF, and the solid lines (post-BH) show each mass after its collapse to a BH.  The black diamonds denote the binaries listed in Table~\ref{T:table1}.}\label{F:ScenBMasses}
\end{figure*}

This section demonstrates how the various processes of stellar-binary evolution can conspire to suppress the emergence of highly precessing BBHs in Scenario A. For maximal core-envelope coupling, tidal synchronization is the only mechanism available for producing significant BBH spins, but the accompanying tidal alignment leaves only a narrow sliver of parameter space for these spins to be misaligned as shown in Fig.~\ref{F:tidaltimes}. Significant misalignments are possible for $\sigma \gtrsim 30$~km/s (in A1) as shown in the fifth column of Table~\ref{T:table1}, but the spin magnitudes will be small. Conversely, if $a_{\rm ZAMS}$ is chosen to be small enough for tidal synchronization to occur, as in the sixth column of Table~\ref{T:table1}, the spin misalignments will be modest for $\sigma \lesssim 200$~km/s (in A2). For minimal core-envelope coupling in Scenario A, as shown in the first two columns of Table~\ref{T:table1}, we can obtain highly precessing BBHs by choosing $a_{\rm ZAMS}$ large enough to avoid tidal alignment and $f_{\rm B}$ large enough so that the WR stars are still highly spinning even after strong wind mass-loss.

\subsubsection{Scenario B}
\label{subsec:ScenB}

\begin{figure*}[!t] 
  \centering
  \includegraphics[width=1.0\linewidth]{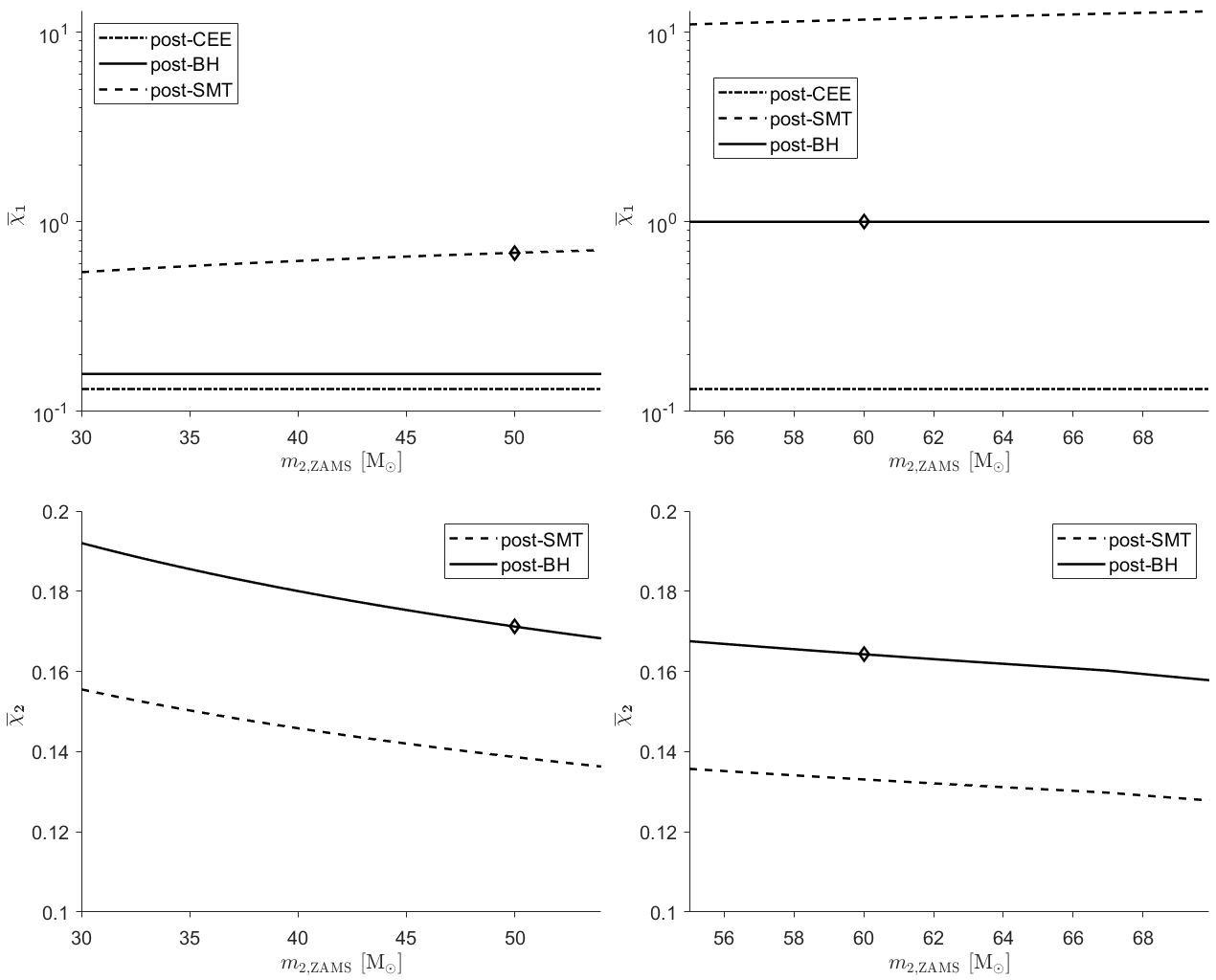}
   \caption{
   The average dimensionless spin $\overline{\rchi}_1$ ($\overline{\rchi}_2$) of the primary (secondary) at different stages of their evolution as a function of the secondary ZAMS mass $m_{2,\rm ZAMS}$ in Pathways B1 (left panels) and B2 (right panels).  The initial parameters are the same as in Fig.~\ref{F:ScenBMasses}. The dot-dashed lines (post-CEE) show the primary mass after CEE following its RLOF, the dashed lines (post-SMT) show each mass after SMT following secondary RLOF, and the solid lines (post-BH) show each mass after its collapse to a BH.  The black diamonds denote the binaries listed in Table~\ref{T:table1}.}\label{F:ScenBChi}
\end{figure*}

In Scenario B, RLOF of the primary leads to CEE and RLOF of the secondary leads to SMT as shown in Fig.~\ref{F:Diagram}. To avoid RLOF of the secondary occurring on the MS before the formation of a well-defined stellar core, the initial separation $a_{\rm ZAMS}$ must be chosen larger in Scenario B than Scenario A, i.e. $\gtrsim 10,000$~R$_{\odot}$ for $Z = 0.0002$ and $\gtrsim 15,000$~R$_{\odot}$ for $Z = 0.02$.

\begin{figure*}[!t] 
  \centering
  \includegraphics[width=1.0\linewidth]{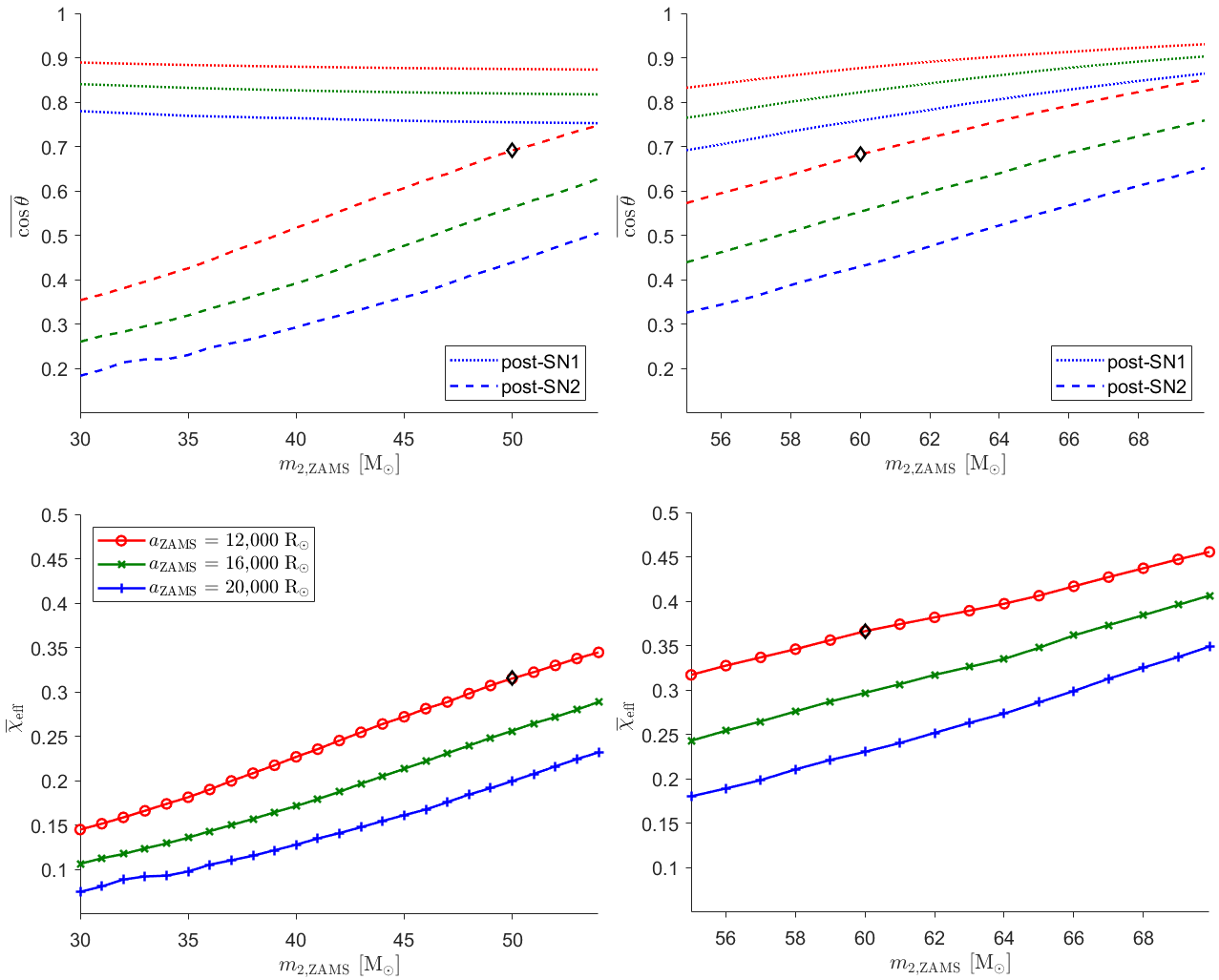}
   \caption{The average of the cosines of the misalignment angles $\theta_i$ (top panels) and the BBH aligned effective spin $\rchi_{\rm eff}$ (bottom panels) as a function of secondary ZAMS mass in Pathways B1 (left panels) and B2 (right panels) for the same initial parameters and separations as in Fig.~\ref{F:ScenBMasses}.  The red, green, and blue lines (marked by circles, X's, and crosses in the bottom panels) correspond to ZAMS separations  $a_{\rm ZAMS} = 12{,}000$~R$_\odot,~16{,}000$~R$_\odot$ and $20{,}000$~R$_\odot$.  In the top panels, the dotted lines (post-SN1) show the values after the primary SN, while the dashed lines (post-SN2) show the values after secondary SN.}\label{F:ScenBChiEffTilts}
\end{figure*}

To illustrate binary evolution in Scenario B, we choose initial parameters $m_{1, \rm ZAMS} = 70$~M$_{\odot}, Z = 0.0002, f_{\rm a} = 0.2$ with $m_{2, \rm ZAMS} \in (30, 69.9)$~M$_{\odot}$ as in Scenario A. However, differences between Scenarios A and B require three key changes to produce highly precessing BBHs:
\begin{enumerate}[label=(\arabic*)]

\item We choose wider initial separations $a_{ \rm ZAMS} \in \{12{,}000, 16{,}000, 25{,}000\}$~R$_{\odot}$ to avoid RLOF of the secondary on the MS.

\item We assume minimal core-envelope coupling with a WR breakup spin fraction $f_{\rm B} = 0.01$ as a lower bound for producing a significantly spinning secondary BH in the absence of tidal synchronization - the primary BH receives a high spin regardless of $f_{\rm B}$ from accretion.

\item As in Pathway A2, we choose larger natal kicks ($\sigma = 200$~km/s) to provide significant spin misalignments for SN that occur after CEE.

\end{enumerate}
The transition between Pathways 1 and 2 occurs at a slightly smaller ZAMS mass ratio ($q_{\rm trans} \approx 0.77$) than in Scenario A for $Z=0.0002$, because the wider initial separations delay the RLOF of the primary and the start of its WR stage.  This gives less massive secondaries additional time to evolve to fill their Roche lobes before the primary core collapse, the order of events that defines Pathway 2 as seen in Fig.~\ref{F:Diagram}. In the figures of this section, the left panels ($m_{2, \rm ZAMS} < q_{\rm trans}m_{1, \rm ZAMS} \approx 53.9~{\rm M}_\odot$) correspond to Pathway B1, while the right panels ($m_{2, \rm ZAMS} > q_{\rm trans}m_{1, \rm ZAMS}$) correspond to Pathway B2.

Fig.~\ref{F:ScenBMasses} displays the mass evolution of the primary and secondary as functions of $m_{2, \rm ZAMS}$. Unlike in Scenario A, where tidal synchronization leads to separation-dependent mass loss during core collapse to preserve the Kerr spin limit, the mass evolution is nearly independent of $a_{ \rm ZAMS}$ in our model in Scenario B. In both pathways, the primary loses mass during CEE following its RLOF. In Pathway B1, it loses an additional 10\% of its mass to neutrino emission in core collapse, then gains a fraction $f_{\rm a}$ of the mass of the secondary's envelope during SMT following secondary RLOF. In Pathway B2, this SMT precedes the core collapse of the primary leading to a much greater increase in its spin because of the higher specific angular momentum of the accreted gas at the surface of a WR star compared to the innermost stable circular orbit (ISCO) of a comparable-mass BH. This leads greater isotropic mass loss ($\gtrsim 8$~M$_{\odot}$) during core collapse to preserve the Kerr spin limit and correspondingly lower primary BH masses, despite the larger amount of accreted gas compared to Pathway B1. Enough mass is lost during core collapse to cause a mass-ratio reversal (MRR), in which the primary evolves into the less massive BH, for $m_{2,\rm ZAMS} \gtrsim 64$~M$_{\odot}$. The evolution of the secondary is comparatively straightforward: it loses its envelope during SMT following RLOF and then loses an additional 10\% of its mass to neutrino emission during core collapse.

Fig.~\ref{F:ScenBChi} shows the evolution of the dimensionless spin in Scenario B.  The WR stars are born following RLOF with spins $\rchi \approx 0.1$ from our choice of minimal core-envelope coupling with a WR breakup spin fraction $f_{\rm B} = 0.01$.  Accretion by the primary during SMT following secondary RLOF spins it up to $\rchi_1 \approx 0.6$ in Pathway B1 as seen in the top left panel, with the values of the specific energy and angular momentum of the accreted gas at the ISCO naturally imposing the Kerr spin limit.  In Pathway B2 (top right panel), the primary is still a WR star at the time of SMT and can therefore attain a dimensionless spin $\rchi_1 \approx 10$ by accreting gas at its surface which is much less compact than the ISCO of a comparable-mass BH.  Neutrino emission of 10\% of the final WR rest mass during core collapse does not carry away angular momentum in our model, increasing dimensionless spins below the Kerr limit by a factor of ($0.9^{-2} \approx 1.23$).  For our choice of initial parameters, this applies to the primary in Pathway B1 and the secondary in both pathways; however, the large spin of the primary after SMT in Pathway B2 implies that it is spun down to the Kerr limit by mass loss during core collapse.  There is negligible wind-driven loss of mass and angular momentum for the low metallicity $Z = 0.0002$ we have chosen in this section.

\begin{figure*}[!t] 
  \centering
  \includegraphics[width=1.0\linewidth]{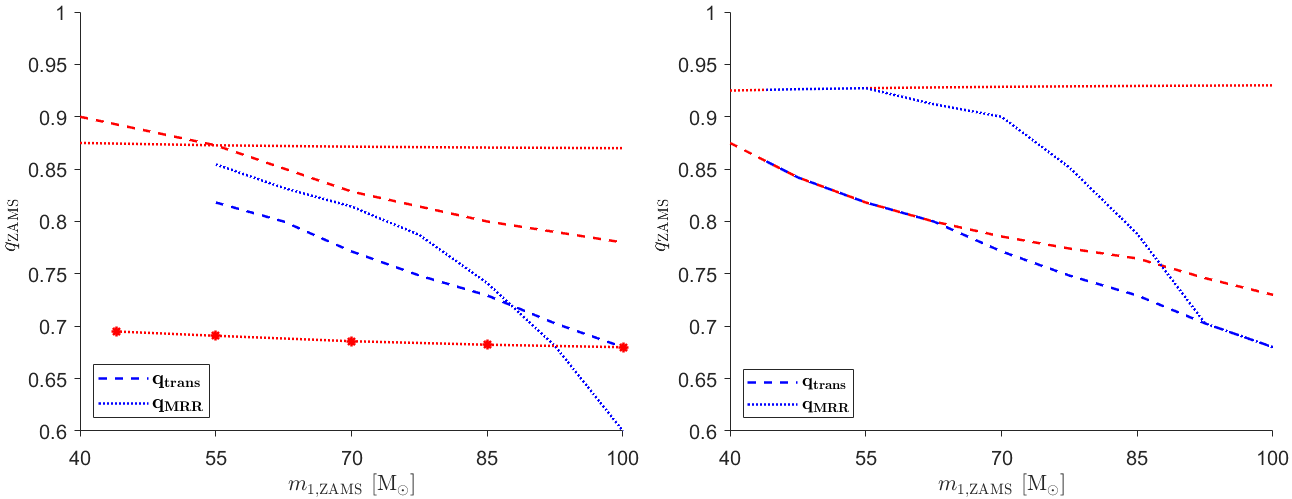}
   \caption{The mass ratios of the pathway transition $q_{\rm trans}$ (dashed lines) and mass-ratio reversal $q_{\rm MRR}$ (dotted lines) as functions of primary ZAMS mass $m_{1,\rm ZAMS}$ in Scenario A (left) with ZAMS separation $a_{\rm ZAMS} = 10{,}000$~R$_{\odot}$ and in Scenario B (right) with ZAMS separation $a_{\rm ZAMS} = 20{,}000$~R$_{\odot}$. The red (blue) lines correspond to a metallicity of $Z = 0.0002$ ($Z = 0.02$). The lines that are unmarked correspond to an accreted fraction $f_{\rm a} = 0.2$, and the red line in the left panel marked by filled circles corresponds to $f_{\rm a} = 0.5$. The blue lines ($Z = 0.02$) do not extend to $m_{1,\rm ZAMS} = 40~{\rm M}_\odot$ because secondary (primary) RLOF fails to occur for low primary ZAMS masses in Scenario A (B). The red line in the left panel marked by filled circles ($f_{\rm a} = 0.5$) does not extend below $m_{1,\rm ZAMS} \approx 43~{\rm M}_\odot$ because $q_{\rm MRR}m_{1,\rm ZAMS} \leq 30~{\rm M}_\odot$ for these values. In Scenario B, MRR does not occur when we assume that negligible mass is lost in primary BH-formation due to the Kerr spin limit. 
   }\label{F:QtransQmrr}
\end{figure*}

The top panels of Fig.~\ref{F:ScenBChiEffTilts} show the evolution of the misalignment angles $\theta_i$ as functions of the secondary ZAMS mass $m_{2,\rm ZAMS}$.  Larger natal kicks strengths ($\sigma = 200$~km/s) are needed to produce significant misalignments in Scenario B because both SN occur after CEE following primary RLOF as shown in Fig.~\ref{F:Diagram}.  As in Scenario A shown in the left panel of Fig.~\ref{F:ScenATiltsFbound}, larger ZAMS separations $a_{\rm ZAMS}$ lead to smaller orbital velocities $v_{\rm orb} \equiv (GM/a)^{1/2}$ at the time of each SN and thus larger normalized kicks $u_{\rm k}  = v_{\rm k}/v_{\rm orb}$ and larger misalignments given by Eqs.~(\ref{E:SN1}) and (\ref{E:SN2}).  However, unlike in Scenario A, the increase in the misalignments is larger following the secondary SN than the primary SN.  This occurs because the decrease in the total binary mass $M$ following secondary RLOF reduces $v_{\rm orb}$ and thus increases $u_{\rm  k}$ in Eq.~(\ref{E:SN2}).
Misalignments increase more following the primary SN than the secondary SN in Pathway A1 because CEE between the two SN shrinks the binary separation $a$, increasing $v_{\rm orb}$ and thus reducing $u_{\rm k}$ and the misalignments.

Another difference is that the mean misalignment following the secondary SN decreases with secondary ZAMS mass $m_{2,\rm ZAMS}$ in Scenario B, whereas the mean misalignment following the primary SN increases with secondary ZAMS mass $m_{2,\rm ZAMS}$ in Scenario A.  This occurs because the secondary is the donor rather than the accretor during SMT in Scenario B, and therefore its mass appears in the numerator rather than the denominator of the exponent in Eq.~(\ref{E:SMT}) for the decrease in binary separation during SMT.  Smaller separations following SMT lead to larger $v_{\rm orb}$ and smaller misalignments.  The dependence of the BBH aligned effective spin $\rchi_{\rm eff}$ on $m_{2,\rm ZAMS}$ and $a_{\rm ZAMS}$ shown in the bottom panels of Fig.~\ref{F:ScenBChiEffTilts} mirrors that of $\cos\theta_i$ following the secondary SN, as the spin magnitudes shown in Fig.~\ref{F:ScenBChi} are almost independent of $m_{2,\rm ZAMS}$ and $a_{\rm ZAMS}$.  The absence of tidal alignment in Scenario B suggests that this could be a promising source of highly precessing BBHs provided that minimal core-envelope coupling leaves significant residual spins in the WR stars born following mass transfer.

\subsection{Transition Mass Ratio}
\label{subsec:Trans}

As depicted in Fig.~\ref{F:Diagram}, binary stellar evolution is classified as belonging to Pathway 1 (2) if the core collapse of the primary, SN1, occurs before (after) the secondary experiences RLOF. A binary will evolve along Pathway 1 (2) if the primary WR lifetime $t_{1, \rm WR}$ is less (greater) than the time $\Delta t_{\rm RLOF}$ between the primary and secondary RLOFs.  Binaries with a ZAMS mass ratio $m_{2,\rm ZAMS}/m_{1,\rm ZAMS}$ below (above) $q_{\rm trans}$ will evolve along Pathway 1 (2).  In our model, this transition mass ratio depends on the scenario (A or B), the primary ZAMS mass $m_{1,\rm ZAMS}$, the metallicity $Z$, the ZAMS binary separation $a_{\rm ZAMS}$, and the accreted fraction $f_{\rm a}$ during SMT.

The transition mass ratio $q_{\rm trans}$ is distinct from the mass ratio $q_{\rm MRR}$ above which a mass-ratio reversal (MRR) occurs, i.e. the primary star evolves into the less massive BH.  The asymmetry between the primary and secondary spins as discussed in the previous subsection implies that MRR has profound consequences for the BBH spin-precession morphologies prior to merger \cite{Gerosa2013} and the masses, spins, and gravitational recoils of the final BH produced in BBH mergers \cite{Kesden2010,Kesden2010b}.

Fig.~\ref{F:QtransQmrr} displays $q_{\rm trans}$ and $q_{\rm MRR}$ as functions of the primary ZAMS mass $m_{1, \rm ZAMS}$.  According to the fitting formulae for stellar evolution \cite{Hurley2000} that we adopt in our model, the difference $\Delta t_{\rm MS}$ between the primary and secondary MS lifetimes (a good proxy for $\Delta t_{\rm RLOF}$) scales as $m_{1, \rm ZAMS}^{-1.5}$ for fixed $q_{\rm ZAMS}$ in the large-mass limit.  However, the primary WR lifetime $t_{1, \rm WR}$ has a much weaker dependence on $m_{1, \rm ZAMS}$ since $t_{\rm WR} \propto m_{\rm WR}^{-0.5}$ in the large-mass limit according to Eq.~(\ref{E:WRlifetime}) and $m_{\rm WR}$ scales almost linearly with $m_{\rm ZAMS}$ as seen in Fig.~\ref{F:CoreMVSzamsM}.  This implies that 
$\Delta t_{\rm RLOF} < t_{1, \rm WR}$ for smaller values of $q_{\rm ZAMS}$ at larger values of $m_{1, \rm ZAMS}$, i.e. $q_{\rm trans}$ is a monotonically decreasing function of $m_{1, \rm ZAMS}$ as seen in Fig.~\ref{F:QtransQmrr}.

At high metallicity ($Z = 0.02$), primary RLOF occurs on the Hertzsprung gap (HG) shortly after the end of the MS lifetime.  In Scenario A, SMT following primary RLOF increases the binary separation for low $m_{1, \rm ZAMS}$ because of the prefactors to the exponential in Eq.~(\ref{E:SMT}).  This increase implies that secondary RLOF will fail to occur for $m_{1, \rm ZAMS} \lesssim 52~{\rm M}_\odot$ and the binary will never experience the CEE needed to bring the BBHs close enough to merge within the age of the Universe.  The blue lines corresponding to $Z = 0.02$ therefore do not extend below this value in the left panel of Fig.~\ref{F:QtransQmrr}.  In Scenario B, the larger ZAMS separation $a_{\rm ZAMS} = 20{,}000$~R$_{\odot}$ needed to avoid secondary RLOF on the MS implies that, for $m_{1, \rm ZAMS} \lesssim 44~{\rm M}_\odot$, stellar winds can drive the primary mass below that for which primary RLOF occurs at this separation as seen in Fig.~\ref{F:RLOF}.  The blue lines corresponding to $Z = 0.02$ therefore do not extend below this value of $m_{1, \rm ZAMS}$ in the right panel of Fig.~\ref{F:QtransQmrr}.

At low metallicity ($Z = 0.0002$), the smaller stellar radii on the HG imply that primary RLOF does not occur until the core helium-burning (CHeB) stage where the larger stellar radii allow primary RLOF to occur for $m_{1, \rm ZAMS} > 40~{\rm M}_\odot$ as shown by the dashed red lines in the left panel of Fig.~\ref{F:QtransQmrr}.  The longer MS lifetimes for $Z = 0.0002$ and the delay of RLOF until the CHeB stage imply that higher ZAMS mass ratio are needed if the primary WR star is to survive until secondary RLOF in Scenario A.  This explains why the dashed red curve is above the dashed blue curve in the left panel of Fig.~\ref{F:QtransQmrr}.  In Scenario B, CEE following primary RLOF shrinks the binary separation and causes secondary RLOF to occur on the HG.  For $m_{1, \rm ZAMS} \lesssim 60~{\rm M}_\odot$, this decreases the value of $q_{\rm trans}$ for $Z = 0.0002$ close to its value for $Z = 0.02$ where secondary RLOF also occurs on the HG.  For $m_{1, \rm ZAMS} \gtrsim 60~{\rm M}_\odot$, the reduced wind mass loss for $Z = 0.0002$ leads to heavier, more short-lived primary WR stars which require larger values of $q_{\rm ZAMS}$ if they are to survive until secondary RLOF.  This explains the gap between the dashed curves in the right panel of Fig.~\ref{F:QtransQmrr}.

\subsection{Mass-Ratio Reversal}
\label{subsec:MRR}

Fig.~\ref{F:QtransQmrr} also shows how $q_{\rm MRR}$, the ZAMS mass ratio above which MRR occurs, depends on the primary ZAMS mass $m_{1,\rm ZAMS}$.  In Scenario A (left panel), SMT from the primary to the secondary following primary RLOF leads to MRR for $q_{\rm ZAMS} > q_{\rm MRR} \approx 0.88$ for low metallicity ($Z = 0.0002$) as shown by the dotted red line.  As winds are negligible for this metallicity and the mass of each star's envelope is roughly proportional to its ZAMS mass, $q_{\rm MRR}$ is nearly independent of $m_{1,\rm ZAMS}$.  Increasing the fraction $f_{\rm a}$ of the primary envelope accreted during SMT reduces $q_{\rm MRR}$ to $\approx 0.7$ as shown by the dashed red line marked by filled circles.  The mass loss due to winds is a nonlinear function of the ZAMS mass for high metallicity ($Z = 0.02$) as shown in Fig.~\ref{F:SSEmass}.  This implies that the heavier primary loses more mass than the lighter secondary, making $q_{\rm MRR}$ a decreasing function of $m_{1,\rm ZAMS}$ as shown by the dotted blue line in Fig.~\ref{F:QtransQmrr}.

In Scenario B (right panel of Fig.~\ref{F:QtransQmrr}), SMT from the secondary to the primary occurs after secondary RLOF, so one might assume that MRR is not possible at low metallicity ($Z = 0.0002$) where winds are negligible. However, MRR can indeed occur because of the asymmetry between accretion and mass loss. Our model assumes that accretion during SMT occurs through a thin disk in the equatorial plane of the accretor. In Pathway B2 ($q_{\rm ZAMS} > q_{\rm trans}$, i.e. above the dashed lines), this implies that the change in dimensionless spin of the accretor per unit accreted mass is given by
\begin{equation} \label{E:accB2}
\frac{d\rchi}{dm} = \left( \frac{c^2R}{Gm^3} \right)^{1/2} - \frac{2\rchi}{m}  = \left( \frac{1}{f_{\rm B}r_{\rm g}^2} - 2 \right) \frac{\rchi}{m}~,
\end{equation}
where $f_{\rm B}$ is the fraction of breakup at which the accreting WR star is spinning and $r_{\rm g}$ is its radius of gyration.  Comparing Eq.~(\ref{E:accB2}) to the second term on the right-hand side of Eq.~(\ref{E:TWalignODE}), we see that, for $f_{\rm B, WR} \ll 1$, accretion is far more efficient than isotropic mass loss at changing the spin of a star.  If accretion during SMT spins up the primary above the Kerr limit, a greater amount of mass must be lost isotropically during core collapse to preserve the Kerr limit for the final black hole.  This mass loss leads to MRR for $q_{\rm ZAMS} > q_{\rm MRR} \approx 0.93$ for $Z = 0.0002$ as shown by the dotted red line.  Although the exact values of the dimensionless prefactors in Eqs.~(\ref{E:TWspinODE}) and (\ref{E:accB2}) depend on the simplifying assumptions of our model, accretion and mass loss are genuinely asymmetric so MRR through this channel could potentially occur. 

At high metallicity ($Z = 0.02$), mass loss by the primary prior to its RLOF reduces $q_{\rm MRR}$, as in Scenario A. It is interesting to note in Pathway B1 that for $m_{1,\rm ZAMS} \gtrsim 90~{\rm M}_\odot$, $q_{\rm trans}$ acts as floor for $q_{\rm MRR}$ since primary core collapse precedes SMT and therefore isotropic mass loss is not needed to shed the angular momentum gained during accretion.

\section{Discussion}
\label{sec:Disc}

We have presented a simplified model of BBH formation in the isolated channel that can produce BBHs with large misaligned spins. This model, summarized in Fig.~\ref{F:Diagram}, is divided into Scenarios A or B depending on whether the CEE needed to produce small binary separations occurs after RLOF of the secondary or primary star. Each scenario can be further subdivided into Pathways 1 or 2 depending on whether the core collapse of the primary occurs before or after the RLOF of the secondary.

For these misaligned spins to produce detectable modulation of the GWs emitted at frequencies observed by ground-based detectors, they must be both large in magnitude and sufficiently misaligned with the orbital angular momentum. In our model, BHs can acquire large spins through three mechanisms:
\begin{enumerate}[label=(\arabic*)]

\item Inheritance: The BH inherits a significant portion of the natal rotational angular momentum of its ZAMS stellar progenitor. 

\item Tides: The WR progenitor is spun up by tides exerted by its companion at the small binary separations that follow CEE.

\item Accretion: The BH or its WR progenitor gains angular momentum by accreting gas from its companion during SMT.

\end{enumerate}

The first of these mechanisms requires weak coupling between the stellar core and its envelope \cite{Maeder2004,Maeder2008,Maeder2012,Belczynski2020}. In the limit that the core and envelope are entirely decoupled, the WR star that forms from the stellar core could in principle inherit a dimensionless natal spin as large as the breakup value $\rchi_{\rm B}$ given by Eq.~(\ref{E:Break}). In practice, the WR star will only inherit a spin that is a fraction $f_{\rm B} < 1$ of this value, but stellar models suggest that this fraction could be large \cite{Meynet2005,Crowther2007}. The first through fourth columns of Table~\ref{T:table1} indicate that WR stars born with a spin fraction $f_{\rm B} = 0.01$ of their breakup value yield BHs with spins $\rchi_1 \approx 0.15$ in the absence of further spin up by tides or accretion. The BH spin increases linearly with $f_{\rm B}$ up to the Kerr limit, but higher values of $f_{\rm B}$ are needed to saturate this limit at high metallicities because of spin down due to WR winds as shown in Fig.~\ref{F:WRSpin}. The tight upper bounds on the aligned effective spin $\rchi_{\rm eff}$ observed in many LIGO/Virgo events set upper bounds on the value of $f_{\rm B}$ but do not eliminate the possibility of minimal coupling.

Strong core-envelope coupling causes stars to lose almost all of their rotational angular momentum when they lose their envelopes. It produces BH spins $\rchi_i \lesssim 0.01$, as shown in the fifth through eighth columns of Table~\ref{T:table1}, for BHs for which neither tides nor accretion provide alternative sources of spin. We neglect the possibility that additional mass and rotational angular momentum could be lost in the collapse of the WR stars into BHs \cite{Batta2019}, but this would further suppress the BH spins and must be avoided to produce BBHs with large misaligned spins.

Tidal synchronization during the WR stage provides a second mechanism for attaining large BH spins, but its effectiveness varies with our scenarios and pathways of binary stellar evolution. In Scenario A, CEE following RLOF of the secondary can decrease the binary separation below the synchronization separation $a_{\rm sync}$ given by Eq.~(\ref{E:async}). At such small separations, tides can spin up WR stars leading to the enhanced secondary BH spin $\rchi_2$ seen in the bottom panels of Fig.~\ref{F:ScenAChi} and listed in the fifth and sixth columns of Table~\ref{T:table1}. The primary BH spin $\rchi_1$ depends on the pathway. In Pathway 1, the primary collapses into a BH prior to CEE and therefore never has a chance to be tidally synchronized. This leads to the small values of $\rchi_1$ seen in the upper left panel of Fig.~\ref{F:ScenAChi} and listed in the fifth column of Table~\ref{T:table1}. In Pathway 2, the primary remains a WR star at the time of CEE and can therefore be tidally synchronized to the large values of $\rchi_1$ seen in the upper right panel of Fig.~\ref{F:ScenAChi} and listed in the sixth column of Table~\ref{T:table1}. In Scenario B, the large ZAMS separations needed to avoid RLOF of the secondary on the MS ($12{,}000~{\rm R}_\odot$ in the third, fourth, seventh, and eighth columns of Table~\ref{T:table1}) imply that tides are too weak to synchronize the WR stars prior to their collapse.

The effectiveness of accretion in generating large BH spins also depends on the scenario and pathway of stellar evolution.  In Scenario A, the secondary accretes during SMT following RLOF of the primary, but the angular momentum it gains is subsequently lost when the secondary loses its envelope.  In Scenario B, the primary has already lost its envelope prior to the SMT which follows RLOF of the secondary.  In Pathway 1, it has also collapsed into a BH by this time.  If super-Eddington accretion is allowed during SMT, the primary can attain a large spin for accreted fractions $f_{\rm a} \gtrsim 0.2$ as seen in the upper left panel of Fig.~\ref{F:ScenBChi} and listed in the third and seventh columns of Table~\ref{T:table1}.  In Pathway 2, the primary remains a WR star at the time of SMT.  The larger specific angular momentum at the surface of a WR star compared to a BH of similar mass implies that the same amount of accreted mass can provide spins well above the Kerr limit as seen in the upper right panel of Fig.~\ref{F:ScenBChi}.  We assume that the excess angular momentum is lost in the collapse to a BH, but this can be accompanied by mass loss as seen in the upper right panel of Fig.~\ref{F:ScenBMasses} and listed in the fourth and eighth columns of Table~\ref{T:table1}.  We see that tides and accretion are thus complementary sources of BH spin; tides operate exclusively in Scenario A, while accretion is only effective in Scenario B.  Both sources have greater effects on the primary in Pathway 2, because the primary is more readily spun up by both tides and accretion during its WR stage than as a BH.

The second ingredient needed to make precessing BBHs is a source of misalignment between the BH spins and the orbital angular momentum. In our model, natal kicks during core collapse are this source of misalignment. As shown in Fig.~\ref{F:TiltsFbound}, the ratio of the natal kick velocity dispersion to the orbital velocity must satisfy $\sigma/v_{\rm orb} \lesssim 1$ to provide appreciable misalignments without unbinding the majority of binaries.  As $v_{\rm orb} \propto a^{-1/2}$, the kick dispersion $\sigma$ required to produce precessing BBHs decreases with increasing binary separation prior to core collapse. In Pathway A1, the primary core collapse occurs before CEE shrinks the binary separation implying that $\sigma \approx 30~{\rm km/s}$ can produce significant misalignments as seen in the first and fifth columns of Table~\ref{T:table1}. In the other pathways, both core collapses occur after CEE and $\sigma \approx 150~{\rm km/s}$ is needed to produce significant misalignments.

Our model also predicts correlations between spin magnitude and misalignment. Spins inherited from the ZAMS stellar progenitors remain oriented along the direction of the initial orbital angular momentum.  Unless $\sigma \gtrsim v_{\rm orb}$, these spins will remain biased towards $\cos\theta_i > 0$ even after both SN as seen in the left panel of Fig.~\ref{F:TiltsFbound}. However, the magnitude of the inherited spins is uncorrelated with their misalignment. In contrast, the similarity of the tidal synchronization and alignment timescales given by Eqs.~(\ref{E:tsync1}) and (\ref{E:talign1}) imply that spins sourced by tides only remain misaligned in the narrow band of parameter space between the dash-dotted brown and solid black lines in Fig.~\ref{F:tidaltimes}. We can see this anti-correlation by comparing the fifth and sixth columns of Table~\ref{T:table1}, where the larger spin magnitudes in Pathway A2 ($a_{\rm ZAMS} = 1{,}500~{\rm R}_\odot$) compared to Pathway A1 ($a_{\rm ZAMS} = 6{,}000~{\rm R}_\odot$) come at the expense of smaller misalignments. Although not included in our model, the Bardeen-Petterson effect could generate a similar anti-correlation between the magnitudes and misalignments of spins sourced by accretion during SMT.

We also explored how stellar metallicity $Z$ affects BBH formation, with additional details provided in the Appendix. The strong winds at high ZAMS mass $m_{\rm ZAMS}$ imply that the highest metallicity stars ($Z = 0.02$) produce the least massive BHs as seen in Fig.~\ref{F:SSEmass}. These strong winds can also reduce the dimensionless spin $\rchi$ by a factor $\lesssim 2$ during the WR stage as seen in Figs.~\ref{F:WRSpin} and \ref{F:SSEspin}. This is particularly important when inheritance is the only source of spin as it is for the primary in Pathway A1 or the secondary in Scenario B. The increasing wind strength with $m_{\rm ZAMS}$ at high $Z$ also drives the BBH mass ratio $q$ above the ZAMS mass ratio $q_{\rm ZAMS}$, increasing the possibility of a mass-ratio reversal (MRR). 

MRR was a focus of \citeauthor{Gerosa2015}~\cite{Gerosa2015} because of its consequences for BBH spin precession. Fig.~\ref{F:ScenATiltsFbound} of this paper reproduces a key result of that work, that the primary spin misalignment is usually greater than that of the secondary in Pathway A1.  MRR determines whether the more misaligned primary evolves into the more or less massive BH, and thus whether the BBH spins librate about $\Delta\Phi = \pm 180^\circ$ or $0^\circ$ near merger. As illustrated in its Fig.~3, \citeauthor{Gerosa2015}~\cite{Gerosa2015} dealt exclusively with Pathway A1 and neglected the possibility that the primary WR star could survive until after CEE as in Pathway 2 of this paper. Although MRR can indeed occur in Pathway A1 for the large accreted fraction $f_{\rm a} = 0.5$ adopted in \citeauthor{Gerosa2015}~\cite{Gerosa2015}, the left panel of Fig.~\ref{F:QtransQmrr} of this paper shows that MRR in Scenario A occurs exclusively in Pathway 2 (dotted line above dashed line) except at the lowest values of $m_{1, \rm ZAMS}$ for $Z= 0.0002$ and the highest values of $m_{1, \rm ZAMS}$ for $Z= 0.02$. This is critical because tides can realign both the primary and secondary spins following the primary SN in Pathway A2, eliminating the difference in their misalignments that determines the morphology of spin precession near merger.  Table~\ref{T:table1} indeed shows that only in Pathway A1 do the mean primary and secondary spin misalignments differ.

We have also found that mass loss by the highly spinning primary star during core collapse can induce MRR in Pathway B2, as shown in the right panel of Fig.~\ref{F:QtransQmrr}. As SMT transfers mass from the secondary to the primary in Scenario B, MRR can be prevented in this scenario by an accreted fraction $f_{\rm a} \gtrsim 0.5$. As tides are inefficient in Scenario B, no asymmetry in the spin misalignments which could be inverted by a MRR will be produced in this scenario. Increased magnetization of outflows during core collapse could reduce the amount of mass loss needed to preserve the Kerr limit in the resulting BH, preventing MRR in Pathway B2.

One aspect of binary stellar evolution not included in our model is the possibility that highly spinning MS stars may experience enhanced rotational mixing leading to chemically homogeneous evolution \cite{deMink2010,Marchant2016,Mandel2016,Song2016,Cui2018}. Such binaries would avoid RLOF and must therefore originate at very small $a_{\rm ZAMS}$ if they are to produce BBHs that merge within the age of the Universe. Binaries with such small initial separations could potentially evolve into BBHs with large misaligned spins, but if sourced by natal kicks this misalignment would require large kick velocity dispersion $\sigma \gtrsim 150\,{\rm km/s}$ comparable to that in Pathways A2, B1, and B2 of our model.

Our study of the formation of precessing BBHs is timely as two GW events observed during the third observing run of LIGO/Virgo show marginal evidence of spin precession.  GW190412 \cite{GW190412} has an effective precession parameter $\rchi_{\rm p} = 0.30_{-0.15}^{+0.19}$ \cite{Schmidt2015} leading to a precession signal-to-noise ratio \cite{Fairhurst2019} $\rho_{\rm p} = 2.86_{-1.56}^{+3.43}$. The small mass ratio $q = 0.28_{-0.07}^{+0.13}$ of this system suggests that it was produced in Pathway 1, but the constraints on the individual spins ($\rchi_1 = 0.43_{-0.26}^{+0.16}$) are too weak to determine whether the system was produced in Scenario A or B. GW190521 \cite{GW190521} has an extremely large total mass $150_{-17}^{+29}~{\rm M}_\odot$ implying that only a few GW cycles are observed in band. The constraint of the effective precession parameter $\rchi_{\rm p} = 0.68_{-0.25}^{+0.37}$ is quite broad, but there is nonetheless a weak preference for a precessing orbital plane ($\log_{10}$ Bayes factor of $1.06_{-0.06}^{+0.06}$ in favor of precessing versus non-precessing spins) \cite{GW190521}. This preference is driven by a slight amplitude suppression of the lowest-frequency part of the waveform which could arise from the orbital angular momentum precessing into the line of sight later in the inspiral. The mass ratio $q = 0.79_{-0.29}^{+0.19}$ of GW190521 is consistent with either Pathway 1 or 2, and the individual spins are essentially unconstrained ($\rchi_1 = 0.69_{-0.62}^{+0.27}, \rchi_2 = 0.73_{-0.64}^{+0.24}$). If this system was produced in the isolated formation channel, the precession of its orbital plane suggests that even massive BHs experience nonzero natal kicks and these kicks must be quite large unless the spins were inherited through weak core-envelope coupling in Pathway A1.

In light of these discoveries, we are currently developing a new model of the orbital-plane precession \cite{SteinleGangardt2020}. We will use this model to explore the spin precession of the astrophysical binaries generated here \cite{SteinleFuture1}. We hope that this analysis can help to reveal the astrophysical origin of individual high signal-to-noise events and the large samples of lower signal-to-noise events \cite{Stevenson2017,Vitale2017} that will be found by future GW detectors 
\cite{Punturo2010,Abbott2017,Baibhav2019,Kuns2020}.

\bibliography{bibme}{}

\begin{thebibliography}{166}%
\makeatletter
\providecommand \@ifxundefined [1]{%
 \@ifx{#1\undefined}
}%
\providecommand \@ifnum [1]{%
 \ifnum #1\expandafter \@firstoftwo
 \else \expandafter \@secondoftwo
 \fi
}%
\providecommand \@ifx [1]{%
 \ifx #1\expandafter \@firstoftwo
 \else \expandafter \@secondoftwo
 \fi
}%
\providecommand \natexlab [1]{#1}%
\providecommand \enquote  [1]{``#1''}%
\providecommand \bibnamefont  [1]{#1}%
\providecommand \bibfnamefont [1]{#1}%
\providecommand \citenamefont [1]{#1}%
\providecommand \href@noop [0]{\@secondoftwo}%
\providecommand \href [0]{\begingroup \@sanitize@url \@href}%
\providecommand \@href[1]{\@@startlink{#1}\@@href}%
\providecommand \@@href[1]{\endgroup#1\@@endlink}%
\providecommand \@sanitize@url [0]{\catcode `\\12\catcode `\$12\catcode
  `\&12\catcode `\#12\catcode `\^12\catcode `\_12\catcode `\%12\relax}%
\providecommand \@@startlink[1]{}%
\providecommand \@@endlink[0]{}%
\providecommand \url  [0]{\begingroup\@sanitize@url \@url }%
\providecommand \@url [1]{\endgroup\@href {#1}{\urlprefix }}%
\providecommand \urlprefix  [0]{URL }%
\providecommand \Eprint [0]{\href }%
\providecommand \doibase [0]{http://dx.doi.org/}%
\providecommand \selectlanguage [0]{\@gobble}%
\providecommand \bibinfo  [0]{\@secondoftwo}%
\providecommand \bibfield  [0]{\@secondoftwo}%
\providecommand \translation [1]{[#1]}%
\providecommand \BibitemOpen [0]{}%
\providecommand \bibitemStop [0]{}%
\providecommand \bibitemNoStop [0]{.\EOS\space}%
\providecommand \EOS [0]{\spacefactor3000\relax}%
\providecommand \BibitemShut  [1]{\csname bibitem#1\endcsname}%
\let\auto@bib@innerbib\@empty
\bibitem [{\citenamefont {{Abbott}}\ \emph
  {et~al.}(2019{\natexlab{a}})\citenamefont {{Abbott}} \emph
  {et~al.}}]{LIGO2019catalog}%
  \BibitemOpen
  \bibfield  {author} {\bibinfo {author} {\bibfnamefont {B.~P.}\ \bibnamefont
  {{Abbott}}} \emph {et~al.} (\bibinfo {collaboration} {LIGO and Virgo
  Collaborations}),\ }\href {\doibase 10.1103/PhysRevX.9.031040} {\bibfield
  {journal} {\bibinfo  {journal} {Physical Review X}\ }\textbf {\bibinfo
  {volume} {9}},\ \bibinfo {eid} {031040} (\bibinfo {year}
  {2019}{\natexlab{a}})},\ \Eprint {http://arxiv.org/abs/1811.12907}
  {arXiv:1811.12907 [astro-ph.HE]} \BibitemShut {NoStop}%
\bibitem [{\citenamefont {{Punturo}}\ \emph {et~al.}(2010)\citenamefont
  {{Punturo}} \emph {et~al.}}]{Punturo2010}%
  \BibitemOpen
  \bibfield  {author} {\bibinfo {author} {\bibfnamefont {M.}~\bibnamefont
  {{Punturo}}} \emph {et~al.},\ }\href {\doibase
  10.1088/0264-9381/27/19/194002} {\bibfield  {journal} {\bibinfo  {journal}
  {Classical and Quantum Gravity}\ }\textbf {\bibinfo {volume} {27}},\ \bibinfo
  {pages} {194002} (\bibinfo {year} {2010})}\BibitemShut {NoStop}%
\bibitem [{\citenamefont {{Abbott}}\ \emph {et~al.}(2017)\citenamefont
  {{Abbott}} \emph {et~al.}}]{Abbott2017}%
  \BibitemOpen
  \bibfield  {author} {\bibinfo {author} {\bibfnamefont {B.~P.}\ \bibnamefont
  {{Abbott}}} \emph {et~al.} (\bibinfo {collaboration} {LIGO and Virgo
  Collaborations}),\ }\href {\doibase 10.1088/1361-6382/aa51f4} {\bibfield
  {journal} {\bibinfo  {journal} {Classical and Quantum Gravity}\ }\textbf
  {\bibinfo {volume} {34}},\ \bibinfo {eid} {044001} (\bibinfo {year}
  {2017})},\ \Eprint {http://arxiv.org/abs/1607.08697} {arXiv:1607.08697
  [astro-ph.IM]} \BibitemShut {NoStop}%
\bibitem [{\citenamefont {{Baibhav}}\ \emph {et~al.}(2019)\citenamefont
  {{Baibhav}}, \citenamefont {{Berti}}, \citenamefont {{Gerosa}}, \citenamefont
  {{Mapelli}}, \citenamefont {{Giacobbo}}, \citenamefont {{Bouffanais}},\ and\
  \citenamefont {{Di Carlo}}}]{Baibhav2019}%
  \BibitemOpen
  \bibfield  {author} {\bibinfo {author} {\bibfnamefont {V.}~\bibnamefont
  {{Baibhav}}}, \bibinfo {author} {\bibfnamefont {E.}~\bibnamefont {{Berti}}},
  \bibinfo {author} {\bibfnamefont {D.}~\bibnamefont {{Gerosa}}}, \bibinfo
  {author} {\bibfnamefont {M.}~\bibnamefont {{Mapelli}}}, \bibinfo {author}
  {\bibfnamefont {N.}~\bibnamefont {{Giacobbo}}}, \bibinfo {author}
  {\bibfnamefont {Y.}~\bibnamefont {{Bouffanais}}}, \ and\ \bibinfo {author}
  {\bibfnamefont {U.~N.}\ \bibnamefont {{Di Carlo}}},\ }\href {\doibase
  10.1103/PhysRevD.100.064060} {\bibfield  {journal} {\bibinfo  {journal}
  {\prd}\ }\textbf {\bibinfo {volume} {100}},\ \bibinfo {eid} {064060}
  (\bibinfo {year} {2019})},\ \Eprint {http://arxiv.org/abs/1906.04197}
  {arXiv:1906.04197 [gr-qc]} \BibitemShut {NoStop}%
\bibitem [{\citenamefont {{Kuns}}\ \emph {et~al.}(2020)\citenamefont {{Kuns}},
  \citenamefont {{Yu}}, \citenamefont {{Chen}},\ and\ \citenamefont
  {{Adhikari}}}]{Kuns2020}%
  \BibitemOpen
  \bibfield  {author} {\bibinfo {author} {\bibfnamefont {K.~A.}\ \bibnamefont
  {{Kuns}}}, \bibinfo {author} {\bibfnamefont {H.}~\bibnamefont {{Yu}}},
  \bibinfo {author} {\bibfnamefont {Y.}~\bibnamefont {{Chen}}}, \ and\ \bibinfo
  {author} {\bibfnamefont {R.~X.}\ \bibnamefont {{Adhikari}}},\ }\href
  {\doibase 10.1103/PhysRevD.102.043001} {\bibfield  {journal} {\bibinfo
  {journal} {\prd}\ }\textbf {\bibinfo {volume} {102}},\ \bibinfo {eid}
  {043001} (\bibinfo {year} {2020})},\ \Eprint
  {http://arxiv.org/abs/1908.06004} {arXiv:1908.06004 [gr-qc]} \BibitemShut
  {NoStop}%
\bibitem [{\citenamefont {{Breivik}}\ \emph {et~al.}(2016)\citenamefont
  {{Breivik}}, \citenamefont {{Rodriguez}}, \citenamefont {{Larson}},
  \citenamefont {{Kalogera}},\ and\ \citenamefont {{Rasio}}}]{Breivik2016}%
  \BibitemOpen
  \bibfield  {author} {\bibinfo {author} {\bibfnamefont {K.}~\bibnamefont
  {{Breivik}}}, \bibinfo {author} {\bibfnamefont {C.~L.}\ \bibnamefont
  {{Rodriguez}}}, \bibinfo {author} {\bibfnamefont {S.~L.}\ \bibnamefont
  {{Larson}}}, \bibinfo {author} {\bibfnamefont {V.}~\bibnamefont
  {{Kalogera}}}, \ and\ \bibinfo {author} {\bibfnamefont {F.~A.}\ \bibnamefont
  {{Rasio}}},\ }\href {\doibase 10.3847/2041-8205/830/1/L18} {\bibfield
  {journal} {\bibinfo  {journal} {\apjl}\ }\textbf {\bibinfo {volume} {830}},\
  \bibinfo {eid} {L18} (\bibinfo {year} {2016})},\ \Eprint
  {http://arxiv.org/abs/1606.09558} {arXiv:1606.09558 [astro-ph.GA]}
  \BibitemShut {NoStop}%
\bibitem [{\citenamefont {{Nishizawa}}\ \emph {et~al.}(2016)\citenamefont
  {{Nishizawa}}, \citenamefont {{Berti}}, \citenamefont {{Klein}},\ and\
  \citenamefont {{Sesana}}}]{Nishizawa2016}%
  \BibitemOpen
  \bibfield  {author} {\bibinfo {author} {\bibfnamefont {A.}~\bibnamefont
  {{Nishizawa}}}, \bibinfo {author} {\bibfnamefont {E.}~\bibnamefont
  {{Berti}}}, \bibinfo {author} {\bibfnamefont {A.}~\bibnamefont {{Klein}}}, \
  and\ \bibinfo {author} {\bibfnamefont {A.}~\bibnamefont {{Sesana}}},\ }\href
  {\doibase 10.1103/PhysRevD.94.064020} {\bibfield  {journal} {\bibinfo
  {journal} {\prd}\ }\textbf {\bibinfo {volume} {94}},\ \bibinfo {eid} {064020}
  (\bibinfo {year} {2016})},\ \Eprint {http://arxiv.org/abs/1605.01341}
  {arXiv:1605.01341 [gr-qc]} \BibitemShut {NoStop}%
\bibitem [{\citenamefont {{Nishizawa}}\ \emph {et~al.}(2017)\citenamefont
  {{Nishizawa}}, \citenamefont {{Sesana}}, \citenamefont {{Berti}},\ and\
  \citenamefont {{Klein}}}]{Nishizawa2017}%
  \BibitemOpen
  \bibfield  {author} {\bibinfo {author} {\bibfnamefont {A.}~\bibnamefont
  {{Nishizawa}}}, \bibinfo {author} {\bibfnamefont {A.}~\bibnamefont
  {{Sesana}}}, \bibinfo {author} {\bibfnamefont {E.}~\bibnamefont {{Berti}}}, \
  and\ \bibinfo {author} {\bibfnamefont {A.}~\bibnamefont {{Klein}}},\ }\href
  {\doibase 10.1093/mnras/stw2993} {\bibfield  {journal} {\bibinfo  {journal}
  {\mnras}\ }\textbf {\bibinfo {volume} {465}},\ \bibinfo {pages} {4375}
  (\bibinfo {year} {2017})},\ \Eprint {http://arxiv.org/abs/1606.09295}
  {arXiv:1606.09295 [astro-ph.HE]} \BibitemShut {NoStop}%
\bibitem [{\citenamefont {{Zevin}}\ \emph {et~al.}(2017)\citenamefont
  {{Zevin}}, \citenamefont {{Pankow}}, \citenamefont {{Rodriguez}},
  \citenamefont {{Sampson}}, \citenamefont {{Chase}}, \citenamefont
  {{Kalogera}},\ and\ \citenamefont {{Rasio}}}]{Zevin2017}%
  \BibitemOpen
  \bibfield  {author} {\bibinfo {author} {\bibfnamefont {M.}~\bibnamefont
  {{Zevin}}}, \bibinfo {author} {\bibfnamefont {C.}~\bibnamefont {{Pankow}}},
  \bibinfo {author} {\bibfnamefont {C.~L.}\ \bibnamefont {{Rodriguez}}},
  \bibinfo {author} {\bibfnamefont {L.}~\bibnamefont {{Sampson}}}, \bibinfo
  {author} {\bibfnamefont {E.}~\bibnamefont {{Chase}}}, \bibinfo {author}
  {\bibfnamefont {V.}~\bibnamefont {{Kalogera}}}, \ and\ \bibinfo {author}
  {\bibfnamefont {F.~A.}\ \bibnamefont {{Rasio}}},\ }\href {\doibase
  10.3847/1538-4357/aa8408} {\bibfield  {journal} {\bibinfo  {journal} {\apj}\
  }\textbf {\bibinfo {volume} {846}},\ \bibinfo {eid} {82} (\bibinfo {year}
  {2017})},\ \Eprint {http://arxiv.org/abs/1704.07379} {arXiv:1704.07379
  [astro-ph.HE]} \BibitemShut {NoStop}%
\bibitem [{\citenamefont {{Vitale}}\ \emph
  {et~al.}(2017{\natexlab{a}})\citenamefont {{Vitale}}, \citenamefont
  {{Lynch}}, \citenamefont {{Sturani}},\ and\ \citenamefont
  {{Graff}}}]{Vitale2017}%
  \BibitemOpen
  \bibfield  {author} {\bibinfo {author} {\bibfnamefont {S.}~\bibnamefont
  {{Vitale}}}, \bibinfo {author} {\bibfnamefont {R.}~\bibnamefont {{Lynch}}},
  \bibinfo {author} {\bibfnamefont {R.}~\bibnamefont {{Sturani}}}, \ and\
  \bibinfo {author} {\bibfnamefont {P.}~\bibnamefont {{Graff}}},\ }\href
  {\doibase 10.1088/1361-6382/aa552e} {\bibfield  {journal} {\bibinfo
  {journal} {Classical and Quantum Gravity}\ }\textbf {\bibinfo {volume}
  {34}},\ \bibinfo {eid} {03LT01} (\bibinfo {year} {2017}{\natexlab{a}})},\
  \Eprint {http://arxiv.org/abs/1503.04307} {arXiv:1503.04307 [gr-qc]}
  \BibitemShut {NoStop}%
\bibitem [{\citenamefont {{Bouffanais}}\ \emph {et~al.}(2019)\citenamefont
  {{Bouffanais}}, \citenamefont {{Mapelli}}, \citenamefont {{Gerosa}},
  \citenamefont {{Di Carlo}}, \citenamefont {{Giacobbo}}, \citenamefont
  {{Berti}},\ and\ \citenamefont {{Baibhav}}}]{Bouffanais2019}%
  \BibitemOpen
  \bibfield  {author} {\bibinfo {author} {\bibfnamefont {Y.}~\bibnamefont
  {{Bouffanais}}}, \bibinfo {author} {\bibfnamefont {M.}~\bibnamefont
  {{Mapelli}}}, \bibinfo {author} {\bibfnamefont {D.}~\bibnamefont {{Gerosa}}},
  \bibinfo {author} {\bibfnamefont {U.~N.}\ \bibnamefont {{Di Carlo}}},
  \bibinfo {author} {\bibfnamefont {N.}~\bibnamefont {{Giacobbo}}}, \bibinfo
  {author} {\bibfnamefont {E.}~\bibnamefont {{Berti}}}, \ and\ \bibinfo
  {author} {\bibfnamefont {V.}~\bibnamefont {{Baibhav}}},\ }\href {\doibase
  10.3847/1538-4357/ab4a79} {\bibfield  {journal} {\bibinfo  {journal} {\apj}\
  }\textbf {\bibinfo {volume} {886}},\ \bibinfo {eid} {25} (\bibinfo {year}
  {2019})},\ \Eprint {http://arxiv.org/abs/1905.11054} {arXiv:1905.11054
  [astro-ph.HE]} \BibitemShut {NoStop}%
\bibitem [{\citenamefont {{Abbott}}\ \emph
  {et~al.}(2019{\natexlab{b}})\citenamefont {{Abbott}} \emph
  {et~al.}}]{LIGO2019astro}%
  \BibitemOpen
  \bibfield  {author} {\bibinfo {author} {\bibfnamefont {B.~P.}\ \bibnamefont
  {{Abbott}}} \emph {et~al.} (\bibinfo {collaboration} {LIGO Collaboration}),\
  }\href {\doibase 10.3847/2041-8213/ab3800} {\bibfield  {journal} {\bibinfo
  {journal} {\apjl}\ }\textbf {\bibinfo {volume} {882}},\ \bibinfo {eid} {L24}
  (\bibinfo {year} {2019}{\natexlab{b}})},\ \Eprint
  {http://arxiv.org/abs/1811.12940} {arXiv:1811.12940 [astro-ph.HE]}
  \BibitemShut {NoStop}%
\bibitem [{\citenamefont {{Zevin}}\ \emph
  {et~al.}(2020{\natexlab{a}})\citenamefont {{Zevin}}, \citenamefont
  {{Bavera}}, \citenamefont {{Berry}}, \citenamefont {{Kalogera}},
  \citenamefont {{Fragos}}, \citenamefont {{Marchant}}, \citenamefont
  {{Rodriguez}}, \citenamefont {{Antonini}}, \citenamefont {{Holz}},\ and\
  \citenamefont {{Pankow}}}]{Zevin2020b}%
  \BibitemOpen
  \bibfield  {author} {\bibinfo {author} {\bibfnamefont {M.}~\bibnamefont
  {{Zevin}}}, \bibinfo {author} {\bibfnamefont {S.~S.}\ \bibnamefont
  {{Bavera}}}, \bibinfo {author} {\bibfnamefont {C.~P.~L.}\ \bibnamefont
  {{Berry}}}, \bibinfo {author} {\bibfnamefont {V.}~\bibnamefont {{Kalogera}}},
  \bibinfo {author} {\bibfnamefont {T.}~\bibnamefont {{Fragos}}}, \bibinfo
  {author} {\bibfnamefont {P.}~\bibnamefont {{Marchant}}}, \bibinfo {author}
  {\bibfnamefont {C.~L.}\ \bibnamefont {{Rodriguez}}}, \bibinfo {author}
  {\bibfnamefont {F.}~\bibnamefont {{Antonini}}}, \bibinfo {author}
  {\bibfnamefont {D.~E.}\ \bibnamefont {{Holz}}}, \ and\ \bibinfo {author}
  {\bibfnamefont {C.}~\bibnamefont {{Pankow}}},\ }\href@noop {} {\bibfield
  {journal} {\bibinfo  {journal} {arXiv e-prints}\ ,\ \bibinfo {eid}
  {arXiv:2011.10057}} (\bibinfo {year} {2020}{\natexlab{a}})}\BibitemShut
  {NoStop}%
\bibitem [{\citenamefont {{Wong}}\ \emph {et~al.}(2020)\citenamefont {{Wong}},
  \citenamefont {{Breivik}}, \citenamefont {{Kremer}},\ and\ \citenamefont
  {{Callister}}}]{Wong2020}%
  \BibitemOpen
  \bibfield  {author} {\bibinfo {author} {\bibfnamefont {K.~W.~K.}\
  \bibnamefont {{Wong}}}, \bibinfo {author} {\bibfnamefont {K.}~\bibnamefont
  {{Breivik}}}, \bibinfo {author} {\bibfnamefont {K.}~\bibnamefont {{Kremer}}},
  \ and\ \bibinfo {author} {\bibfnamefont {T.}~\bibnamefont {{Callister}}},\
  }\href@noop {} {\bibfield  {journal} {\bibinfo  {journal} {arXiv e-prints}\
  ,\ \bibinfo {eid} {arXiv:2011.03564}} (\bibinfo {year} {2020})},\ \Eprint
  {http://arxiv.org/abs/2011.03564} {arXiv:2011.03564 [astro-ph.HE]}
  \BibitemShut {NoStop}%
\bibitem [{\citenamefont {{Mandel}}\ and\ \citenamefont
  {{O'Shaughnessy}}(2010)}]{Mandel2010}%
  \BibitemOpen
  \bibfield  {author} {\bibinfo {author} {\bibfnamefont {I.}~\bibnamefont
  {{Mandel}}}\ and\ \bibinfo {author} {\bibfnamefont {R.}~\bibnamefont
  {{O'Shaughnessy}}},\ }\href {\doibase 10.1088/0264-9381/27/11/114007}
  {\bibfield  {journal} {\bibinfo  {journal} {Classical and Quantum Gravity}\
  }\textbf {\bibinfo {volume} {27}},\ \bibinfo {eid} {114007} (\bibinfo {year}
  {2010})},\ \Eprint {http://arxiv.org/abs/0912.1074} {arXiv:0912.1074
  [astro-ph.HE]} \BibitemShut {NoStop}%
\bibitem [{\citenamefont {{Gerosa}}\ \emph {et~al.}(2013)\citenamefont
  {{Gerosa}}, \citenamefont {{Kesden}}, \citenamefont {{Berti}}, \citenamefont
  {{O'Shaughnessy}},\ and\ \citenamefont {{Sperhake}}}]{Gerosa2013}%
  \BibitemOpen
  \bibfield  {author} {\bibinfo {author} {\bibfnamefont {D.}~\bibnamefont
  {{Gerosa}}}, \bibinfo {author} {\bibfnamefont {M.}~\bibnamefont {{Kesden}}},
  \bibinfo {author} {\bibfnamefont {E.}~\bibnamefont {{Berti}}}, \bibinfo
  {author} {\bibfnamefont {R.}~\bibnamefont {{O'Shaughnessy}}}, \ and\ \bibinfo
  {author} {\bibfnamefont {U.}~\bibnamefont {{Sperhake}}},\ }\href {\doibase
  10.1103/PhysRevD.87.104028} {\bibfield  {journal} {\bibinfo  {journal}
  {\prd}\ }\textbf {\bibinfo {volume} {87}},\ \bibinfo {eid} {104028} (\bibinfo
  {year} {2013})},\ \Eprint {http://arxiv.org/abs/1302.4442} {arXiv:1302.4442
  [gr-qc]} \BibitemShut {NoStop}%
\bibitem [{\citenamefont {{Rodriguez}}\ \emph {et~al.}(2016)\citenamefont
  {{Rodriguez}}, \citenamefont {{Zevin}}, \citenamefont {{Pankow}},
  \citenamefont {{Kalogera}},\ and\ \citenamefont {{Rasio}}}]{Rodriguez2016}%
  \BibitemOpen
  \bibfield  {author} {\bibinfo {author} {\bibfnamefont {C.~L.}\ \bibnamefont
  {{Rodriguez}}}, \bibinfo {author} {\bibfnamefont {M.}~\bibnamefont
  {{Zevin}}}, \bibinfo {author} {\bibfnamefont {C.}~\bibnamefont {{Pankow}}},
  \bibinfo {author} {\bibfnamefont {V.}~\bibnamefont {{Kalogera}}}, \ and\
  \bibinfo {author} {\bibfnamefont {F.~A.}\ \bibnamefont {{Rasio}}},\ }\href
  {\doibase 10.3847/2041-8205/832/1/L2} {\bibfield  {journal} {\bibinfo
  {journal} {\apjl}\ }\textbf {\bibinfo {volume} {832}},\ \bibinfo {eid} {L2}
  (\bibinfo {year} {2016})},\ \Eprint {http://arxiv.org/abs/1609.05916}
  {arXiv:1609.05916 [astro-ph.HE]} \BibitemShut {NoStop}%
\bibitem [{\citenamefont {{Talbot}}\ and\ \citenamefont
  {{Thrane}}(2017)}]{Talbot2017}%
  \BibitemOpen
  \bibfield  {author} {\bibinfo {author} {\bibfnamefont {C.}~\bibnamefont
  {{Talbot}}}\ and\ \bibinfo {author} {\bibfnamefont {E.}~\bibnamefont
  {{Thrane}}},\ }\href {\doibase 10.1103/PhysRevD.96.023012} {\bibfield
  {journal} {\bibinfo  {journal} {\prd}\ }\textbf {\bibinfo {volume} {96}},\
  \bibinfo {eid} {023012} (\bibinfo {year} {2017})},\ \Eprint
  {http://arxiv.org/abs/1704.08370} {arXiv:1704.08370 [astro-ph.HE]}
  \BibitemShut {NoStop}%
\bibitem [{\citenamefont {{Farr}}\ \emph {et~al.}(2017)\citenamefont {{Farr}},
  \citenamefont {{Stevenson}}, \citenamefont {{Miller}}, \citenamefont
  {{Mandel}}, \citenamefont {{Farr}},\ and\ \citenamefont
  {{Vecchio}}}]{Farr2017}%
  \BibitemOpen
  \bibfield  {author} {\bibinfo {author} {\bibfnamefont {W.~M.}\ \bibnamefont
  {{Farr}}}, \bibinfo {author} {\bibfnamefont {S.}~\bibnamefont {{Stevenson}}},
  \bibinfo {author} {\bibfnamefont {M.~C.}\ \bibnamefont {{Miller}}}, \bibinfo
  {author} {\bibfnamefont {I.}~\bibnamefont {{Mandel}}}, \bibinfo {author}
  {\bibfnamefont {B.}~\bibnamefont {{Farr}}}, \ and\ \bibinfo {author}
  {\bibfnamefont {A.}~\bibnamefont {{Vecchio}}},\ }\href {\doibase
  10.1038/nature23453} {\bibfield  {journal} {\bibinfo  {journal} {\nat}\
  }\textbf {\bibinfo {volume} {548}},\ \bibinfo {pages} {426} (\bibinfo {year}
  {2017})},\ \Eprint {http://arxiv.org/abs/1706.01385} {arXiv:1706.01385
  [astro-ph.HE]} \BibitemShut {NoStop}%
\bibitem [{\citenamefont {{Stevenson}}\ \emph
  {et~al.}(2017{\natexlab{a}})\citenamefont {{Stevenson}}, \citenamefont
  {{Berry}},\ and\ \citenamefont {{Mandel}}}]{Stevenson2017}%
  \BibitemOpen
  \bibfield  {author} {\bibinfo {author} {\bibfnamefont {S.}~\bibnamefont
  {{Stevenson}}}, \bibinfo {author} {\bibfnamefont {C.~P.~L.}\ \bibnamefont
  {{Berry}}}, \ and\ \bibinfo {author} {\bibfnamefont {I.}~\bibnamefont
  {{Mandel}}},\ }\href {\doibase 10.1093/mnras/stx1764} {\bibfield  {journal}
  {\bibinfo  {journal} {\mnras}\ }\textbf {\bibinfo {volume} {471}},\ \bibinfo
  {pages} {2801} (\bibinfo {year} {2017}{\natexlab{a}})},\ \Eprint
  {http://arxiv.org/abs/1703.06873} {arXiv:1703.06873 [astro-ph.HE]}
  \BibitemShut {NoStop}%
\bibitem [{\citenamefont {{Gerosa}}\ \emph {et~al.}(2018)\citenamefont
  {{Gerosa}}, \citenamefont {{Berti}}, \citenamefont {{O'Shaughnessy}},
  \citenamefont {{Belczynski}}, \citenamefont {{Kesden}}, \citenamefont
  {{Wysocki}},\ and\ \citenamefont {{Gladysz}}}]{Gerosa2018}%
  \BibitemOpen
  \bibfield  {author} {\bibinfo {author} {\bibfnamefont {D.}~\bibnamefont
  {{Gerosa}}}, \bibinfo {author} {\bibfnamefont {E.}~\bibnamefont {{Berti}}},
  \bibinfo {author} {\bibfnamefont {R.}~\bibnamefont {{O'Shaughnessy}}},
  \bibinfo {author} {\bibfnamefont {K.}~\bibnamefont {{Belczynski}}}, \bibinfo
  {author} {\bibfnamefont {M.}~\bibnamefont {{Kesden}}}, \bibinfo {author}
  {\bibfnamefont {D.}~\bibnamefont {{Wysocki}}}, \ and\ \bibinfo {author}
  {\bibfnamefont {W.}~\bibnamefont {{Gladysz}}},\ }\href {\doibase
  10.1103/PhysRevD.98.084036} {\bibfield  {journal} {\bibinfo  {journal}
  {\prd}\ }\textbf {\bibinfo {volume} {98}},\ \bibinfo {eid} {084036} (\bibinfo
  {year} {2018})},\ \Eprint {http://arxiv.org/abs/1808.02491} {arXiv:1808.02491
  [astro-ph.HE]} \BibitemShut {NoStop}%
\bibitem [{\citenamefont {{Farr}}\ \emph {et~al.}(2018)\citenamefont {{Farr}},
  \citenamefont {{Holz}},\ and\ \citenamefont {{Farr}}}]{Farr2018}%
  \BibitemOpen
  \bibfield  {author} {\bibinfo {author} {\bibfnamefont {B.}~\bibnamefont
  {{Farr}}}, \bibinfo {author} {\bibfnamefont {D.~E.}\ \bibnamefont {{Holz}}},
  \ and\ \bibinfo {author} {\bibfnamefont {W.~M.}\ \bibnamefont {{Farr}}},\
  }\href {\doibase 10.3847/2041-8213/aaaa64} {\bibfield  {journal} {\bibinfo
  {journal} {\apjl}\ }\textbf {\bibinfo {volume} {854}},\ \bibinfo {eid} {L9}
  (\bibinfo {year} {2018})},\ \Eprint {http://arxiv.org/abs/1709.07896}
  {arXiv:1709.07896 [astro-ph.HE]} \BibitemShut {NoStop}%
\bibitem [{\citenamefont {{Talbot}}\ and\ \citenamefont
  {{Thrane}}(2018)}]{Talbot2018}%
  \BibitemOpen
  \bibfield  {author} {\bibinfo {author} {\bibfnamefont {C.}~\bibnamefont
  {{Talbot}}}\ and\ \bibinfo {author} {\bibfnamefont {E.}~\bibnamefont
  {{Thrane}}},\ }\href {\doibase 10.3847/1538-4357/aab34c} {\bibfield
  {journal} {\bibinfo  {journal} {\apj}\ }\textbf {\bibinfo {volume} {856}},\
  \bibinfo {eid} {173} (\bibinfo {year} {2018})},\ \Eprint
  {http://arxiv.org/abs/1801.02699} {arXiv:1801.02699 [astro-ph.HE]}
  \BibitemShut {NoStop}%
\bibitem [{\citenamefont {{Wysocki}}\ \emph {et~al.}(2018)\citenamefont
  {{Wysocki}}, \citenamefont {{Gerosa}}, \citenamefont {{O'Shaughnessy}},
  \citenamefont {{Belczynski}}, \citenamefont {{Gladysz}}, \citenamefont
  {{Berti}}, \citenamefont {{Kesden}},\ and\ \citenamefont
  {{Holz}}}]{Wysocki2018}%
  \BibitemOpen
  \bibfield  {author} {\bibinfo {author} {\bibfnamefont {D.}~\bibnamefont
  {{Wysocki}}}, \bibinfo {author} {\bibfnamefont {D.}~\bibnamefont {{Gerosa}}},
  \bibinfo {author} {\bibfnamefont {R.}~\bibnamefont {{O'Shaughnessy}}},
  \bibinfo {author} {\bibfnamefont {K.}~\bibnamefont {{Belczynski}}}, \bibinfo
  {author} {\bibfnamefont {W.}~\bibnamefont {{Gladysz}}}, \bibinfo {author}
  {\bibfnamefont {E.}~\bibnamefont {{Berti}}}, \bibinfo {author} {\bibfnamefont
  {M.}~\bibnamefont {{Kesden}}}, \ and\ \bibinfo {author} {\bibfnamefont
  {D.~E.}\ \bibnamefont {{Holz}}},\ }\href {\doibase
  10.1103/PhysRevD.97.043014} {\bibfield  {journal} {\bibinfo  {journal}
  {\prd}\ }\textbf {\bibinfo {volume} {97}},\ \bibinfo {eid} {043014} (\bibinfo
  {year} {2018})},\ \Eprint {http://arxiv.org/abs/1709.01943} {arXiv:1709.01943
  [astro-ph.HE]} \BibitemShut {NoStop}%
\bibitem [{\citenamefont {{Miller}}\ \emph {et~al.}(2020)\citenamefont
  {{Miller}}, \citenamefont {{Callister}},\ and\ \citenamefont
  {{Farr}}}]{Miller2020}%
  \BibitemOpen
  \bibfield  {author} {\bibinfo {author} {\bibfnamefont {S.}~\bibnamefont
  {{Miller}}}, \bibinfo {author} {\bibfnamefont {T.~A.}\ \bibnamefont
  {{Callister}}}, \ and\ \bibinfo {author} {\bibfnamefont {W.~M.}\ \bibnamefont
  {{Farr}}},\ }\href {\doibase 10.3847/1538-4357/ab80c0} {\bibfield  {journal}
  {\bibinfo  {journal} {\apj}\ }\textbf {\bibinfo {volume} {895}},\ \bibinfo
  {eid} {128} (\bibinfo {year} {2020})},\ \Eprint
  {http://arxiv.org/abs/2001.06051} {arXiv:2001.06051 [astro-ph.HE]}
  \BibitemShut {NoStop}%
\bibitem [{\citenamefont {{Callister}}\ \emph {et~al.}(2020)\citenamefont
  {{Callister}}, \citenamefont {{Farr}},\ and\ \citenamefont
  {{Renzo}}}]{Callister2020}%
  \BibitemOpen
  \bibfield  {author} {\bibinfo {author} {\bibfnamefont {T.~A.}\ \bibnamefont
  {{Callister}}}, \bibinfo {author} {\bibfnamefont {W.~M.}\ \bibnamefont
  {{Farr}}}, \ and\ \bibinfo {author} {\bibfnamefont {M.}~\bibnamefont
  {{Renzo}}},\ }\href@noop {} {\bibfield  {journal} {\bibinfo  {journal} {arXiv
  e-prints}\ ,\ \bibinfo {eid} {arXiv:2011.09570}} (\bibinfo {year} {2020})},\
  \Eprint {http://arxiv.org/abs/2011.09570} {arXiv:2011.09570 [astro-ph.HE]}
  \BibitemShut {NoStop}%
\bibitem [{\citenamefont {{Kobulnicky}}\ and\ \citenamefont
  {{Fryer}}(2007)}]{Kobulnicky2007}%
  \BibitemOpen
  \bibfield  {author} {\bibinfo {author} {\bibfnamefont {H.~A.}\ \bibnamefont
  {{Kobulnicky}}}\ and\ \bibinfo {author} {\bibfnamefont {C.~L.}\ \bibnamefont
  {{Fryer}}},\ }\href {\doibase 10.1086/522073} {\bibfield  {journal} {\bibinfo
   {journal} {\apj}\ }\textbf {\bibinfo {volume} {670}},\ \bibinfo {pages}
  {747} (\bibinfo {year} {2007})}\BibitemShut {NoStop}%
\bibitem [{\citenamefont {{Sana}}\ \emph {et~al.}(2012)\citenamefont {{Sana}},
  \citenamefont {{de Mink}}, \citenamefont {{de Koter}}, \citenamefont
  {{Langer}}, \citenamefont {{Evans}}, \citenamefont {{Gieles}}, \citenamefont
  {{Gosset}}, \citenamefont {{Izzard}}, \citenamefont {{Le Bouquin}},\ and\
  \citenamefont {{Schneider}}}]{Sana2012}%
  \BibitemOpen
  \bibfield  {author} {\bibinfo {author} {\bibfnamefont {H.}~\bibnamefont
  {{Sana}}}, \bibinfo {author} {\bibfnamefont {S.~E.}\ \bibnamefont {{de
  Mink}}}, \bibinfo {author} {\bibfnamefont {A.}~\bibnamefont {{de Koter}}},
  \bibinfo {author} {\bibfnamefont {N.}~\bibnamefont {{Langer}}}, \bibinfo
  {author} {\bibfnamefont {C.~J.}\ \bibnamefont {{Evans}}}, \bibinfo {author}
  {\bibfnamefont {M.}~\bibnamefont {{Gieles}}}, \bibinfo {author}
  {\bibfnamefont {E.}~\bibnamefont {{Gosset}}}, \bibinfo {author}
  {\bibfnamefont {R.~G.}\ \bibnamefont {{Izzard}}}, \bibinfo {author}
  {\bibfnamefont {J.~B.}\ \bibnamefont {{Le Bouquin}}}, \ and\ \bibinfo
  {author} {\bibfnamefont {F.~R.~N.}\ \bibnamefont {{Schneider}}},\ }\href
  {\doibase 10.1126/science.1223344} {\bibfield  {journal} {\bibinfo  {journal}
  {Science}\ }\textbf {\bibinfo {volume} {337}},\ \bibinfo {pages} {444}
  (\bibinfo {year} {2012})},\ \Eprint {http://arxiv.org/abs/1207.6397}
  {arXiv:1207.6397 [astro-ph.SR]} \BibitemShut {NoStop}%
\bibitem [{\citenamefont {{Moe}}\ and\ \citenamefont {{Di
  Stefano}}(2017)}]{Moe2017}%
  \BibitemOpen
  \bibfield  {author} {\bibinfo {author} {\bibfnamefont {M.}~\bibnamefont
  {{Moe}}}\ and\ \bibinfo {author} {\bibfnamefont {R.}~\bibnamefont {{Di
  Stefano}}},\ }\href {\doibase 10.3847/1538-4365/aa6fb6} {\bibfield  {journal}
  {\bibinfo  {journal} {\apjs}\ }\textbf {\bibinfo {volume} {230}},\ \bibinfo
  {pages} {15} (\bibinfo {year} {2017})},\ \Eprint
  {http://arxiv.org/abs/1606.05347} {arXiv:1606.05347 [astro-ph.SR]}
  \BibitemShut {NoStop}%
\bibitem [{\citenamefont {{Ivanova}}\ \emph {et~al.}(2013)\citenamefont
  {{Ivanova}}, \citenamefont {{Justham}}, \citenamefont {{Chen}}, \citenamefont
  {{De Marco}}, \citenamefont {{Fryer}}, \citenamefont {{Gaburov}},
  \citenamefont {{Ge}}, \citenamefont {{Glebbeek}}, \citenamefont {{Han}},
  \citenamefont {{Li}}, \citenamefont {{Lu}}, \citenamefont {{Marsh}},
  \citenamefont {{Podsiadlowski}}, \citenamefont {{Potter}}, \citenamefont
  {{Soker}}, \citenamefont {{Taam}}, \citenamefont {{Tauris}}, \citenamefont
  {{van den Heuvel}},\ and\ \citenamefont {{Webbink}}}]{Ivanova2013}%
  \BibitemOpen
  \bibfield  {author} {\bibinfo {author} {\bibfnamefont {N.}~\bibnamefont
  {{Ivanova}}}, \bibinfo {author} {\bibfnamefont {S.}~\bibnamefont
  {{Justham}}}, \bibinfo {author} {\bibfnamefont {X.}~\bibnamefont {{Chen}}},
  \bibinfo {author} {\bibfnamefont {O.}~\bibnamefont {{De Marco}}}, \bibinfo
  {author} {\bibfnamefont {C.~L.}\ \bibnamefont {{Fryer}}}, \bibinfo {author}
  {\bibfnamefont {E.}~\bibnamefont {{Gaburov}}}, \bibinfo {author}
  {\bibfnamefont {H.}~\bibnamefont {{Ge}}}, \bibinfo {author} {\bibfnamefont
  {E.}~\bibnamefont {{Glebbeek}}}, \bibinfo {author} {\bibfnamefont
  {Z.}~\bibnamefont {{Han}}}, \bibinfo {author} {\bibfnamefont {X.~D.}\
  \bibnamefont {{Li}}}, \bibinfo {author} {\bibfnamefont {G.}~\bibnamefont
  {{Lu}}}, \bibinfo {author} {\bibfnamefont {T.}~\bibnamefont {{Marsh}}},
  \bibinfo {author} {\bibfnamefont {P.}~\bibnamefont {{Podsiadlowski}}},
  \bibinfo {author} {\bibfnamefont {A.}~\bibnamefont {{Potter}}}, \bibinfo
  {author} {\bibfnamefont {N.}~\bibnamefont {{Soker}}}, \bibinfo {author}
  {\bibfnamefont {R.}~\bibnamefont {{Taam}}}, \bibinfo {author} {\bibfnamefont
  {T.~M.}\ \bibnamefont {{Tauris}}}, \bibinfo {author} {\bibfnamefont
  {E.~P.~J.}\ \bibnamefont {{van den Heuvel}}}, \ and\ \bibinfo {author}
  {\bibfnamefont {R.~F.}\ \bibnamefont {{Webbink}}},\ }\href {\doibase
  10.1007/s00159-013-0059-2} {\bibfield  {journal} {\bibinfo  {journal}
  {\aapr}\ }\textbf {\bibinfo {volume} {21}},\ \bibinfo {pages} {59} (\bibinfo
  {year} {2013})}\BibitemShut {NoStop}%
\bibitem [{\citenamefont {{Postnov}}\ and\ \citenamefont
  {{Yungelson}}(2014)}]{Postnov2014}%
  \BibitemOpen
  \bibfield  {author} {\bibinfo {author} {\bibfnamefont {K.~A.}\ \bibnamefont
  {{Postnov}}}\ and\ \bibinfo {author} {\bibfnamefont {L.~R.}\ \bibnamefont
  {{Yungelson}}},\ }\href {\doibase 10.12942/lrr-2014-3} {\bibfield  {journal}
  {\bibinfo  {journal} {Living Reviews in Relativity}\ }\textbf {\bibinfo
  {volume} {17}},\ \bibinfo {eid} {3} (\bibinfo {year} {2014})},\ \Eprint
  {http://arxiv.org/abs/1403.4754} {arXiv:1403.4754 [astro-ph.HE]} \BibitemShut
  {NoStop}%
\bibitem [{\citenamefont {{Schr{\o}der}}\ \emph {et~al.}(2020)\citenamefont
  {{Schr{\o}der}}, \citenamefont {{MacLeod}}, \citenamefont {{Loeb}},
  \citenamefont {{Vigna-G{\'o}mez}},\ and\ \citenamefont
  {{Mandel}}}]{Schroder2020}%
  \BibitemOpen
  \bibfield  {author} {\bibinfo {author} {\bibfnamefont {S.~L.}\ \bibnamefont
  {{Schr{\o}der}}}, \bibinfo {author} {\bibfnamefont {M.}~\bibnamefont
  {{MacLeod}}}, \bibinfo {author} {\bibfnamefont {A.}~\bibnamefont {{Loeb}}},
  \bibinfo {author} {\bibfnamefont {A.}~\bibnamefont {{Vigna-G{\'o}mez}}}, \
  and\ \bibinfo {author} {\bibfnamefont {I.}~\bibnamefont {{Mandel}}},\ }\href
  {\doibase 10.3847/1538-4357/ab7014} {\bibfield  {journal} {\bibinfo
  {journal} {\apj}\ }\textbf {\bibinfo {volume} {892}},\ \bibinfo {eid} {13}
  (\bibinfo {year} {2020})},\ \Eprint {http://arxiv.org/abs/1906.04189}
  {arXiv:1906.04189 [astro-ph.HE]} \BibitemShut {NoStop}%
\bibitem [{\citenamefont {{Mosqueda}}\ and\ \citenamefont
  {{Koenigsberger}}(1990)}]{WRprogenitors1990}%
  \BibitemOpen
  \bibfield  {author} {\bibinfo {author} {\bibfnamefont {A.}~\bibnamefont
  {{Mosqueda}}}\ and\ \bibinfo {author} {\bibfnamefont {G.}~\bibnamefont
  {{Koenigsberger}}},\ }\href@noop {} {\bibfield  {journal} {\bibinfo
  {journal} {\rmxaa}\ }\textbf {\bibinfo {volume} {20}},\ \bibinfo {pages} {79}
  (\bibinfo {year} {1990})}\BibitemShut {NoStop}%
\bibitem [{\citenamefont {{Tutukov}}\ and\ \citenamefont
  {{Cherepashchuk}}(2003)}]{Tutukov2003}%
  \BibitemOpen
  \bibfield  {author} {\bibinfo {author} {\bibfnamefont {A.~V.}\ \bibnamefont
  {{Tutukov}}}\ and\ \bibinfo {author} {\bibfnamefont {A.~M.}\ \bibnamefont
  {{Cherepashchuk}}},\ }\href {\doibase 10.1134/1.1575853} {\bibfield
  {journal} {\bibinfo  {journal} {Astronomy Reports}\ }\textbf {\bibinfo
  {volume} {47}},\ \bibinfo {pages} {386} (\bibinfo {year} {2003})}\BibitemShut
  {NoStop}%
\bibitem [{\citenamefont {{Hainich}}\ \emph {et~al.}(2018)\citenamefont
  {{Hainich}}, \citenamefont {{Oskinova}}, \citenamefont {{Shenar}},
  \citenamefont {{Marchant}}, \citenamefont {{Eldridge}}, \citenamefont
  {{Sander}}, \citenamefont {{Hamann}}, \citenamefont {{Langer}},\ and\
  \citenamefont {{Todt}}}]{WRprogenitors2018}%
  \BibitemOpen
  \bibfield  {author} {\bibinfo {author} {\bibfnamefont {R.}~\bibnamefont
  {{Hainich}}}, \bibinfo {author} {\bibfnamefont {L.~M.}\ \bibnamefont
  {{Oskinova}}}, \bibinfo {author} {\bibfnamefont {T.}~\bibnamefont
  {{Shenar}}}, \bibinfo {author} {\bibfnamefont {P.}~\bibnamefont
  {{Marchant}}}, \bibinfo {author} {\bibfnamefont {J.~J.}\ \bibnamefont
  {{Eldridge}}}, \bibinfo {author} {\bibfnamefont {A.~A.~C.}\ \bibnamefont
  {{Sander}}}, \bibinfo {author} {\bibfnamefont {W.~R.}\ \bibnamefont
  {{Hamann}}}, \bibinfo {author} {\bibfnamefont {N.}~\bibnamefont {{Langer}}},
  \ and\ \bibinfo {author} {\bibfnamefont {H.}~\bibnamefont {{Todt}}},\ }\href
  {\doibase 10.1051/0004-6361/201731449} {\bibfield  {journal} {\bibinfo
  {journal} {\aap}\ }\textbf {\bibinfo {volume} {609}},\ \bibinfo {eid} {A94}
  (\bibinfo {year} {2018})},\ \Eprint {http://arxiv.org/abs/1707.01912}
  {arXiv:1707.01912 [astro-ph.SR]} \BibitemShut {NoStop}%
\bibitem [{\citenamefont {{Meynet}}\ and\ \citenamefont
  {{Maeder}}(2005)}]{Meynet2005}%
  \BibitemOpen
  \bibfield  {author} {\bibinfo {author} {\bibfnamefont {G.}~\bibnamefont
  {{Meynet}}}\ and\ \bibinfo {author} {\bibfnamefont {A.}~\bibnamefont
  {{Maeder}}},\ }\href {\doibase 10.1051/0004-6361:20047106} {\bibfield
  {journal} {\bibinfo  {journal} {\aap}\ }\textbf {\bibinfo {volume} {429}},\
  \bibinfo {pages} {581} (\bibinfo {year} {2005})},\ \Eprint
  {http://arxiv.org/abs/astro-ph/0408319} {arXiv:astro-ph/0408319 [astro-ph]}
  \BibitemShut {NoStop}%
\bibitem [{\citenamefont {{Maeder}}\ and\ \citenamefont
  {{Meynet}}(2012)}]{Maeder2012}%
  \BibitemOpen
  \bibfield  {author} {\bibinfo {author} {\bibfnamefont {A.}~\bibnamefont
  {{Maeder}}}\ and\ \bibinfo {author} {\bibfnamefont {G.}~\bibnamefont
  {{Meynet}}},\ }\href {\doibase 10.1103/RevModPhys.84.25} {\bibfield
  {journal} {\bibinfo  {journal} {Reviews of Modern Physics}\ }\textbf
  {\bibinfo {volume} {84}},\ \bibinfo {pages} {25} (\bibinfo {year}
  {2012})}\BibitemShut {NoStop}%
\bibitem [{\citenamefont {{Meynet}}\ \emph {et~al.}(2013)\citenamefont
  {{Meynet}}, \citenamefont {{Ekstrom}}, \citenamefont {{Maeder}},
  \citenamefont {{Eggenberger}}, \citenamefont {{Saio}}, \citenamefont
  {{Chomienne}},\ and\ \citenamefont {{Haemmerl{\'e}}}}]{Meynet2013}%
  \BibitemOpen
  \bibfield  {author} {\bibinfo {author} {\bibfnamefont {G.}~\bibnamefont
  {{Meynet}}}, \bibinfo {author} {\bibfnamefont {S.}~\bibnamefont {{Ekstrom}}},
  \bibinfo {author} {\bibfnamefont {A.}~\bibnamefont {{Maeder}}}, \bibinfo
  {author} {\bibfnamefont {P.}~\bibnamefont {{Eggenberger}}}, \bibinfo {author}
  {\bibfnamefont {H.}~\bibnamefont {{Saio}}}, \bibinfo {author} {\bibfnamefont
  {V.}~\bibnamefont {{Chomienne}}}, \ and\ \bibinfo {author} {\bibfnamefont
  {L.}~\bibnamefont {{Haemmerl{\'e}}}},\ }\enquote {\bibinfo {title} {{Models
  of Rotating Massive Stars: Impacts of Various Prescriptions}},}\ in\ \href
  {\doibase 10.1007/978-3-642-33380-4_1} {\emph {\bibinfo {booktitle} {Lecture
  Notes in Physics, Berlin Springer Verlag}}},\ Vol.\ \bibinfo {volume} {865},\
  \bibinfo {editor} {edited by\ \bibinfo {editor} {\bibfnamefont
  {M.}~\bibnamefont {{Goupil}}}, \bibinfo {editor} {\bibfnamefont
  {K.}~\bibnamefont {{Belkacem}}}, \bibinfo {editor} {\bibfnamefont
  {C.}~\bibnamefont {{Neiner}}}, \bibinfo {editor} {\bibfnamefont
  {F.}~\bibnamefont {{Ligni{\`e}res}}}, \ and\ \bibinfo {editor} {\bibfnamefont
  {J.~J.}\ \bibnamefont {{Green}}}}\ (\bibinfo {year} {2013})\ p.~\bibinfo
  {pages} {3}\BibitemShut {NoStop}%
\bibitem [{\citenamefont {{Kalogera}}(2000)}]{Kalogera2000}%
  \BibitemOpen
  \bibfield  {author} {\bibinfo {author} {\bibfnamefont {V.}~\bibnamefont
  {{Kalogera}}},\ }\href {\doibase 10.1086/309400} {\bibfield  {journal}
  {\bibinfo  {journal} {\apj}\ }\textbf {\bibinfo {volume} {541}},\ \bibinfo
  {pages} {319} (\bibinfo {year} {2000})},\ \Eprint
  {http://arxiv.org/abs/astro-ph/9911417} {arXiv:astro-ph/9911417 [astro-ph]}
  \BibitemShut {NoStop}%
\bibitem [{\citenamefont {{Penrose}}(1969)}]{Penrose1969}%
  \BibitemOpen
  \bibfield  {author} {\bibinfo {author} {\bibfnamefont {R.}~\bibnamefont
  {{Penrose}}},\ }\href@noop {} {\bibfield  {journal} {\bibinfo  {journal}
  {Nuovo Cimento Rivista Serie}\ }\textbf {\bibinfo {volume} {1}},\ \bibinfo
  {pages} {252} (\bibinfo {year} {1969})}\BibitemShut {NoStop}%
\bibitem [{\citenamefont {{Heger}}\ \emph {et~al.}(2003)\citenamefont
  {{Heger}}, \citenamefont {{Fryer}}, \citenamefont {{Woosley}}, \citenamefont
  {{Langer}},\ and\ \citenamefont {{Hartmann}}}]{Heger2003}%
  \BibitemOpen
  \bibfield  {author} {\bibinfo {author} {\bibfnamefont {A.}~\bibnamefont
  {{Heger}}}, \bibinfo {author} {\bibfnamefont {C.~L.}\ \bibnamefont
  {{Fryer}}}, \bibinfo {author} {\bibfnamefont {S.~E.}\ \bibnamefont
  {{Woosley}}}, \bibinfo {author} {\bibfnamefont {N.}~\bibnamefont {{Langer}}},
  \ and\ \bibinfo {author} {\bibfnamefont {D.~H.}\ \bibnamefont {{Hartmann}}},\
  }\href {\doibase 10.1086/375341} {\bibfield  {journal} {\bibinfo  {journal}
  {\apj}\ }\textbf {\bibinfo {volume} {591}},\ \bibinfo {pages} {288} (\bibinfo
  {year} {2003})},\ \Eprint {http://arxiv.org/abs/astro-ph/0212469}
  {arXiv:astro-ph/0212469 [astro-ph]} \BibitemShut {NoStop}%
\bibitem [{\citenamefont {{O'Connor}}(2017)}]{OConnor2017}%
  \BibitemOpen
  \bibfield  {author} {\bibinfo {author} {\bibfnamefont {E.}~\bibnamefont
  {{O'Connor}}},\ }\enquote {\bibinfo {title} {{The Core-Collapse
  Supernova-Black Hole Connection}},}\ in\ \href {\doibase
  10.1007/978-3-319-21846-5_129} {\emph {\bibinfo {booktitle} {Handbook of
  Supernovae}}},\ \bibinfo {editor} {edited by\ \bibinfo {editor}
  {\bibfnamefont {A.~W.}\ \bibnamefont {{Alsabti}}}\ and\ \bibinfo {editor}
  {\bibfnamefont {P.}~\bibnamefont {{Murdin}}}}\ (\bibinfo {year} {2017})\ p.\
  \bibinfo {pages} {1555}\BibitemShut {NoStop}%
\bibitem [{\citenamefont {{Damour}}(2001)}]{Damour2001}%
  \BibitemOpen
  \bibfield  {author} {\bibinfo {author} {\bibfnamefont {T.}~\bibnamefont
  {{Damour}}},\ }\href {\doibase 10.1103/PhysRevD.64.124013} {\bibfield
  {journal} {\bibinfo  {journal} {\prd}\ }\textbf {\bibinfo {volume} {64}},\
  \bibinfo {pages} {124013} (\bibinfo {year} {2001})},\ \Eprint
  {http://arxiv.org/abs/gr-qc/0103018} {arXiv:gr-qc/0103018 [gr-qc]}
  \BibitemShut {NoStop}%
\bibitem [{\citenamefont {{Racine}}(2008)}]{Racine2008}%
  \BibitemOpen
  \bibfield  {author} {\bibinfo {author} {\bibfnamefont {{\'E}.}~\bibnamefont
  {{Racine}}},\ }\href {\doibase 10.1103/PhysRevD.78.044021} {\bibfield
  {journal} {\bibinfo  {journal} {\prd}\ }\textbf {\bibinfo {volume} {78}},\
  \bibinfo {eid} {044021} (\bibinfo {year} {2008})},\ \Eprint
  {http://arxiv.org/abs/0803.1820} {arXiv:0803.1820 [gr-qc]} \BibitemShut
  {NoStop}%
\bibitem [{\citenamefont {{Kesden}}\ \emph {et~al.}(2015)\citenamefont
  {{Kesden}}, \citenamefont {{Gerosa}}, \citenamefont {{O'Shaughnessy}},
  \citenamefont {{Berti}},\ and\ \citenamefont {{Sperhake}}}]{Kesden2015}%
  \BibitemOpen
  \bibfield  {author} {\bibinfo {author} {\bibfnamefont {M.}~\bibnamefont
  {{Kesden}}}, \bibinfo {author} {\bibfnamefont {D.}~\bibnamefont {{Gerosa}}},
  \bibinfo {author} {\bibfnamefont {R.}~\bibnamefont {{O'Shaughnessy}}},
  \bibinfo {author} {\bibfnamefont {E.}~\bibnamefont {{Berti}}}, \ and\
  \bibinfo {author} {\bibfnamefont {U.}~\bibnamefont {{Sperhake}}},\ }\href
  {\doibase 10.1103/PhysRevLett.114.081103} {\bibfield  {journal} {\bibinfo
  {journal} {\prl}\ }\textbf {\bibinfo {volume} {114}},\ \bibinfo {eid}
  {081103} (\bibinfo {year} {2015})},\ \Eprint {http://arxiv.org/abs/1411.0674}
  {arXiv:1411.0674 [gr-qc]} \BibitemShut {NoStop}%
\bibitem [{\citenamefont {{Gerosa}}\ \emph {et~al.}(2015)\citenamefont
  {{Gerosa}}, \citenamefont {{Kesden}}, \citenamefont {{Sperhake}},
  \citenamefont {{Berti}},\ and\ \citenamefont {{O'Shaughnessy}}}]{Gerosa2015}%
  \BibitemOpen
  \bibfield  {author} {\bibinfo {author} {\bibfnamefont {D.}~\bibnamefont
  {{Gerosa}}}, \bibinfo {author} {\bibfnamefont {M.}~\bibnamefont {{Kesden}}},
  \bibinfo {author} {\bibfnamefont {U.}~\bibnamefont {{Sperhake}}}, \bibinfo
  {author} {\bibfnamefont {E.}~\bibnamefont {{Berti}}}, \ and\ \bibinfo
  {author} {\bibfnamefont {R.}~\bibnamefont {{O'Shaughnessy}}},\ }\href
  {\doibase 10.1103/PhysRevD.92.064016} {\bibfield  {journal} {\bibinfo
  {journal} {\prd}\ }\textbf {\bibinfo {volume} {92}},\ \bibinfo {eid} {064016}
  (\bibinfo {year} {2015})},\ \Eprint {http://arxiv.org/abs/1506.03492}
  {arXiv:1506.03492 [gr-qc]} \BibitemShut {NoStop}%
\bibitem [{\citenamefont {{Vitale}}\ \emph
  {et~al.}(2017{\natexlab{b}})\citenamefont {{Vitale}}, \citenamefont
  {{Gerosa}}, \citenamefont {{Haster}}, \citenamefont {{Chatziioannou}},\ and\
  \citenamefont {{Zimmerman}}}]{Vitale2017b}%
  \BibitemOpen
  \bibfield  {author} {\bibinfo {author} {\bibfnamefont {S.}~\bibnamefont
  {{Vitale}}}, \bibinfo {author} {\bibfnamefont {D.}~\bibnamefont {{Gerosa}}},
  \bibinfo {author} {\bibfnamefont {C.-J.}\ \bibnamefont {{Haster}}}, \bibinfo
  {author} {\bibfnamefont {K.}~\bibnamefont {{Chatziioannou}}}, \ and\ \bibinfo
  {author} {\bibfnamefont {A.}~\bibnamefont {{Zimmerman}}},\ }\href {\doibase
  10.1103/PhysRevLett.119.251103} {\bibfield  {journal} {\bibinfo  {journal}
  {\prl}\ }\textbf {\bibinfo {volume} {119}},\ \bibinfo {eid} {251103}
  (\bibinfo {year} {2017}{\natexlab{b}})},\ \Eprint
  {http://arxiv.org/abs/1707.04637} {arXiv:1707.04637 [gr-qc]} \BibitemShut
  {NoStop}%
\bibitem [{\citenamefont {{Zevin}}\ \emph
  {et~al.}(2020{\natexlab{b}})\citenamefont {{Zevin}}, \citenamefont {{Berry}},
  \citenamefont {{Coughlin}}, \citenamefont {{Chatziioannou}},\ and\
  \citenamefont {{Vitale}}}]{Zevin2020}%
  \BibitemOpen
  \bibfield  {author} {\bibinfo {author} {\bibfnamefont {M.}~\bibnamefont
  {{Zevin}}}, \bibinfo {author} {\bibfnamefont {C.~P.~L.}\ \bibnamefont
  {{Berry}}}, \bibinfo {author} {\bibfnamefont {S.}~\bibnamefont {{Coughlin}}},
  \bibinfo {author} {\bibfnamefont {K.}~\bibnamefont {{Chatziioannou}}}, \ and\
  \bibinfo {author} {\bibfnamefont {S.}~\bibnamefont {{Vitale}}},\ }\href
  {\doibase 10.3847/2041-8213/aba8ef} {\bibfield  {journal} {\bibinfo
  {journal} {\apjl}\ }\textbf {\bibinfo {volume} {899}},\ \bibinfo {eid} {L17}
  (\bibinfo {year} {2020}{\natexlab{b}})},\ \Eprint
  {http://arxiv.org/abs/2006.11293} {arXiv:2006.11293 [astro-ph.HE]}
  \BibitemShut {NoStop}%
\bibitem [{\citenamefont {{Abbott}}\ \emph
  {et~al.}(2020{\natexlab{a}})\citenamefont {{Abbott}} \emph
  {et~al.}}]{GW190412}%
  \BibitemOpen
  \bibfield  {author} {\bibinfo {author} {\bibfnamefont {B.~P.}\ \bibnamefont
  {{Abbott}}} \emph {et~al.} (\bibinfo {collaboration} {LIGO and Virgo
  Collaborations}),\ }\href {\doibase 10.1103/PhysRevD.102.043015} {\bibfield
  {journal} {\bibinfo  {journal} {\prd}\ }\textbf {\bibinfo {volume} {102}},\
  \bibinfo {eid} {043015} (\bibinfo {year} {2020}{\natexlab{a}})}\BibitemShut
  {NoStop}%
\bibitem [{\citenamefont {{Abbott}}\ \emph
  {et~al.}(2020{\natexlab{b}})\citenamefont {{Abbott}} \emph
  {et~al.}}]{GW190521}%
  \BibitemOpen
  \bibfield  {author} {\bibinfo {author} {\bibfnamefont {B.~P.}\ \bibnamefont
  {{Abbott}}} \emph {et~al.} (\bibinfo {collaboration} {LIGO and Virgo
  Collaborations}),\ }\href {\doibase 10.1103/PhysRevLett.125.101102}
  {\bibfield  {journal} {\bibinfo  {journal} {\prl}\ }\textbf {\bibinfo
  {volume} {125}},\ \bibinfo {eid} {101102} (\bibinfo {year}
  {2020}{\natexlab{b}})}\BibitemShut {NoStop}%
\bibitem [{\citenamefont {{O'Shaughnessy}}\ \emph {et~al.}(2005)\citenamefont
  {{O'Shaughnessy}}, \citenamefont {{Kaplan}}, \citenamefont {{Kalogera}},\
  and\ \citenamefont {{Belczynski}}}]{OShaughnessy2005}%
  \BibitemOpen
  \bibfield  {author} {\bibinfo {author} {\bibfnamefont {R.}~\bibnamefont
  {{O'Shaughnessy}}}, \bibinfo {author} {\bibfnamefont {J.}~\bibnamefont
  {{Kaplan}}}, \bibinfo {author} {\bibfnamefont {V.}~\bibnamefont
  {{Kalogera}}}, \ and\ \bibinfo {author} {\bibfnamefont {K.}~\bibnamefont
  {{Belczynski}}},\ }\href {\doibase 10.1086/444346} {\bibfield  {journal}
  {\bibinfo  {journal} {\apj}\ }\textbf {\bibinfo {volume} {632}},\ \bibinfo
  {pages} {1035} (\bibinfo {year} {2005})},\ \Eprint
  {http://arxiv.org/abs/astro-ph/0503219} {arXiv:astro-ph/0503219 [astro-ph]}
  \BibitemShut {NoStop}%
\bibitem [{\citenamefont {{Fairhurst}}\ \emph {et~al.}(2020)\citenamefont
  {{Fairhurst}}, \citenamefont {{Green}}, \citenamefont {{Hannam}},\ and\
  \citenamefont {{Hoy}}}]{Fairhurst2019}%
  \BibitemOpen
  \bibfield  {author} {\bibinfo {author} {\bibfnamefont {S.}~\bibnamefont
  {{Fairhurst}}}, \bibinfo {author} {\bibfnamefont {R.}~\bibnamefont
  {{Green}}}, \bibinfo {author} {\bibfnamefont {M.}~\bibnamefont {{Hannam}}}, \
  and\ \bibinfo {author} {\bibfnamefont {C.}~\bibnamefont {{Hoy}}},\ }\href
  {\doibase 10.1103/PhysRevD.102.041302} {\bibfield  {journal} {\bibinfo
  {journal} {\prd}\ }\textbf {\bibinfo {volume} {102}},\ \bibinfo {eid}
  {041302} (\bibinfo {year} {2020})},\ \Eprint
  {http://arxiv.org/abs/1908.00555} {arXiv:1908.00555 [gr-qc]} \BibitemShut
  {NoStop}%
\bibitem [{\citenamefont {{Postnov}}\ and\ \citenamefont
  {{Kuranov}}(2019)}]{Postnov2018}%
  \BibitemOpen
  \bibfield  {author} {\bibinfo {author} {\bibfnamefont {K.~A.}\ \bibnamefont
  {{Postnov}}}\ and\ \bibinfo {author} {\bibfnamefont {A.~G.}\ \bibnamefont
  {{Kuranov}}},\ }\href {\doibase 10.1093/mnras/sty3313} {\bibfield  {journal}
  {\bibinfo  {journal} {\mnras}\ }\textbf {\bibinfo {volume} {483}},\ \bibinfo
  {pages} {3288} (\bibinfo {year} {2019})},\ \Eprint
  {http://arxiv.org/abs/1706.00369} {arXiv:1706.00369 [astro-ph.HE]}
  \BibitemShut {NoStop}%
\bibitem [{\citenamefont {{Arca Sedda}}\ and\ \citenamefont
  {{Benacquista}}(2019)}]{Sedda2018}%
  \BibitemOpen
  \bibfield  {author} {\bibinfo {author} {\bibfnamefont {M.}~\bibnamefont
  {{Arca Sedda}}}\ and\ \bibinfo {author} {\bibfnamefont {M.}~\bibnamefont
  {{Benacquista}}},\ }\href {\doibase 10.1093/mnras/sty2764} {\bibfield
  {journal} {\bibinfo  {journal} {\mnras}\ }\textbf {\bibinfo {volume} {482}},\
  \bibinfo {pages} {2991} (\bibinfo {year} {2019})},\ \Eprint
  {http://arxiv.org/abs/1806.01285} {arXiv:1806.01285 [astro-ph.GA]}
  \BibitemShut {NoStop}%
\bibitem [{\citenamefont {{Hotokezaka}}\ and\ \citenamefont
  {{Piran}}(2017{\natexlab{a}})}]{Hotokezaka2017}%
  \BibitemOpen
  \bibfield  {author} {\bibinfo {author} {\bibfnamefont {K.}~\bibnamefont
  {{Hotokezaka}}}\ and\ \bibinfo {author} {\bibfnamefont {T.}~\bibnamefont
  {{Piran}}},\ }\href@noop {} {\bibfield  {journal} {\bibinfo  {journal} {arXiv
  e-prints}\ ,\ \bibinfo {eid} {arXiv:1707.08978}} (\bibinfo {year}
  {2017}{\natexlab{a}})},\ \Eprint {http://arxiv.org/abs/1707.08978}
  {arXiv:1707.08978 [astro-ph.HE]} \BibitemShut {NoStop}%
\bibitem [{\citenamefont {{Qin}}\ \emph {et~al.}(2018)\citenamefont {{Qin}},
  \citenamefont {{Fragos}}, \citenamefont {{Meynet}}, \citenamefont
  {{Andrews}}, \citenamefont {{S{\o}rensen}},\ and\ \citenamefont
  {{Song}}}]{Qin2018}%
  \BibitemOpen
  \bibfield  {author} {\bibinfo {author} {\bibfnamefont {Y.}~\bibnamefont
  {{Qin}}}, \bibinfo {author} {\bibfnamefont {T.}~\bibnamefont {{Fragos}}},
  \bibinfo {author} {\bibfnamefont {G.}~\bibnamefont {{Meynet}}}, \bibinfo
  {author} {\bibfnamefont {J.}~\bibnamefont {{Andrews}}}, \bibinfo {author}
  {\bibfnamefont {M.}~\bibnamefont {{S{\o}rensen}}}, \ and\ \bibinfo {author}
  {\bibfnamefont {H.~F.}\ \bibnamefont {{Song}}},\ }\href {\doibase
  10.1051/0004-6361/201832839} {\bibfield  {journal} {\bibinfo  {journal}
  {\aap}\ }\textbf {\bibinfo {volume} {616}},\ \bibinfo {eid} {A28} (\bibinfo
  {year} {2018})},\ \Eprint {http://arxiv.org/abs/1802.05738} {arXiv:1802.05738
  [astro-ph.SR]} \BibitemShut {NoStop}%
\bibitem [{\citenamefont {{Fuller}}\ and\ \citenamefont
  {{Ma}}(2019)}]{Fuller2019}%
  \BibitemOpen
  \bibfield  {author} {\bibinfo {author} {\bibfnamefont {J.}~\bibnamefont
  {{Fuller}}}\ and\ \bibinfo {author} {\bibfnamefont {L.}~\bibnamefont
  {{Ma}}},\ }\href {\doibase 10.3847/2041-8213/ab339b} {\bibfield  {journal}
  {\bibinfo  {journal} {\apjl}\ }\textbf {\bibinfo {volume} {881}},\ \bibinfo
  {eid} {L1} (\bibinfo {year} {2019})},\ \Eprint
  {http://arxiv.org/abs/1907.03714} {arXiv:1907.03714 [astro-ph.SR]}
  \BibitemShut {NoStop}%
\bibitem [{\citenamefont {{Belczynski}}\ \emph
  {et~al.}(2008{\natexlab{a}})\citenamefont {{Belczynski}}, \citenamefont
  {{Kalogera}}, \citenamefont {{Rasio}}, \citenamefont {{Taam}}, \citenamefont
  {{Zezas}}, \citenamefont {{Bulik}}, \citenamefont {{Maccarone}},\ and\
  \citenamefont {{Ivanova}}}]{BK2008}%
  \BibitemOpen
  \bibfield  {author} {\bibinfo {author} {\bibfnamefont {K.}~\bibnamefont
  {{Belczynski}}}, \bibinfo {author} {\bibfnamefont {V.}~\bibnamefont
  {{Kalogera}}}, \bibinfo {author} {\bibfnamefont {F.~A.}\ \bibnamefont
  {{Rasio}}}, \bibinfo {author} {\bibfnamefont {R.~E.}\ \bibnamefont {{Taam}}},
  \bibinfo {author} {\bibfnamefont {A.}~\bibnamefont {{Zezas}}}, \bibinfo
  {author} {\bibfnamefont {T.}~\bibnamefont {{Bulik}}}, \bibinfo {author}
  {\bibfnamefont {T.~J.}\ \bibnamefont {{Maccarone}}}, \ and\ \bibinfo {author}
  {\bibfnamefont {N.}~\bibnamefont {{Ivanova}}},\ }\href {\doibase
  10.1086/521026} {\bibfield  {journal} {\bibinfo  {journal} {\apjs}\ }\textbf
  {\bibinfo {volume} {174}},\ \bibinfo {pages} {223} (\bibinfo {year}
  {2008}{\natexlab{a}})},\ \Eprint {http://arxiv.org/abs/astro-ph/0511811}
  {arXiv:astro-ph/0511811 [astro-ph]} \BibitemShut {NoStop}%
\bibitem [{\citenamefont {{Wong}}\ and\ \citenamefont
  {{Gerosa}}(2019)}]{Wong2019}%
  \BibitemOpen
  \bibfield  {author} {\bibinfo {author} {\bibfnamefont {K.~W.~K.}\
  \bibnamefont {{Wong}}}\ and\ \bibinfo {author} {\bibfnamefont
  {D.}~\bibnamefont {{Gerosa}}},\ }\href {\doibase 10.1103/PhysRevD.100.083015}
  {\bibfield  {journal} {\bibinfo  {journal} {\prd}\ }\textbf {\bibinfo
  {volume} {100}},\ \bibinfo {eid} {083015} (\bibinfo {year} {2019})},\ \Eprint
  {http://arxiv.org/abs/1909.06373} {arXiv:1909.06373 [astro-ph.HE]}
  \BibitemShut {NoStop}%
\bibitem [{\citenamefont {{Belczynski}}\ \emph {et~al.}(2020)\citenamefont
  {{Belczynski}}, \citenamefont {{Klencki}}, \citenamefont {{Fields}},
  \citenamefont {{Olejak}}, \citenamefont {{Berti}}, \citenamefont {{Meynet}},
  \citenamefont {{Fryer}}, \citenamefont {{Holz}}, \citenamefont
  {{O'Shaughnessy}} \emph {et~al.}}]{Belczynski2020}%
  \BibitemOpen
  \bibfield  {author} {\bibinfo {author} {\bibfnamefont {K.}~\bibnamefont
  {{Belczynski}}}, \bibinfo {author} {\bibfnamefont {J.}~\bibnamefont
  {{Klencki}}}, \bibinfo {author} {\bibfnamefont {C.~E.}\ \bibnamefont
  {{Fields}}}, \bibinfo {author} {\bibfnamefont {A.}~\bibnamefont {{Olejak}}},
  \bibinfo {author} {\bibfnamefont {E.}~\bibnamefont {{Berti}}}, \bibinfo
  {author} {\bibfnamefont {G.}~\bibnamefont {{Meynet}}}, \bibinfo {author}
  {\bibfnamefont {C.~L.}\ \bibnamefont {{Fryer}}}, \bibinfo {author}
  {\bibfnamefont {D.~E.}\ \bibnamefont {{Holz}}}, \bibinfo {author}
  {\bibfnamefont {R.}~\bibnamefont {{O'Shaughnessy}}},  \emph {et~al.},\ }\href
  {\doibase 10.1051/0004-6361/201936528} {\bibfield  {journal} {\bibinfo
  {journal} {\aap}\ }\textbf {\bibinfo {volume} {636}},\ \bibinfo {eid} {A104}
  (\bibinfo {year} {2020})},\ \Eprint {http://arxiv.org/abs/1706.07053}
  {arXiv:1706.07053 [astro-ph.HE]} \BibitemShut {NoStop}%
\bibitem [{\citenamefont {{Belczynski}}\ \emph
  {et~al.}(2008{\natexlab{b}})\citenamefont {{Belczynski}}, \citenamefont
  {{Taam}}, \citenamefont {{Rantsiou}},\ and\ \citenamefont {{van der
  Sluys}}}]{BKspins2008}%
  \BibitemOpen
  \bibfield  {author} {\bibinfo {author} {\bibfnamefont {K.}~\bibnamefont
  {{Belczynski}}}, \bibinfo {author} {\bibfnamefont {R.~E.}\ \bibnamefont
  {{Taam}}}, \bibinfo {author} {\bibfnamefont {E.}~\bibnamefont {{Rantsiou}}},
  \ and\ \bibinfo {author} {\bibfnamefont {M.}~\bibnamefont {{van der
  Sluys}}},\ }\href {\doibase 10.1086/589609} {\bibfield  {journal} {\bibinfo
  {journal} {\apj}\ }\textbf {\bibinfo {volume} {682}},\ \bibinfo {pages} {474}
  (\bibinfo {year} {2008}{\natexlab{b}})},\ \Eprint
  {http://arxiv.org/abs/astro-ph/0703131} {arXiv:astro-ph/0703131 [astro-ph]}
  \BibitemShut {NoStop}%
\bibitem [{\citenamefont {{Fragos}}\ \emph {et~al.}(2010)\citenamefont
  {{Fragos}}, \citenamefont {{Tremmel}}, \citenamefont {{Rantsiou}},\ and\
  \citenamefont {{Belczynski}}}]{Fragos2010}%
  \BibitemOpen
  \bibfield  {author} {\bibinfo {author} {\bibfnamefont {T.}~\bibnamefont
  {{Fragos}}}, \bibinfo {author} {\bibfnamefont {M.}~\bibnamefont {{Tremmel}}},
  \bibinfo {author} {\bibfnamefont {E.}~\bibnamefont {{Rantsiou}}}, \ and\
  \bibinfo {author} {\bibfnamefont {K.}~\bibnamefont {{Belczynski}}},\ }\href
  {\doibase 10.1088/2041-8205/719/1/L79} {\bibfield  {journal} {\bibinfo
  {journal} {\apjl}\ }\textbf {\bibinfo {volume} {719}},\ \bibinfo {pages}
  {L79} (\bibinfo {year} {2010})},\ \Eprint {http://arxiv.org/abs/1001.1107}
  {arXiv:1001.1107 [astro-ph.HE]} \BibitemShut {NoStop}%
\bibitem [{\citenamefont {{Breivik}}\ \emph {et~al.}(2017)\citenamefont
  {{Breivik}}, \citenamefont {{Chatterjee}},\ and\ \citenamefont
  {{Larson}}}]{Breivik2017}%
  \BibitemOpen
  \bibfield  {author} {\bibinfo {author} {\bibfnamefont {K.}~\bibnamefont
  {{Breivik}}}, \bibinfo {author} {\bibfnamefont {S.}~\bibnamefont
  {{Chatterjee}}}, \ and\ \bibinfo {author} {\bibfnamefont {S.~L.}\
  \bibnamefont {{Larson}}},\ }\href {\doibase 10.3847/2041-8213/aa97d5}
  {\bibfield  {journal} {\bibinfo  {journal} {\apjl}\ }\textbf {\bibinfo
  {volume} {850}},\ \bibinfo {eid} {L13} (\bibinfo {year} {2017})},\ \Eprint
  {http://arxiv.org/abs/1710.04657} {arXiv:1710.04657 [astro-ph.SR]}
  \BibitemShut {NoStop}%
\bibitem [{\citenamefont {{Ivanova}}\ and\ \citenamefont
  {{Taam}}(2004)}]{Ivanova2004}%
  \BibitemOpen
  \bibfield  {author} {\bibinfo {author} {\bibfnamefont {N.}~\bibnamefont
  {{Ivanova}}}\ and\ \bibinfo {author} {\bibfnamefont {R.~E.}\ \bibnamefont
  {{Taam}}},\ }\href {\doibase 10.1086/380561} {\bibfield  {journal} {\bibinfo
  {journal} {\apj}\ }\textbf {\bibinfo {volume} {601}},\ \bibinfo {pages}
  {1058} (\bibinfo {year} {2004})},\ \Eprint
  {http://arxiv.org/abs/astro-ph/0310126} {arXiv:astro-ph/0310126 [astro-ph]}
  \BibitemShut {NoStop}%
\bibitem [{\citenamefont {{Belczynski}}\ \emph {et~al.}(2007)\citenamefont
  {{Belczynski}}, \citenamefont {{Taam}}, \citenamefont {{Kalogera}},
  \citenamefont {{Rasio}},\ and\ \citenamefont {{Bulik}}}]{BK2007}%
  \BibitemOpen
  \bibfield  {author} {\bibinfo {author} {\bibfnamefont {K.}~\bibnamefont
  {{Belczynski}}}, \bibinfo {author} {\bibfnamefont {R.~E.}\ \bibnamefont
  {{Taam}}}, \bibinfo {author} {\bibfnamefont {V.}~\bibnamefont {{Kalogera}}},
  \bibinfo {author} {\bibfnamefont {F.~A.}\ \bibnamefont {{Rasio}}}, \ and\
  \bibinfo {author} {\bibfnamefont {T.}~\bibnamefont {{Bulik}}},\ }\href
  {\doibase 10.1086/513562} {\bibfield  {journal} {\bibinfo  {journal} {\apj}\
  }\textbf {\bibinfo {volume} {662}},\ \bibinfo {pages} {504} (\bibinfo {year}
  {2007})},\ \Eprint {http://arxiv.org/abs/astro-ph/0612032}
  {arXiv:astro-ph/0612032 [astro-ph]} \BibitemShut {NoStop}%
\bibitem [{\citenamefont {{Steinle}}\ and\ \citenamefont
  {{Kesden}}()}]{SteinleFuture1}%
  \BibitemOpen
  \bibfield  {author} {\bibinfo {author} {\bibfnamefont {N.}~\bibnamefont
  {{Steinle}}}\ and\ \bibinfo {author} {\bibfnamefont {M.}~\bibnamefont
  {{Kesden}}},\ }\href@noop {} {\enquote {\bibinfo {title} {{Precessing binary
  black-holes from the isolated formation channel}},}\ }\bibinfo {note} {(in
  preparation)}\BibitemShut {NoStop}%
\bibitem [{\citenamefont {{Gerosa}}\ and\ \citenamefont
  {{Kesden}}(2016)}]{Gerosa2016}%
  \BibitemOpen
  \bibfield  {author} {\bibinfo {author} {\bibfnamefont {D.}~\bibnamefont
  {{Gerosa}}}\ and\ \bibinfo {author} {\bibfnamefont {M.}~\bibnamefont
  {{Kesden}}},\ }\href {\doibase 10.1103/PhysRevD.93.124066} {\bibfield
  {journal} {\bibinfo  {journal} {\prd}\ }\textbf {\bibinfo {volume} {93}},\
  \bibinfo {eid} {124066} (\bibinfo {year} {2016})},\ \Eprint
  {http://arxiv.org/abs/1605.01067} {arXiv:1605.01067 [astro-ph.HE]}
  \BibitemShut {NoStop}%
\bibitem [{\citenamefont {{Belczynski}}\ \emph
  {et~al.}(2016{\natexlab{a}})\citenamefont {{Belczynski}}, \citenamefont
  {{Holz}}, \citenamefont {{Bulik}},\ and\ \citenamefont
  {{O'Shaughnessy}}}]{Belczynski2016}%
  \BibitemOpen
  \bibfield  {author} {\bibinfo {author} {\bibfnamefont {K.}~\bibnamefont
  {{Belczynski}}}, \bibinfo {author} {\bibfnamefont {D.~E.}\ \bibnamefont
  {{Holz}}}, \bibinfo {author} {\bibfnamefont {T.}~\bibnamefont {{Bulik}}}, \
  and\ \bibinfo {author} {\bibfnamefont {R.}~\bibnamefont {{O'Shaughnessy}}},\
  }\href {\doibase 10.1038/nature18322} {\bibfield  {journal} {\bibinfo
  {journal} {\nat}\ }\textbf {\bibinfo {volume} {534}},\ \bibinfo {pages} {512}
  (\bibinfo {year} {2016}{\natexlab{a}})},\ \Eprint
  {http://arxiv.org/abs/1602.04531} {arXiv:1602.04531 [astro-ph.HE]}
  \BibitemShut {NoStop}%
\bibitem [{\citenamefont {{Stevenson}}\ \emph
  {et~al.}(2017{\natexlab{b}})\citenamefont {{Stevenson}}, \citenamefont
  {{Vigna-G{\'o}mez}}, \citenamefont {{Mandel}}, \citenamefont {{Barrett}},
  \citenamefont {{Neijssel}}, \citenamefont {{Perkins}},\ and\ \citenamefont
  {{de Mink}}}]{Stevenson2017b}%
  \BibitemOpen
  \bibfield  {author} {\bibinfo {author} {\bibfnamefont {S.}~\bibnamefont
  {{Stevenson}}}, \bibinfo {author} {\bibfnamefont {A.}~\bibnamefont
  {{Vigna-G{\'o}mez}}}, \bibinfo {author} {\bibfnamefont {I.}~\bibnamefont
  {{Mandel}}}, \bibinfo {author} {\bibfnamefont {J.~W.}\ \bibnamefont
  {{Barrett}}}, \bibinfo {author} {\bibfnamefont {C.~J.}\ \bibnamefont
  {{Neijssel}}}, \bibinfo {author} {\bibfnamefont {D.}~\bibnamefont
  {{Perkins}}}, \ and\ \bibinfo {author} {\bibfnamefont {S.~E.}\ \bibnamefont
  {{de Mink}}},\ }\href {\doibase 10.1038/ncomms14906} {\bibfield  {journal}
  {\bibinfo  {journal} {Nature Communications}\ }\textbf {\bibinfo {volume}
  {8}},\ \bibinfo {eid} {14906} (\bibinfo {year} {2017}{\natexlab{b}})},\
  \Eprint {http://arxiv.org/abs/1704.01352} {arXiv:1704.01352 [astro-ph.HE]}
  \BibitemShut {NoStop}%
\bibitem [{\citenamefont {{Giacobbo}}\ and\ \citenamefont
  {{Mapelli}}(2018)}]{Giacobbo2018}%
  \BibitemOpen
  \bibfield  {author} {\bibinfo {author} {\bibfnamefont {N.}~\bibnamefont
  {{Giacobbo}}}\ and\ \bibinfo {author} {\bibfnamefont {M.}~\bibnamefont
  {{Mapelli}}},\ }\href {\doibase 10.1093/mnras/sty1999} {\bibfield  {journal}
  {\bibinfo  {journal} {\mnras}\ }\textbf {\bibinfo {volume} {480}},\ \bibinfo
  {pages} {2011} (\bibinfo {year} {2018})},\ \Eprint
  {http://arxiv.org/abs/1806.00001} {arXiv:1806.00001 [astro-ph.HE]}
  \BibitemShut {NoStop}%
\bibitem [{\citenamefont {{Bogomazov}}\ \emph {et~al.}(2018)\citenamefont
  {{Bogomazov}}, \citenamefont {{Lipunov}}, \citenamefont {{Tutukov}},\ and\
  \citenamefont {{Cherepashchuk}}}]{Bogomazov2018}%
  \BibitemOpen
  \bibfield  {author} {\bibinfo {author} {\bibfnamefont {A.~I.}\ \bibnamefont
  {{Bogomazov}}}, \bibinfo {author} {\bibfnamefont {V.~M.}\ \bibnamefont
  {{Lipunov}}}, \bibinfo {author} {\bibfnamefont {A.~V.}\ \bibnamefont
  {{Tutukov}}}, \ and\ \bibinfo {author} {\bibfnamefont {A.~M.}\ \bibnamefont
  {{Cherepashchuk}}},\ }\href@noop {} {\bibfield  {journal} {\bibinfo
  {journal} {arXiv e-prints}\ ,\ \bibinfo {eid} {arXiv:1811.02294}} (\bibinfo
  {year} {2018})},\ \Eprint {http://arxiv.org/abs/1811.02294} {arXiv:1811.02294
  [astro-ph.HE]} \BibitemShut {NoStop}%
\bibitem [{\citenamefont {{Kruckow}}\ \emph {et~al.}(2018)\citenamefont
  {{Kruckow}}, \citenamefont {{Tauris}}, \citenamefont {{Langer}},
  \citenamefont {{Kramer}},\ and\ \citenamefont {{Izzard}}}]{Kruckow2018}%
  \BibitemOpen
  \bibfield  {author} {\bibinfo {author} {\bibfnamefont {M.~U.}\ \bibnamefont
  {{Kruckow}}}, \bibinfo {author} {\bibfnamefont {T.~M.}\ \bibnamefont
  {{Tauris}}}, \bibinfo {author} {\bibfnamefont {N.}~\bibnamefont {{Langer}}},
  \bibinfo {author} {\bibfnamefont {M.}~\bibnamefont {{Kramer}}}, \ and\
  \bibinfo {author} {\bibfnamefont {R.~G.}\ \bibnamefont {{Izzard}}},\ }\href
  {\doibase 10.1093/mnras/sty2190} {\bibfield  {journal} {\bibinfo  {journal}
  {\mnras}\ }\textbf {\bibinfo {volume} {481}},\ \bibinfo {pages} {1908}
  (\bibinfo {year} {2018})},\ \Eprint {http://arxiv.org/abs/1801.05433}
  {arXiv:1801.05433 [astro-ph.SR]} \BibitemShut {NoStop}%
\bibitem [{\citenamefont {{Vigna-G{\'o}mez}}\ \emph {et~al.}(2018)\citenamefont
  {{Vigna-G{\'o}mez}}, \citenamefont {{Neijssel}}, \citenamefont {{Stevenson}},
  \citenamefont {{Barrett}}, \citenamefont {{Belczynski}}, \citenamefont
  {{Justham}}, \citenamefont {{de Mink}}, \citenamefont {{M{\"u}ller}},
  \citenamefont {{Podsiadlowski}}, \citenamefont {{Renzo}}, \citenamefont
  {{Sz{\'e}csi}},\ and\ \citenamefont {{Mandel}}}]{Vigna2018}%
  \BibitemOpen
  \bibfield  {author} {\bibinfo {author} {\bibfnamefont {A.}~\bibnamefont
  {{Vigna-G{\'o}mez}}}, \bibinfo {author} {\bibfnamefont {C.~J.}\ \bibnamefont
  {{Neijssel}}}, \bibinfo {author} {\bibfnamefont {S.}~\bibnamefont
  {{Stevenson}}}, \bibinfo {author} {\bibfnamefont {J.~W.}\ \bibnamefont
  {{Barrett}}}, \bibinfo {author} {\bibfnamefont {K.}~\bibnamefont
  {{Belczynski}}}, \bibinfo {author} {\bibfnamefont {S.}~\bibnamefont
  {{Justham}}}, \bibinfo {author} {\bibfnamefont {S.~E.}\ \bibnamefont {{de
  Mink}}}, \bibinfo {author} {\bibfnamefont {B.}~\bibnamefont {{M{\"u}ller}}},
  \bibinfo {author} {\bibfnamefont {P.}~\bibnamefont {{Podsiadlowski}}},
  \bibinfo {author} {\bibfnamefont {M.}~\bibnamefont {{Renzo}}}, \bibinfo
  {author} {\bibfnamefont {D.}~\bibnamefont {{Sz{\'e}csi}}}, \ and\ \bibinfo
  {author} {\bibfnamefont {I.}~\bibnamefont {{Mandel}}},\ }\href {\doibase
  10.1093/mnras/sty2463} {\bibfield  {journal} {\bibinfo  {journal} {\mnras}\
  }\textbf {\bibinfo {volume} {481}},\ \bibinfo {pages} {4009} (\bibinfo {year}
  {2018})},\ \Eprint {http://arxiv.org/abs/1805.07974} {arXiv:1805.07974
  [astro-ph.SR]} \BibitemShut {NoStop}%
\bibitem [{\citenamefont {{Breivik}}\ \emph {et~al.}(2020)\citenamefont
  {{Breivik}}, \citenamefont {{Coughlin}}, \citenamefont {{Zevin}},
  \citenamefont {{Rodriguez}}, \citenamefont {{Kremer}}, \citenamefont {{Ye}},
  \citenamefont {{Andrews}}, \citenamefont {{Kurkowski}}, \citenamefont
  {{Digman}}, \citenamefont {{Larson}},\ and\ \citenamefont
  {{Rasio}}}]{Breivik2020}%
  \BibitemOpen
  \bibfield  {author} {\bibinfo {author} {\bibfnamefont {K.}~\bibnamefont
  {{Breivik}}}, \bibinfo {author} {\bibfnamefont {S.}~\bibnamefont
  {{Coughlin}}}, \bibinfo {author} {\bibfnamefont {M.}~\bibnamefont {{Zevin}}},
  \bibinfo {author} {\bibfnamefont {C.~L.}\ \bibnamefont {{Rodriguez}}},
  \bibinfo {author} {\bibfnamefont {K.}~\bibnamefont {{Kremer}}}, \bibinfo
  {author} {\bibfnamefont {C.~S.}\ \bibnamefont {{Ye}}}, \bibinfo {author}
  {\bibfnamefont {J.~J.}\ \bibnamefont {{Andrews}}}, \bibinfo {author}
  {\bibfnamefont {M.}~\bibnamefont {{Kurkowski}}}, \bibinfo {author}
  {\bibfnamefont {M.~C.}\ \bibnamefont {{Digman}}}, \bibinfo {author}
  {\bibfnamefont {S.~L.}\ \bibnamefont {{Larson}}}, \ and\ \bibinfo {author}
  {\bibfnamefont {F.~A.}\ \bibnamefont {{Rasio}}},\ }\href {\doibase
  10.3847/1538-4357/ab9d85} {\bibfield  {journal} {\bibinfo  {journal} {\apj}\
  }\textbf {\bibinfo {volume} {898}},\ \bibinfo {eid} {71} (\bibinfo {year}
  {2020})},\ \Eprint {http://arxiv.org/abs/1911.00903} {arXiv:1911.00903
  [astro-ph.HE]} \BibitemShut {NoStop}%
\bibitem [{\citenamefont {{Hotokezaka}}\ and\ \citenamefont
  {{Piran}}(2017{\natexlab{b}})}]{Hotokezaka2017Implications}%
  \BibitemOpen
  \bibfield  {author} {\bibinfo {author} {\bibfnamefont {K.}~\bibnamefont
  {{Hotokezaka}}}\ and\ \bibinfo {author} {\bibfnamefont {T.}~\bibnamefont
  {{Piran}}},\ }\href {\doibase 10.3847/1538-4357/aa6f61} {\bibfield  {journal}
  {\bibinfo  {journal} {\apj}\ }\textbf {\bibinfo {volume} {842}},\ \bibinfo
  {eid} {111} (\bibinfo {year} {2017}{\natexlab{b}})},\ \Eprint
  {http://arxiv.org/abs/1702.03952} {arXiv:1702.03952 [astro-ph.HE]}
  \BibitemShut {NoStop}%
\bibitem [{\citenamefont {{Bavera}}\ \emph
  {et~al.}(2020{\natexlab{a}})\citenamefont {{Bavera}}, \citenamefont
  {{Fragos}}, \citenamefont {{Qin}}, \citenamefont {{Zapartas}}, \citenamefont
  {{Neijssel}}, \citenamefont {{Mandel}}, \citenamefont {{Batta}},
  \citenamefont {{Gaebel}}, \citenamefont {{Kimball}},\ and\ \citenamefont
  {{Stevenson}}}]{Bavera2020}%
  \BibitemOpen
  \bibfield  {author} {\bibinfo {author} {\bibfnamefont {S.~S.}\ \bibnamefont
  {{Bavera}}}, \bibinfo {author} {\bibfnamefont {T.}~\bibnamefont {{Fragos}}},
  \bibinfo {author} {\bibfnamefont {Y.}~\bibnamefont {{Qin}}}, \bibinfo
  {author} {\bibfnamefont {E.}~\bibnamefont {{Zapartas}}}, \bibinfo {author}
  {\bibfnamefont {C.~J.}\ \bibnamefont {{Neijssel}}}, \bibinfo {author}
  {\bibfnamefont {I.}~\bibnamefont {{Mandel}}}, \bibinfo {author}
  {\bibfnamefont {A.}~\bibnamefont {{Batta}}}, \bibinfo {author} {\bibfnamefont
  {S.~M.}\ \bibnamefont {{Gaebel}}}, \bibinfo {author} {\bibfnamefont
  {C.}~\bibnamefont {{Kimball}}}, \ and\ \bibinfo {author} {\bibfnamefont
  {S.}~\bibnamefont {{Stevenson}}},\ }\href {\doibase
  10.1051/0004-6361/201936204} {\bibfield  {journal} {\bibinfo  {journal}
  {\aap}\ }\textbf {\bibinfo {volume} {635}},\ \bibinfo {eid} {A97} (\bibinfo
  {year} {2020}{\natexlab{a}})},\ \Eprint {http://arxiv.org/abs/1906.12257}
  {arXiv:1906.12257 [astro-ph.HE]} \BibitemShut {NoStop}%
\bibitem [{\citenamefont {{Neijssel}}\ \emph {et~al.}(2019)\citenamefont
  {{Neijssel}}, \citenamefont {{Vigna-G{\'o}mez}}, \citenamefont {{Stevenson}},
  \citenamefont {{Barrett}}, \citenamefont {{Gaebel}}, \citenamefont
  {{Broekgaarden}}, \citenamefont {{de Mink}}, \citenamefont {{Sz{\'e}csi}},
  \citenamefont {{Vinciguerra}},\ and\ \citenamefont
  {{Mandel}}}]{Neijssel2020}%
  \BibitemOpen
  \bibfield  {author} {\bibinfo {author} {\bibfnamefont {C.~J.}\ \bibnamefont
  {{Neijssel}}}, \bibinfo {author} {\bibfnamefont {A.}~\bibnamefont
  {{Vigna-G{\'o}mez}}}, \bibinfo {author} {\bibfnamefont {S.}~\bibnamefont
  {{Stevenson}}}, \bibinfo {author} {\bibfnamefont {J.~W.}\ \bibnamefont
  {{Barrett}}}, \bibinfo {author} {\bibfnamefont {S.~M.}\ \bibnamefont
  {{Gaebel}}}, \bibinfo {author} {\bibfnamefont {F.~S.}\ \bibnamefont
  {{Broekgaarden}}}, \bibinfo {author} {\bibfnamefont {S.~E.}\ \bibnamefont
  {{de Mink}}}, \bibinfo {author} {\bibfnamefont {D.}~\bibnamefont
  {{Sz{\'e}csi}}}, \bibinfo {author} {\bibfnamefont {S.}~\bibnamefont
  {{Vinciguerra}}}, \ and\ \bibinfo {author} {\bibfnamefont {I.}~\bibnamefont
  {{Mandel}}},\ }\href {\doibase 10.1093/mnras/stz2840} {\bibfield  {journal}
  {\bibinfo  {journal} {\mnras}\ }\textbf {\bibinfo {volume} {490}},\ \bibinfo
  {pages} {3740} (\bibinfo {year} {2019})},\ \Eprint
  {http://arxiv.org/abs/1906.08136} {arXiv:1906.08136 [astro-ph.SR]}
  \BibitemShut {NoStop}%
\bibitem [{\citenamefont {{Hurley}}\ \emph {et~al.}(2000)\citenamefont
  {{Hurley}}, \citenamefont {{Pols}},\ and\ \citenamefont
  {{Tout}}}]{Hurley2000}%
  \BibitemOpen
  \bibfield  {author} {\bibinfo {author} {\bibfnamefont {J.~R.}\ \bibnamefont
  {{Hurley}}}, \bibinfo {author} {\bibfnamefont {O.~R.}\ \bibnamefont
  {{Pols}}}, \ and\ \bibinfo {author} {\bibfnamefont {C.~A.}\ \bibnamefont
  {{Tout}}},\ }\href {\doibase 10.1046/j.1365-8711.2000.03426.x} {\bibfield
  {journal} {\bibinfo  {journal} {\mnras}\ }\textbf {\bibinfo {volume} {315}},\
  \bibinfo {pages} {543} (\bibinfo {year} {2000})},\ \Eprint
  {http://arxiv.org/abs/astro-ph/0001295} {arXiv:astro-ph/0001295 [astro-ph]}
  \BibitemShut {NoStop}%
\bibitem [{\citenamefont {Lang}(1992)}]{Lang1992}%
  \BibitemOpen
  \bibfield  {author} {\bibinfo {author} {\bibfnamefont {K.}~\bibnamefont
  {Lang}},\ }\href@noop {} {\emph {\bibinfo {title} {Astrophysical Data:
  Planets and Stars}}},\ \bibinfo {edition} {1st}\ ed.\ (\bibinfo  {publisher}
  {Springer-Verlag New York},\ \bibinfo {year} {1992})\ p.\ \bibinfo {pages}
  {135}\BibitemShut {NoStop}%
\bibitem [{\citenamefont {{Tout}}\ \emph {et~al.}(1996)\citenamefont {{Tout}},
  \citenamefont {{Pols}}, \citenamefont {{Eggleton}},\ and\ \citenamefont
  {{Han}}}]{Tout1996}%
  \BibitemOpen
  \bibfield  {author} {\bibinfo {author} {\bibfnamefont {C.~A.}\ \bibnamefont
  {{Tout}}}, \bibinfo {author} {\bibfnamefont {O.~R.}\ \bibnamefont {{Pols}}},
  \bibinfo {author} {\bibfnamefont {P.~P.}\ \bibnamefont {{Eggleton}}}, \ and\
  \bibinfo {author} {\bibfnamefont {Z.}~\bibnamefont {{Han}}},\ }\href
  {\doibase 10.1093/mnras/281.1.257} {\bibfield  {journal} {\bibinfo  {journal}
  {\mnras}\ }\textbf {\bibinfo {volume} {281}},\ \bibinfo {pages} {257}
  (\bibinfo {year} {1996})}\BibitemShut {NoStop}%
\bibitem [{\citenamefont {{Crowther}}(2007)}]{Crowther2007}%
  \BibitemOpen
  \bibfield  {author} {\bibinfo {author} {\bibfnamefont {P.~A.}\ \bibnamefont
  {{Crowther}}},\ }\href {\doibase 10.1146/annurev.astro.45.051806.110615}
  {\bibfield  {journal} {\bibinfo  {journal} {\araa}\ }\textbf {\bibinfo
  {volume} {45}},\ \bibinfo {pages} {177} (\bibinfo {year} {2007})},\ \Eprint
  {http://arxiv.org/abs/astro-ph/0610356} {arXiv:astro-ph/0610356 [astro-ph]}
  \BibitemShut {NoStop}%
\bibitem [{\citenamefont {Sweet}(1950)}]{Sweet1950}%
  \BibitemOpen
  \bibfield  {author} {\bibinfo {author} {\bibfnamefont {P.~A.}\ \bibnamefont
  {Sweet}},\ }\href {\doibase 10.1093/mnras/110.6.548} {\bibfield  {journal}
  {\bibinfo  {journal} {Monthly Notices of the Royal Astronomical Society}\
  }\textbf {\bibinfo {volume} {110}},\ \bibinfo {pages} {548} (\bibinfo {year}
  {1950})}\BibitemShut {NoStop}%
\bibitem [{\citenamefont {{Zahn}}(1992)}]{Zahn1992}%
  \BibitemOpen
  \bibfield  {author} {\bibinfo {author} {\bibfnamefont {J.~P.}\ \bibnamefont
  {{Zahn}}},\ }\href@noop {} {\bibfield  {journal} {\bibinfo  {journal} {\aap}\
  }\textbf {\bibinfo {volume} {265}},\ \bibinfo {pages} {115} (\bibinfo {year}
  {1992})}\BibitemShut {NoStop}%
\bibitem [{\citenamefont {{Maeder}}(1998)}]{Maeder1998}%
  \BibitemOpen
  \bibfield  {author} {\bibinfo {author} {\bibfnamefont {A.}~\bibnamefont
  {{Maeder}}},\ }in\ \href@noop {} {\emph {\bibinfo {booktitle} {Properties of
  Hot Luminous Stars}}},\ \bibinfo {series} {Astronomical Society of the
  Pacific Conference Series}, Vol.\ \bibinfo {volume} {131},\ \bibinfo {editor}
  {edited by\ \bibinfo {editor} {\bibfnamefont {I.}~\bibnamefont {{Howarth}}}}\
  (\bibinfo {year} {1998})\ p.~\bibinfo {pages} {85}\BibitemShut {NoStop}%
\bibitem [{\citenamefont {{Georgy}}\ \emph {et~al.}(2013)\citenamefont
  {{Georgy}}, \citenamefont {{Ekstr{\"o}m}}, \citenamefont {{Eggenberger}},
  \citenamefont {{Meynet}}, \citenamefont {{Haemmerl{\'e}}}, \citenamefont
  {{Maeder}}, \citenamefont {{Granada}}, \citenamefont {{Groh}}, \citenamefont
  {{Hirschi}}, \citenamefont {{Mowlavi}}, \citenamefont {{Yusof}},
  \citenamefont {{Charbonnel}}, \citenamefont {{Decressin}},\ and\
  \citenamefont {{Barblan}}}]{Georgy2013}%
  \BibitemOpen
  \bibfield  {author} {\bibinfo {author} {\bibfnamefont {C.}~\bibnamefont
  {{Georgy}}}, \bibinfo {author} {\bibfnamefont {S.}~\bibnamefont
  {{Ekstr{\"o}m}}}, \bibinfo {author} {\bibfnamefont {P.}~\bibnamefont
  {{Eggenberger}}}, \bibinfo {author} {\bibfnamefont {G.}~\bibnamefont
  {{Meynet}}}, \bibinfo {author} {\bibfnamefont {L.}~\bibnamefont
  {{Haemmerl{\'e}}}}, \bibinfo {author} {\bibfnamefont {A.}~\bibnamefont
  {{Maeder}}}, \bibinfo {author} {\bibfnamefont {A.}~\bibnamefont {{Granada}}},
  \bibinfo {author} {\bibfnamefont {J.~H.}\ \bibnamefont {{Groh}}}, \bibinfo
  {author} {\bibfnamefont {R.}~\bibnamefont {{Hirschi}}}, \bibinfo {author}
  {\bibfnamefont {N.}~\bibnamefont {{Mowlavi}}}, \bibinfo {author}
  {\bibfnamefont {N.}~\bibnamefont {{Yusof}}}, \bibinfo {author} {\bibfnamefont
  {C.}~\bibnamefont {{Charbonnel}}}, \bibinfo {author} {\bibfnamefont
  {T.}~\bibnamefont {{Decressin}}}, \ and\ \bibinfo {author} {\bibfnamefont
  {F.}~\bibnamefont {{Barblan}}},\ }\href {\doibase
  10.1051/0004-6361/201322178} {\bibfield  {journal} {\bibinfo  {journal}
  {\aap}\ }\textbf {\bibinfo {volume} {558}},\ \bibinfo {eid} {A103} (\bibinfo
  {year} {2013})},\ \Eprint {http://arxiv.org/abs/1308.2914} {arXiv:1308.2914
  [astro-ph.SR]} \BibitemShut {NoStop}%
\bibitem [{\citenamefont {{Postnov}}\ \emph {et~al.}(2016)\citenamefont
  {{Postnov}}, \citenamefont {{Kuranov}}, \citenamefont {{Kolesnikov}},
  \citenamefont {{Popov}},\ and\ \citenamefont {{Porayko}}}]{Postnov2016}%
  \BibitemOpen
  \bibfield  {author} {\bibinfo {author} {\bibfnamefont {K.~A.}\ \bibnamefont
  {{Postnov}}}, \bibinfo {author} {\bibfnamefont {A.~G.}\ \bibnamefont
  {{Kuranov}}}, \bibinfo {author} {\bibfnamefont {D.~A.}\ \bibnamefont
  {{Kolesnikov}}}, \bibinfo {author} {\bibfnamefont {S.~B.}\ \bibnamefont
  {{Popov}}}, \ and\ \bibinfo {author} {\bibfnamefont {N.~K.}\ \bibnamefont
  {{Porayko}}},\ }\href {\doibase 10.1093/mnras/stw2080} {\bibfield  {journal}
  {\bibinfo  {journal} {\mnras}\ }\textbf {\bibinfo {volume} {463}},\ \bibinfo
  {pages} {1642} (\bibinfo {year} {2016})},\ \Eprint
  {http://arxiv.org/abs/1608.04548} {arXiv:1608.04548 [astro-ph.HE]}
  \BibitemShut {NoStop}%
\bibitem [{\citenamefont {{Tayar}}\ \emph {et~al.}(2019)\citenamefont
  {{Tayar}}, \citenamefont {{Beck}}, \citenamefont {{Pinsonneault}},
  \citenamefont {{Garc{\'\i}a}},\ and\ \citenamefont {{Mathur}}}]{Tayar2019}%
  \BibitemOpen
  \bibfield  {author} {\bibinfo {author} {\bibfnamefont {J.}~\bibnamefont
  {{Tayar}}}, \bibinfo {author} {\bibfnamefont {P.~G.}\ \bibnamefont {{Beck}}},
  \bibinfo {author} {\bibfnamefont {M.~H.}\ \bibnamefont {{Pinsonneault}}},
  \bibinfo {author} {\bibfnamefont {R.~A.}\ \bibnamefont {{Garc{\'\i}a}}}, \
  and\ \bibinfo {author} {\bibfnamefont {S.}~\bibnamefont {{Mathur}}},\ }\href
  {\doibase 10.3847/1538-4357/ab558a} {\bibfield  {journal} {\bibinfo
  {journal} {\apj}\ }\textbf {\bibinfo {volume} {887}},\ \bibinfo {eid} {203}
  (\bibinfo {year} {2019})},\ \Eprint {http://arxiv.org/abs/1911.01443}
  {arXiv:1911.01443 [astro-ph.SR]} \BibitemShut {NoStop}%
\bibitem [{\citenamefont {{Tohline}}(2002)}]{Tohline2002}%
  \BibitemOpen
  \bibfield  {author} {\bibinfo {author} {\bibfnamefont {J.~E.}\ \bibnamefont
  {{Tohline}}},\ }\href {\doibase 10.1146/annurev.astro.40.060401.093810}
  {\bibfield  {journal} {\bibinfo  {journal} {\araa}\ }\textbf {\bibinfo
  {volume} {40}},\ \bibinfo {pages} {349} (\bibinfo {year} {2002})}\BibitemShut
  {NoStop}%
\bibitem [{\citenamefont {{Goodwin}}\ \emph {et~al.}(2007)\citenamefont
  {{Goodwin}}, \citenamefont {{Kroupa}}, \citenamefont {{Goodman}},\ and\
  \citenamefont {{Burkert}}}]{Goodwin2007}%
  \BibitemOpen
  \bibfield  {author} {\bibinfo {author} {\bibfnamefont {S.~P.}\ \bibnamefont
  {{Goodwin}}}, \bibinfo {author} {\bibfnamefont {P.}~\bibnamefont {{Kroupa}}},
  \bibinfo {author} {\bibfnamefont {A.}~\bibnamefont {{Goodman}}}, \ and\
  \bibinfo {author} {\bibfnamefont {A.}~\bibnamefont {{Burkert}}},\ }in\
  \href@noop {} {\emph {\bibinfo {booktitle} {Protostars and Planets V}}},\
  \bibinfo {editor} {edited by\ \bibinfo {editor} {\bibfnamefont
  {B.}~\bibnamefont {{Reipurth}}}, \bibinfo {editor} {\bibfnamefont
  {D.}~\bibnamefont {{Jewitt}}}, \ and\ \bibinfo {editor} {\bibfnamefont
  {K.}~\bibnamefont {{Keil}}}}\ (\bibinfo {year} {2007})\ p.\ \bibinfo {pages}
  {133},\ \Eprint {http://arxiv.org/abs/astro-ph/0603233}
  {arXiv:astro-ph/0603233 [astro-ph]} \BibitemShut {NoStop}%
\bibitem [{\citenamefont {{Duch{\^e}ne}}\ and\ \citenamefont
  {{Kraus}}(2013)}]{Duchene2013}%
  \BibitemOpen
  \bibfield  {author} {\bibinfo {author} {\bibfnamefont {G.}~\bibnamefont
  {{Duch{\^e}ne}}}\ and\ \bibinfo {author} {\bibfnamefont {A.}~\bibnamefont
  {{Kraus}}},\ }\href {\doibase 10.1146/annurev-astro-081710-102602} {\bibfield
   {journal} {\bibinfo  {journal} {\araa}\ }\textbf {\bibinfo {volume} {51}},\
  \bibinfo {pages} {269} (\bibinfo {year} {2013})},\ \Eprint
  {http://arxiv.org/abs/1303.3028} {arXiv:1303.3028 [astro-ph.SR]} \BibitemShut
  {NoStop}%
\bibitem [{\citenamefont {{Corsaro}}\ \emph {et~al.}(2017)\citenamefont
  {{Corsaro}}, \citenamefont {{Lee}}, \citenamefont {{Garc{\'\i}a}},
  \citenamefont {{Hennebelle}}, \citenamefont {{Mathur}}, \citenamefont
  {{Beck}}, \citenamefont {{Mathis}}, \citenamefont {{Stello}},\ and\
  \citenamefont {{Bouvier}}}]{Corsaro2017}%
  \BibitemOpen
  \bibfield  {author} {\bibinfo {author} {\bibfnamefont {E.}~\bibnamefont
  {{Corsaro}}}, \bibinfo {author} {\bibfnamefont {Y.-N.}\ \bibnamefont
  {{Lee}}}, \bibinfo {author} {\bibfnamefont {R.~A.}\ \bibnamefont
  {{Garc{\'\i}a}}}, \bibinfo {author} {\bibfnamefont {P.}~\bibnamefont
  {{Hennebelle}}}, \bibinfo {author} {\bibfnamefont {S.}~\bibnamefont
  {{Mathur}}}, \bibinfo {author} {\bibfnamefont {P.~G.}\ \bibnamefont
  {{Beck}}}, \bibinfo {author} {\bibfnamefont {S.}~\bibnamefont {{Mathis}}},
  \bibinfo {author} {\bibfnamefont {D.}~\bibnamefont {{Stello}}}, \ and\
  \bibinfo {author} {\bibfnamefont {J.}~\bibnamefont {{Bouvier}}},\ }\href
  {\doibase 10.1038/s41550-017-0064} {\bibfield  {journal} {\bibinfo  {journal}
  {Nature Astronomy}\ }\textbf {\bibinfo {volume} {1}},\ \bibinfo {eid} {0064}
  (\bibinfo {year} {2017})},\ \Eprint {http://arxiv.org/abs/1703.05588}
  {arXiv:1703.05588 [astro-ph.SR]} \BibitemShut {NoStop}%
\bibitem [{\citenamefont {{Offner}}\ \emph {et~al.}(2016)\citenamefont
  {{Offner}}, \citenamefont {{Dunham}}, \citenamefont {{Lee}}, \citenamefont
  {{Arce}},\ and\ \citenamefont {{Fielding}}}]{Offner2016}%
  \BibitemOpen
  \bibfield  {author} {\bibinfo {author} {\bibfnamefont {S.~S.~R.}\
  \bibnamefont {{Offner}}}, \bibinfo {author} {\bibfnamefont {M.~M.}\
  \bibnamefont {{Dunham}}}, \bibinfo {author} {\bibfnamefont {K.~I.}\
  \bibnamefont {{Lee}}}, \bibinfo {author} {\bibfnamefont {H.~G.}\ \bibnamefont
  {{Arce}}}, \ and\ \bibinfo {author} {\bibfnamefont {D.~B.}\ \bibnamefont
  {{Fielding}}},\ }\href {\doibase 10.3847/2041-8205/827/1/L11} {\bibfield
  {journal} {\bibinfo  {journal} {\apjl}\ }\textbf {\bibinfo {volume} {827}},\
  \bibinfo {eid} {L11} (\bibinfo {year} {2016})},\ \Eprint
  {http://arxiv.org/abs/1606.08445} {arXiv:1606.08445 [astro-ph.SR]}
  \BibitemShut {NoStop}%
\bibitem [{\citenamefont {Podsiadlowski}(2012)}]{Podsiadlowski2012}%
  \BibitemOpen
  \bibfield  {author} {\bibinfo {author} {\bibfnamefont {P.}~\bibnamefont
  {Podsiadlowski}},\ }\href {\doibase 10.1017/CBO9781139343268.003} {\emph
  {\bibinfo {title} {Accretion Processes In Astrophysics: XXI Canary Islands
  Winter School Of Astrophysics}}}\ (\bibinfo {year} {2012})\BibitemShut
  {NoStop}%
\bibitem [{\citenamefont {{Vanbeveren}}\ \emph {et~al.}(2018)\citenamefont
  {{Vanbeveren}}, \citenamefont {{Mennekens}}, \citenamefont {{Shara}},\ and\
  \citenamefont {{Moffat}}}]{Vanbeveren2018}%
  \BibitemOpen
  \bibfield  {author} {\bibinfo {author} {\bibfnamefont {D.}~\bibnamefont
  {{Vanbeveren}}}, \bibinfo {author} {\bibfnamefont {N.}~\bibnamefont
  {{Mennekens}}}, \bibinfo {author} {\bibfnamefont {M.~M.}\ \bibnamefont
  {{Shara}}}, \ and\ \bibinfo {author} {\bibfnamefont {A.~F.~J.}\ \bibnamefont
  {{Moffat}}},\ }\href {\doibase 10.1051/0004-6361/201732212} {\bibfield
  {journal} {\bibinfo  {journal} {\aap}\ }\textbf {\bibinfo {volume} {615}},\
  \bibinfo {eid} {A65} (\bibinfo {year} {2018})},\ \Eprint
  {http://arxiv.org/abs/1711.05989} {arXiv:1711.05989 [astro-ph.SR]}
  \BibitemShut {NoStop}%
\bibitem [{\citenamefont {Campos}(2018)}]{Campos2018}%
  \BibitemOpen
  \bibfield  {author} {\bibinfo {author} {\bibfnamefont {P.~M.}\ \bibnamefont
  {Campos}},\ }\emph {\bibinfo {title} {{The impact of tides and mass transfer
  on the evolution of metal-poor massive binary stars}}},\ \href@noop {} {Ph.D.
  thesis} (\bibinfo {year} {2018})\BibitemShut {NoStop}%
\bibitem [{\citenamefont {{Eggleton}}(1983)}]{Eggleton1983}%
  \BibitemOpen
  \bibfield  {author} {\bibinfo {author} {\bibfnamefont {P.~P.}\ \bibnamefont
  {{Eggleton}}},\ }\href {\doibase 10.1086/160960} {\bibfield  {journal}
  {\bibinfo  {journal} {\apj}\ }\textbf {\bibinfo {volume} {268}},\ \bibinfo
  {pages} {368} (\bibinfo {year} {1983})}\BibitemShut {NoStop}%
\bibitem [{\citenamefont {{Soberman}}\ \emph {et~al.}(1997)\citenamefont
  {{Soberman}}, \citenamefont {{Phinney}},\ and\ \citenamefont {{van den
  Heuvel}}}]{Soberman1997}%
  \BibitemOpen
  \bibfield  {author} {\bibinfo {author} {\bibfnamefont {G.~E.}\ \bibnamefont
  {{Soberman}}}, \bibinfo {author} {\bibfnamefont {E.~S.}\ \bibnamefont
  {{Phinney}}}, \ and\ \bibinfo {author} {\bibfnamefont {E.~P.~J.}\
  \bibnamefont {{van den Heuvel}}},\ }\href@noop {} {\bibfield  {journal}
  {\bibinfo  {journal} {\aap}\ }\textbf {\bibinfo {volume} {327}},\ \bibinfo
  {pages} {620} (\bibinfo {year} {1997})},\ \Eprint
  {http://arxiv.org/abs/astro-ph/9703016} {arXiv:astro-ph/9703016 [astro-ph]}
  \BibitemShut {NoStop}%
\bibitem [{\citenamefont {{Hurley}}\ \emph {et~al.}(2002)\citenamefont
  {{Hurley}}, \citenamefont {{Tout}},\ and\ \citenamefont
  {{Pols}}}]{Hurley2002}%
  \BibitemOpen
  \bibfield  {author} {\bibinfo {author} {\bibfnamefont {J.~R.}\ \bibnamefont
  {{Hurley}}}, \bibinfo {author} {\bibfnamefont {C.~A.}\ \bibnamefont
  {{Tout}}}, \ and\ \bibinfo {author} {\bibfnamefont {O.~R.}\ \bibnamefont
  {{Pols}}},\ }\href {\doibase 10.1046/j.1365-8711.2002.05038.x} {\bibfield
  {journal} {\bibinfo  {journal} {\mnras}\ }\textbf {\bibinfo {volume} {329}},\
  \bibinfo {pages} {897} (\bibinfo {year} {2002})},\ \Eprint
  {http://arxiv.org/abs/astro-ph/0201220} {arXiv:astro-ph/0201220 [astro-ph]}
  \BibitemShut {NoStop}%
\bibitem [{\citenamefont {{Dominik}}\ \emph {et~al.}(2012)\citenamefont
  {{Dominik}}, \citenamefont {{Belczynski}}, \citenamefont {{Fryer}},
  \citenamefont {{Holz}}, \citenamefont {{Berti}}, \citenamefont {{Bulik}},
  \citenamefont {{Mandel}},\ and\ \citenamefont
  {{O'Shaughnessy}}}]{Dominik2012}%
  \BibitemOpen
  \bibfield  {author} {\bibinfo {author} {\bibfnamefont {M.}~\bibnamefont
  {{Dominik}}}, \bibinfo {author} {\bibfnamefont {K.}~\bibnamefont
  {{Belczynski}}}, \bibinfo {author} {\bibfnamefont {C.}~\bibnamefont
  {{Fryer}}}, \bibinfo {author} {\bibfnamefont {D.~E.}\ \bibnamefont {{Holz}}},
  \bibinfo {author} {\bibfnamefont {E.}~\bibnamefont {{Berti}}}, \bibinfo
  {author} {\bibfnamefont {T.}~\bibnamefont {{Bulik}}}, \bibinfo {author}
  {\bibfnamefont {I.}~\bibnamefont {{Mandel}}}, \ and\ \bibinfo {author}
  {\bibfnamefont {R.}~\bibnamefont {{O'Shaughnessy}}},\ }\href {\doibase
  10.1088/0004-637X/759/1/52} {\bibfield  {journal} {\bibinfo  {journal}
  {\apj}\ }\textbf {\bibinfo {volume} {759}},\ \bibinfo {eid} {52} (\bibinfo
  {year} {2012})},\ \Eprint {http://arxiv.org/abs/1202.4901} {arXiv:1202.4901
  [astro-ph.HE]} \BibitemShut {NoStop}%
\bibitem [{\citenamefont {{Pavlovskii}}\ \emph {et~al.}(2017)\citenamefont
  {{Pavlovskii}}, \citenamefont {{Ivanova}}, \citenamefont {{Belczynski}},\
  and\ \citenamefont {{Van}}}]{Pavlovskii2017}%
  \BibitemOpen
  \bibfield  {author} {\bibinfo {author} {\bibfnamefont {K.}~\bibnamefont
  {{Pavlovskii}}}, \bibinfo {author} {\bibfnamefont {N.}~\bibnamefont
  {{Ivanova}}}, \bibinfo {author} {\bibfnamefont {K.}~\bibnamefont
  {{Belczynski}}}, \ and\ \bibinfo {author} {\bibfnamefont {K.~X.}\
  \bibnamefont {{Van}}},\ }\href {\doibase 10.1093/mnras/stw2786} {\bibfield
  {journal} {\bibinfo  {journal} {\mnras}\ }\textbf {\bibinfo {volume} {465}},\
  \bibinfo {pages} {2092} (\bibinfo {year} {2017})},\ \Eprint
  {http://arxiv.org/abs/1606.04921} {arXiv:1606.04921 [astro-ph.HE]}
  \BibitemShut {NoStop}%
\bibitem [{\citenamefont {{Mapelli}}(2018)}]{Mapelli2018}%
  \BibitemOpen
  \bibfield  {author} {\bibinfo {author} {\bibfnamefont {M.}~\bibnamefont
  {{Mapelli}}},\ }\href@noop {} {\bibfield  {journal} {\bibinfo  {journal}
  {arXiv e-prints}\ ,\ \bibinfo {eid} {arXiv:1809.09130}} (\bibinfo {year}
  {2018})},\ \Eprint {http://arxiv.org/abs/1809.09130} {arXiv:1809.09130
  [astro-ph.HE]} \BibitemShut {NoStop}%
\bibitem [{\citenamefont {{Bavera}}\ \emph
  {et~al.}(2020{\natexlab{b}})\citenamefont {{Bavera}}, \citenamefont
  {{Fragos}}, \citenamefont {{Zevin}}, \citenamefont {{Berry}}, \citenamefont
  {{Marchant}}, \citenamefont {{Andrews}}, \citenamefont {{Coughlin}},
  \citenamefont {{Dotter}}, \citenamefont {{Kovlakas}}, \citenamefont
  {{Misra}}, \citenamefont {{Serra-Perez}}, \citenamefont {{Qin}},
  \citenamefont {{Rocha}}, \citenamefont {{Rom{\'a}n-Garza}}, \citenamefont
  {{Tran}},\ and\ \citenamefont {{Zapartas}}}]{Bavera2020b}%
  \BibitemOpen
  \bibfield  {author} {\bibinfo {author} {\bibfnamefont {S.~S.}\ \bibnamefont
  {{Bavera}}}, \bibinfo {author} {\bibfnamefont {T.}~\bibnamefont {{Fragos}}},
  \bibinfo {author} {\bibfnamefont {M.}~\bibnamefont {{Zevin}}}, \bibinfo
  {author} {\bibfnamefont {C.~P.~L.}\ \bibnamefont {{Berry}}}, \bibinfo
  {author} {\bibfnamefont {P.}~\bibnamefont {{Marchant}}}, \bibinfo {author}
  {\bibfnamefont {J.~J.}\ \bibnamefont {{Andrews}}}, \bibinfo {author}
  {\bibfnamefont {S.}~\bibnamefont {{Coughlin}}}, \bibinfo {author}
  {\bibfnamefont {A.}~\bibnamefont {{Dotter}}}, \bibinfo {author}
  {\bibfnamefont {K.}~\bibnamefont {{Kovlakas}}}, \bibinfo {author}
  {\bibfnamefont {D.}~\bibnamefont {{Misra}}}, \bibinfo {author} {\bibfnamefont
  {J.~G.}\ \bibnamefont {{Serra-Perez}}}, \bibinfo {author} {\bibfnamefont
  {Y.}~\bibnamefont {{Qin}}}, \bibinfo {author} {\bibfnamefont {K.~A.}\
  \bibnamefont {{Rocha}}}, \bibinfo {author} {\bibfnamefont {J.}~\bibnamefont
  {{Rom{\'a}n-Garza}}}, \bibinfo {author} {\bibfnamefont {N.~H.}\ \bibnamefont
  {{Tran}}}, \ and\ \bibinfo {author} {\bibfnamefont {E.}~\bibnamefont
  {{Zapartas}}},\ }\href@noop {} {\bibfield  {journal} {\bibinfo  {journal}
  {arXiv e-prints}\ ,\ \bibinfo {eid} {arXiv:2010.16333}} (\bibinfo {year}
  {2020}{\natexlab{b}})},\ \Eprint {http://arxiv.org/abs/2010.16333}
  {arXiv:2010.16333 [astro-ph.HE]} \BibitemShut {NoStop}%
\bibitem [{\citenamefont {{de Mink}}\ and\ \citenamefont
  {{Belczynski}}(2015)}]{deMink2015}%
  \BibitemOpen
  \bibfield  {author} {\bibinfo {author} {\bibfnamefont {S.~E.}\ \bibnamefont
  {{de Mink}}}\ and\ \bibinfo {author} {\bibfnamefont {K.}~\bibnamefont
  {{Belczynski}}},\ }\href {\doibase 10.1088/0004-637X/814/1/58} {\bibfield
  {journal} {\bibinfo  {journal} {\apj}\ }\textbf {\bibinfo {volume} {814}},\
  \bibinfo {eid} {58} (\bibinfo {year} {2015})},\ \Eprint
  {http://arxiv.org/abs/1506.03573} {arXiv:1506.03573 [astro-ph.HE]}
  \BibitemShut {NoStop}%
\bibitem [{\citenamefont {{van den Heuvel}}(1976)}]{Heuvel1976}%
  \BibitemOpen
  \bibfield  {author} {\bibinfo {author} {\bibfnamefont {E.~P.~J.}\
  \bibnamefont {{van den Heuvel}}},\ }in\ \href@noop {} {\emph {\bibinfo
  {booktitle} {Structure and Evolution of Close Binary Systems}}},\
  Vol.~\bibinfo {volume} {73},\ \bibinfo {editor} {edited by\ \bibinfo {editor}
  {\bibfnamefont {P.}~\bibnamefont {{Eggleton}}}, \bibinfo {editor}
  {\bibfnamefont {S.}~\bibnamefont {{Mitton}}}, \ and\ \bibinfo {editor}
  {\bibfnamefont {J.}~\bibnamefont {{Whelan}}}}\ (\bibinfo {year} {1976})\
  p.~\bibinfo {pages} {35}\BibitemShut {NoStop}%
\bibitem [{\citenamefont {{Paczynski}}(1976)}]{Paczynski1976}%
  \BibitemOpen
  \bibfield  {author} {\bibinfo {author} {\bibfnamefont {B.}~\bibnamefont
  {{Paczynski}}},\ }in\ \href@noop {} {\emph {\bibinfo {booktitle} {Structure
  and Evolution of Close Binary Systems}}},\ Vol.~\bibinfo {volume} {73},\
  \bibinfo {editor} {edited by\ \bibinfo {editor} {\bibfnamefont
  {P.}~\bibnamefont {{Eggleton}}}, \bibinfo {editor} {\bibfnamefont
  {S.}~\bibnamefont {{Mitton}}}, \ and\ \bibinfo {editor} {\bibfnamefont
  {J.}~\bibnamefont {{Whelan}}}}\ (\bibinfo {year} {1976})\ p.~\bibinfo {pages}
  {75}\BibitemShut {NoStop}%
\bibitem [{\citenamefont {{Tutukov}}\ and\ \citenamefont
  {{Yungelson}}(1979)}]{Tutukov1979}%
  \BibitemOpen
  \bibfield  {author} {\bibinfo {author} {\bibfnamefont {A.}~\bibnamefont
  {{Tutukov}}}\ and\ \bibinfo {author} {\bibfnamefont {L.}~\bibnamefont
  {{Yungelson}}}\ }(\bibinfo {year} {1979})\ pp.\ \bibinfo {pages}
  {401--406}\BibitemShut {NoStop}%
\bibitem [{\citenamefont {{Webbink}}(1984)}]{Webbink1984}%
  \BibitemOpen
  \bibfield  {author} {\bibinfo {author} {\bibfnamefont {R.~F.}\ \bibnamefont
  {{Webbink}}},\ }\href {\doibase 10.1086/161701} {\bibfield  {journal}
  {\bibinfo  {journal} {\apj}\ }\textbf {\bibinfo {volume} {277}},\ \bibinfo
  {pages} {355} (\bibinfo {year} {1984})}\BibitemShut {NoStop}%
\bibitem [{\citenamefont {{King}}\ and\ \citenamefont
  {{Kolb}}(1999)}]{King1999}%
  \BibitemOpen
  \bibfield  {author} {\bibinfo {author} {\bibfnamefont {A.~R.}\ \bibnamefont
  {{King}}}\ and\ \bibinfo {author} {\bibfnamefont {U.}~\bibnamefont
  {{Kolb}}},\ }\href {\doibase 10.1046/j.1365-8711.1999.02482.x} {\bibfield
  {journal} {\bibinfo  {journal} {\mnras}\ }\textbf {\bibinfo {volume} {305}},\
  \bibinfo {pages} {654} (\bibinfo {year} {1999})},\ \Eprint
  {http://arxiv.org/abs/astro-ph/9901296} {arXiv:astro-ph/9901296 [astro-ph]}
  \BibitemShut {NoStop}%
\bibitem [{\citenamefont {{MacLeod}}\ and\ \citenamefont
  {{Ramirez-Ruiz}}(2015)}]{MacLeod2015}%
  \BibitemOpen
  \bibfield  {author} {\bibinfo {author} {\bibfnamefont {M.}~\bibnamefont
  {{MacLeod}}}\ and\ \bibinfo {author} {\bibfnamefont {E.}~\bibnamefont
  {{Ramirez-Ruiz}}},\ }\href {\doibase 10.1088/0004-637X/803/1/41} {\bibfield
  {journal} {\bibinfo  {journal} {\apj}\ }\textbf {\bibinfo {volume} {803}},\
  \bibinfo {eid} {41} (\bibinfo {year} {2015})},\ \Eprint
  {http://arxiv.org/abs/1410.3823} {arXiv:1410.3823 [astro-ph.SR]} \BibitemShut
  {NoStop}%
\bibitem [{\citenamefont {{Chamandy}}\ \emph {et~al.}(2018)\citenamefont
  {{Chamandy}}, \citenamefont {{Frank}}, \citenamefont {{Blackman}},
  \citenamefont {{Carroll-Nellenback}}, \citenamefont {{Liu}}, \citenamefont
  {{Tu}}, \citenamefont {{Nordhaus}}, \citenamefont {{Chen}},\ and\
  \citenamefont {{Peng}}}]{Chamandy2018}%
  \BibitemOpen
  \bibfield  {author} {\bibinfo {author} {\bibfnamefont {L.}~\bibnamefont
  {{Chamandy}}}, \bibinfo {author} {\bibfnamefont {A.}~\bibnamefont {{Frank}}},
  \bibinfo {author} {\bibfnamefont {E.~G.}\ \bibnamefont {{Blackman}}},
  \bibinfo {author} {\bibfnamefont {J.}~\bibnamefont {{Carroll-Nellenback}}},
  \bibinfo {author} {\bibfnamefont {B.}~\bibnamefont {{Liu}}}, \bibinfo
  {author} {\bibfnamefont {Y.}~\bibnamefont {{Tu}}}, \bibinfo {author}
  {\bibfnamefont {J.}~\bibnamefont {{Nordhaus}}}, \bibinfo {author}
  {\bibfnamefont {Z.}~\bibnamefont {{Chen}}}, \ and\ \bibinfo {author}
  {\bibfnamefont {B.}~\bibnamefont {{Peng}}},\ }\href {\doibase
  10.1093/mnras/sty1950} {\bibfield  {journal} {\bibinfo  {journal} {\mnras}\
  }\textbf {\bibinfo {volume} {480}},\ \bibinfo {pages} {1898} (\bibinfo {year}
  {2018})},\ \Eprint {http://arxiv.org/abs/1805.03607} {arXiv:1805.03607
  [astro-ph.SR]} \BibitemShut {NoStop}%
\bibitem [{\citenamefont {{Meurs}}\ and\ \citenamefont {{van den
  Heuvel}}(1989)}]{Meurs1989}%
  \BibitemOpen
  \bibfield  {author} {\bibinfo {author} {\bibfnamefont {E.~J.~A.}\
  \bibnamefont {{Meurs}}}\ and\ \bibinfo {author} {\bibfnamefont {E.~P.~J.}\
  \bibnamefont {{van den Heuvel}}},\ }\href@noop {} {\bibfield  {journal}
  {\bibinfo  {journal} {\aap}\ }\textbf {\bibinfo {volume} {226}},\ \bibinfo
  {pages} {88} (\bibinfo {year} {1989})}\BibitemShut {NoStop}%
\bibitem [{\citenamefont {{Kushnir}}\ \emph {et~al.}(2016)\citenamefont
  {{Kushnir}}, \citenamefont {{Zaldarriaga}}, \citenamefont {{Kollmeier}},\
  and\ \citenamefont {{Waldman}}}]{Kushnir2016}%
  \BibitemOpen
  \bibfield  {author} {\bibinfo {author} {\bibfnamefont {D.}~\bibnamefont
  {{Kushnir}}}, \bibinfo {author} {\bibfnamefont {M.}~\bibnamefont
  {{Zaldarriaga}}}, \bibinfo {author} {\bibfnamefont {J.~A.}\ \bibnamefont
  {{Kollmeier}}}, \ and\ \bibinfo {author} {\bibfnamefont {R.}~\bibnamefont
  {{Waldman}}},\ }\href {\doibase 10.1093/mnras/stw1684} {\bibfield  {journal}
  {\bibinfo  {journal} {\mnras}\ }\textbf {\bibinfo {volume} {462}},\ \bibinfo
  {pages} {844} (\bibinfo {year} {2016})},\ \Eprint
  {http://arxiv.org/abs/1605.03839} {arXiv:1605.03839 [astro-ph.HE]}
  \BibitemShut {NoStop}%
\bibitem [{\citenamefont {{Zahn}}(2008)}]{Zahn2008}%
  \BibitemOpen
  \bibfield  {author} {\bibinfo {author} {\bibfnamefont {J.~P.}\ \bibnamefont
  {{Zahn}}},\ }\bibfield  {booktitle} {\emph {\bibinfo {booktitle} {EAS
  Publications Series}},\ }\href {\doibase 10.1051/eas:0829002} {\ \bibinfo
  {series} {EAS Publications Series},\ \textbf {\bibinfo {volume} {29}},\
  \bibinfo {pages} {67} (\bibinfo {year} {2008})},\ \Eprint
  {http://arxiv.org/abs/0807.4870} {arXiv:0807.4870 [astro-ph]} \BibitemShut
  {NoStop}%
\bibitem [{\citenamefont {{Zahn}}(1975)}]{Zahn1975}%
  \BibitemOpen
  \bibfield  {author} {\bibinfo {author} {\bibfnamefont {J.~P.}\ \bibnamefont
  {{Zahn}}},\ }\href@noop {} {\bibfield  {journal} {\bibinfo  {journal} {\aap}\
  }\textbf {\bibinfo {volume} {41}},\ \bibinfo {pages} {329} (\bibinfo {year}
  {1975})}\BibitemShut {NoStop}%
\bibitem [{\citenamefont {{Zahn}}(1977)}]{Zahn1977}%
  \BibitemOpen
  \bibfield  {author} {\bibinfo {author} {\bibfnamefont {J.~P.}\ \bibnamefont
  {{Zahn}}},\ }\href@noop {} {\bibfield  {journal} {\bibinfo  {journal} {\aap}\
  }\textbf {\bibinfo {volume} {500}},\ \bibinfo {pages} {121} (\bibinfo {year}
  {1977})}\BibitemShut {NoStop}%
\bibitem [{\citenamefont {{Hut}}(1981)}]{Hut1981}%
  \BibitemOpen
  \bibfield  {author} {\bibinfo {author} {\bibfnamefont {P.}~\bibnamefont
  {{Hut}}},\ }\href@noop {} {\bibfield  {journal} {\bibinfo  {journal} {\aap}\
  }\textbf {\bibinfo {volume} {99}},\ \bibinfo {pages} {126} (\bibinfo {year}
  {1981})}\BibitemShut {NoStop}%
\bibitem [{\citenamefont {{Lecar}}\ \emph {et~al.}(1976)\citenamefont
  {{Lecar}}, \citenamefont {{Wheeler}},\ and\ \citenamefont
  {{McKee}}}]{Lecar1976}%
  \BibitemOpen
  \bibfield  {author} {\bibinfo {author} {\bibfnamefont {M.}~\bibnamefont
  {{Lecar}}}, \bibinfo {author} {\bibfnamefont {J.~C.}\ \bibnamefont
  {{Wheeler}}}, \ and\ \bibinfo {author} {\bibfnamefont {C.~F.}\ \bibnamefont
  {{McKee}}},\ }\href {\doibase 10.1086/154311} {\bibfield  {journal} {\bibinfo
   {journal} {\apj}\ }\textbf {\bibinfo {volume} {205}},\ \bibinfo {pages}
  {556} (\bibinfo {year} {1976})}\BibitemShut {NoStop}%
\bibitem [{\citenamefont {{Siess}}\ \emph {et~al.}(2013)\citenamefont
  {{Siess}}, \citenamefont {{Izzard}}, \citenamefont {{Davis}},\ and\
  \citenamefont {{Deschamps}}}]{Siess2013}%
  \BibitemOpen
  \bibfield  {author} {\bibinfo {author} {\bibfnamefont {L.}~\bibnamefont
  {{Siess}}}, \bibinfo {author} {\bibfnamefont {R.~G.}\ \bibnamefont
  {{Izzard}}}, \bibinfo {author} {\bibfnamefont {P.~J.}\ \bibnamefont
  {{Davis}}}, \ and\ \bibinfo {author} {\bibfnamefont {R.}~\bibnamefont
  {{Deschamps}}},\ }\href {\doibase 10.1051/0004-6361/201220327} {\bibfield
  {journal} {\bibinfo  {journal} {\aap}\ }\textbf {\bibinfo {volume} {550}},\
  \bibinfo {eid} {A100} (\bibinfo {year} {2013})}\BibitemShut {NoStop}%
\bibitem [{\citenamefont {Peters}(1964)}]{Peters1964}%
  \BibitemOpen
  \bibfield  {author} {\bibinfo {author} {\bibfnamefont {P.~C.}\ \bibnamefont
  {Peters}},\ }\href {\doibase 10.1103/PhysRev.136.B1224} {\bibfield  {journal}
  {\bibinfo  {journal} {Phys. Rev.}\ }\textbf {\bibinfo {volume} {136}},\
  \bibinfo {pages} {B1224} (\bibinfo {year} {1964})}\BibitemShut {NoStop}%
\bibitem [{\citenamefont {{Planck Collaboration}}\ \emph
  {et~al.}(2016)\citenamefont {{Planck Collaboration}} \emph
  {et~al.}}]{Planck2016}%
  \BibitemOpen
  \bibfield  {author} {\bibinfo {author} {\bibnamefont {{Planck
  Collaboration}}} \emph {et~al.},\ }\href {\doibase
  10.1051/0004-6361/201525830} {\bibfield  {journal} {\bibinfo  {journal}
  {\aap}\ }\textbf {\bibinfo {volume} {594}},\ \bibinfo {eid} {A13} (\bibinfo
  {year} {2016})},\ \Eprint {http://arxiv.org/abs/1502.01589} {arXiv:1502.01589
  [astro-ph.CO]} \BibitemShut {NoStop}%
\bibitem [{\citenamefont {{Lamers}}\ and\ \citenamefont
  {{Cassinelli}}(1999)}]{Lamers1999}%
  \BibitemOpen
  \bibfield  {author} {\bibinfo {author} {\bibfnamefont {H.~J.~G.~L.~M.}\
  \bibnamefont {{Lamers}}}\ and\ \bibinfo {author} {\bibfnamefont {J.~P.}\
  \bibnamefont {{Cassinelli}}},\ }\href@noop {} {\emph {\bibinfo {title}
  {Introduction to Stellar Winds, by Henny J.~G.~L.~M.~Lamers and Joseph
  P.~Cassinelli, pp.~452.~ISBN 0521593980.~Cambridge, UK: Cambridge University
  Press, June 1999.}}}\ (\bibinfo {year} {1999})\ p.\ \bibinfo {pages}
  {452}\BibitemShut {NoStop}%
\bibitem [{\citenamefont {{Abbott}}(1982)}]{Abbott1982}%
  \BibitemOpen
  \bibfield  {author} {\bibinfo {author} {\bibfnamefont {D.~C.}\ \bibnamefont
  {{Abbott}}},\ }\href {\doibase 10.1086/160166} {\bibfield  {journal}
  {\bibinfo  {journal} {\apj}\ }\textbf {\bibinfo {volume} {259}},\ \bibinfo
  {pages} {282} (\bibinfo {year} {1982})}\BibitemShut {NoStop}%
\bibitem [{\citenamefont {{Shimada}}\ \emph {et~al.}(1994)\citenamefont
  {{Shimada}}, \citenamefont {{Ito}}, \citenamefont {{Hirata}},\ and\
  \citenamefont {{Horaguchi}}}]{Shimada1994}%
  \BibitemOpen
  \bibfield  {author} {\bibinfo {author} {\bibfnamefont {M.~R.}\ \bibnamefont
  {{Shimada}}}, \bibinfo {author} {\bibfnamefont {M.}~\bibnamefont {{Ito}}},
  \bibinfo {author} {\bibfnamefont {B.}~\bibnamefont {{Hirata}}}, \ and\
  \bibinfo {author} {\bibfnamefont {T.}~\bibnamefont {{Horaguchi}}},\ }in\
  \href@noop {} {\emph {\bibinfo {booktitle} {Pulsation; Rotation; and Mass
  Loss in Early-Type Stars}}},\ \bibinfo {series} {IAU Symposium}, Vol.\
  \bibinfo {volume} {162},\ \bibinfo {editor} {edited by\ \bibinfo {editor}
  {\bibfnamefont {L.~A.}\ \bibnamefont {{Balona}}}, \bibinfo {editor}
  {\bibfnamefont {H.~F.}\ \bibnamefont {{Henrichs}}}, \ and\ \bibinfo {editor}
  {\bibfnamefont {J.~M.}\ \bibnamefont {{Le Contel}}}}\ (\bibinfo {year}
  {1994})\ p.\ \bibinfo {pages} {487}\BibitemShut {NoStop}%
\bibitem [{\citenamefont {{Vink}}\ \emph {et~al.}(2001)\citenamefont {{Vink}},
  \citenamefont {{de Koter}},\ and\ \citenamefont {{Lamers}}}]{Vink2001}%
  \BibitemOpen
  \bibfield  {author} {\bibinfo {author} {\bibfnamefont {J.~S.}\ \bibnamefont
  {{Vink}}}, \bibinfo {author} {\bibfnamefont {A.}~\bibnamefont {{de Koter}}},
  \ and\ \bibinfo {author} {\bibfnamefont {H.~J.~G.~L.~M.}\ \bibnamefont
  {{Lamers}}},\ }\href {\doibase 10.1051/0004-6361:20010127} {\bibfield
  {journal} {\bibinfo  {journal} {\aap}\ }\textbf {\bibinfo {volume} {369}},\
  \bibinfo {pages} {574} (\bibinfo {year} {2001})},\ \Eprint
  {http://arxiv.org/abs/astro-ph/0101509} {arXiv:astro-ph/0101509 [astro-ph]}
  \BibitemShut {NoStop}%
\bibitem [{\citenamefont {{Smith}}(2014)}]{Smith2014}%
  \BibitemOpen
  \bibfield  {author} {\bibinfo {author} {\bibfnamefont {N.}~\bibnamefont
  {{Smith}}},\ }\href {\doibase 10.1146/annurev-astro-081913-040025} {\bibfield
   {journal} {\bibinfo  {journal} {\araa}\ }\textbf {\bibinfo {volume} {52}},\
  \bibinfo {pages} {487} (\bibinfo {year} {2014})},\ \Eprint
  {http://arxiv.org/abs/1402.1237} {arXiv:1402.1237 [astro-ph.SR]} \BibitemShut
  {NoStop}%
\bibitem [{\citenamefont {{Renzo}}\ \emph {et~al.}(2017)\citenamefont
  {{Renzo}}, \citenamefont {{Ott}}, \citenamefont {{Shore}},\ and\
  \citenamefont {{de Mink}}}]{Renzo2017}%
  \BibitemOpen
  \bibfield  {author} {\bibinfo {author} {\bibfnamefont {M.}~\bibnamefont
  {{Renzo}}}, \bibinfo {author} {\bibfnamefont {C.~D.}\ \bibnamefont {{Ott}}},
  \bibinfo {author} {\bibfnamefont {S.~N.}\ \bibnamefont {{Shore}}}, \ and\
  \bibinfo {author} {\bibfnamefont {S.~E.}\ \bibnamefont {{de Mink}}},\ }\href
  {\doibase 10.1051/0004-6361/201730698} {\bibfield  {journal} {\bibinfo
  {journal} {\aap}\ }\textbf {\bibinfo {volume} {603}},\ \bibinfo {eid} {A118}
  (\bibinfo {year} {2017})},\ \Eprint {http://arxiv.org/abs/1703.09705}
  {arXiv:1703.09705 [astro-ph.SR]} \BibitemShut {NoStop}%
\bibitem [{\citenamefont {{Pols}}\ and\ \citenamefont
  {{Dewi}}(2002)}]{Pols2002}%
  \BibitemOpen
  \bibfield  {author} {\bibinfo {author} {\bibfnamefont {O.~R.}\ \bibnamefont
  {{Pols}}}\ and\ \bibinfo {author} {\bibfnamefont {J.~D.~M.}\ \bibnamefont
  {{Dewi}}},\ }\href {\doibase 10.1071/AS01121} {\bibfield  {journal} {\bibinfo
   {journal} {\pasa}\ }\textbf {\bibinfo {volume} {19}},\ \bibinfo {pages}
  {233} (\bibinfo {year} {2002})},\ \Eprint
  {http://arxiv.org/abs/astro-ph/0203308} {arXiv:astro-ph/0203308 [astro-ph]}
  \BibitemShut {NoStop}%
\bibitem [{\citenamefont {{Vink}}(2017)}]{Vink2017}%
  \BibitemOpen
  \bibfield  {author} {\bibinfo {author} {\bibfnamefont {J.~S.}\ \bibnamefont
  {{Vink}}},\ }\href {\doibase 10.1051/0004-6361/201731902} {\bibfield
  {journal} {\bibinfo  {journal} {\aap}\ }\textbf {\bibinfo {volume} {607}},\
  \bibinfo {eid} {L8} (\bibinfo {year} {2017})},\ \Eprint
  {http://arxiv.org/abs/1710.02010} {arXiv:1710.02010 [astro-ph.SR]}
  \BibitemShut {NoStop}%
\bibitem [{\citenamefont {{Motz}}(1952)}]{Motz1952}%
  \BibitemOpen
  \bibfield  {author} {\bibinfo {author} {\bibfnamefont {L.}~\bibnamefont
  {{Motz}}},\ }\href {\doibase 10.1086/145570} {\bibfield  {journal} {\bibinfo
  {journal} {\apj}\ }\textbf {\bibinfo {volume} {115}},\ \bibinfo {pages} {562}
  (\bibinfo {year} {1952})}\BibitemShut {NoStop}%
\bibitem [{\citenamefont {{Claret}}\ and\ \citenamefont
  {{Gimenez}}(1989)}]{Claret1989}%
  \BibitemOpen
  \bibfield  {author} {\bibinfo {author} {\bibfnamefont {A.}~\bibnamefont
  {{Claret}}}\ and\ \bibinfo {author} {\bibfnamefont {A.}~\bibnamefont
  {{Gimenez}}},\ }\href@noop {} {\bibfield  {journal} {\bibinfo  {journal}
  {\aaps}\ }\textbf {\bibinfo {volume} {81}},\ \bibinfo {pages} {37} (\bibinfo
  {year} {1989})}\BibitemShut {NoStop}%
\bibitem [{\citenamefont {{Hobbs}}\ \emph {et~al.}(2005)\citenamefont
  {{Hobbs}}, \citenamefont {{Lorimer}}, \citenamefont {{Lyne}},\ and\
  \citenamefont {{Kramer}}}]{Hobbs2005}%
  \BibitemOpen
  \bibfield  {author} {\bibinfo {author} {\bibfnamefont {G.}~\bibnamefont
  {{Hobbs}}}, \bibinfo {author} {\bibfnamefont {D.~R.}\ \bibnamefont
  {{Lorimer}}}, \bibinfo {author} {\bibfnamefont {A.~G.}\ \bibnamefont
  {{Lyne}}}, \ and\ \bibinfo {author} {\bibfnamefont {M.}~\bibnamefont
  {{Kramer}}},\ }\href {\doibase 10.1111/j.1365-2966.2005.09087.x} {\bibfield
  {journal} {\bibinfo  {journal} {\mnras}\ }\textbf {\bibinfo {volume} {360}},\
  \bibinfo {pages} {974} (\bibinfo {year} {2005})},\ \Eprint
  {http://arxiv.org/abs/astro-ph/0504584} {arXiv:astro-ph/0504584 [astro-ph]}
  \BibitemShut {NoStop}%
\bibitem [{\citenamefont {{Fryer}}(1999)}]{Fryer1999}%
  \BibitemOpen
  \bibfield  {author} {\bibinfo {author} {\bibfnamefont {C.~L.}\ \bibnamefont
  {{Fryer}}},\ }\href {\doibase 10.1086/307647} {\bibfield  {journal} {\bibinfo
   {journal} {\apj}\ }\textbf {\bibinfo {volume} {522}},\ \bibinfo {pages}
  {413} (\bibinfo {year} {1999})},\ \Eprint
  {http://arxiv.org/abs/astro-ph/9902315} {arXiv:astro-ph/9902315 [astro-ph]}
  \BibitemShut {NoStop}%
\bibitem [{\citenamefont {{Martin}}\ \emph {et~al.}(2010)\citenamefont
  {{Martin}}, \citenamefont {{Tout}},\ and\ \citenamefont
  {{Pringle}}}]{Martin2010}%
  \BibitemOpen
  \bibfield  {author} {\bibinfo {author} {\bibfnamefont {R.~G.}\ \bibnamefont
  {{Martin}}}, \bibinfo {author} {\bibfnamefont {C.~A.}\ \bibnamefont
  {{Tout}}}, \ and\ \bibinfo {author} {\bibfnamefont {J.~E.}\ \bibnamefont
  {{Pringle}}},\ }\href {\doibase 10.1111/j.1365-2966.2009.15777.x} {\bibfield
  {journal} {\bibinfo  {journal} {\mnras}\ }\textbf {\bibinfo {volume} {401}},\
  \bibinfo {pages} {1514} (\bibinfo {year} {2010})},\ \Eprint
  {http://arxiv.org/abs/0910.0018} {arXiv:0910.0018 [astro-ph.HE]} \BibitemShut
  {NoStop}%
\bibitem [{\citenamefont {{Janka}}(2012)}]{Janka2012}%
  \BibitemOpen
  \bibfield  {author} {\bibinfo {author} {\bibfnamefont {H.-T.}\ \bibnamefont
  {{Janka}}},\ }\href {\doibase 10.1146/annurev-nucl-102711-094901} {\bibfield
  {journal} {\bibinfo  {journal} {Annual Review of Nuclear and Particle
  Science}\ }\textbf {\bibinfo {volume} {62}},\ \bibinfo {pages} {407}
  (\bibinfo {year} {2012})},\ \Eprint {http://arxiv.org/abs/1206.2503}
  {arXiv:1206.2503 [astro-ph.SR]} \BibitemShut {NoStop}%
\bibitem [{\citenamefont {{Repetto}}\ \emph {et~al.}(2012)\citenamefont
  {{Repetto}}, \citenamefont {{Davies}},\ and\ \citenamefont
  {{Sigurdsson}}}]{Repetto2012}%
  \BibitemOpen
  \bibfield  {author} {\bibinfo {author} {\bibfnamefont {S.}~\bibnamefont
  {{Repetto}}}, \bibinfo {author} {\bibfnamefont {M.~B.}\ \bibnamefont
  {{Davies}}}, \ and\ \bibinfo {author} {\bibfnamefont {S.}~\bibnamefont
  {{Sigurdsson}}},\ }\href {\doibase 10.1111/j.1365-2966.2012.21549.x}
  {\bibfield  {journal} {\bibinfo  {journal} {\mnras}\ }\textbf {\bibinfo
  {volume} {425}},\ \bibinfo {pages} {2799} (\bibinfo {year} {2012})},\ \Eprint
  {http://arxiv.org/abs/1203.3077} {arXiv:1203.3077 [astro-ph.GA]} \BibitemShut
  {NoStop}%
\bibitem [{\citenamefont {{Repetto}}\ and\ \citenamefont
  {{Nelemans}}(2015)}]{Repetto2015}%
  \BibitemOpen
  \bibfield  {author} {\bibinfo {author} {\bibfnamefont {S.}~\bibnamefont
  {{Repetto}}}\ and\ \bibinfo {author} {\bibfnamefont {G.}~\bibnamefont
  {{Nelemans}}},\ }\href {\doibase 10.1093/mnras/stv1753} {\bibfield  {journal}
  {\bibinfo  {journal} {\mnras}\ }\textbf {\bibinfo {volume} {453}},\ \bibinfo
  {pages} {3341} (\bibinfo {year} {2015})},\ \Eprint
  {http://arxiv.org/abs/1507.08105} {arXiv:1507.08105 [astro-ph.HE]}
  \BibitemShut {NoStop}%
\bibitem [{\citenamefont {{Mandel}}(2016)}]{Mandel2016}%
  \BibitemOpen
  \bibfield  {author} {\bibinfo {author} {\bibfnamefont {I.}~\bibnamefont
  {{Mandel}}},\ }\href {\doibase 10.1093/mnras/stv2733} {\bibfield  {journal}
  {\bibinfo  {journal} {\mnras}\ }\textbf {\bibinfo {volume} {456}},\ \bibinfo
  {pages} {578} (\bibinfo {year} {2016})},\ \Eprint
  {http://arxiv.org/abs/1510.03871} {arXiv:1510.03871 [astro-ph.HE]}
  \BibitemShut {NoStop}%
\bibitem [{\citenamefont {{Janka}}(2013)}]{Janka2013}%
  \BibitemOpen
  \bibfield  {author} {\bibinfo {author} {\bibfnamefont {H.-T.}\ \bibnamefont
  {{Janka}}},\ }\href {\doibase 10.1093/mnras/stt1106} {\bibfield  {journal}
  {\bibinfo  {journal} {\mnras}\ }\textbf {\bibinfo {volume} {434}},\ \bibinfo
  {pages} {1355} (\bibinfo {year} {2013})},\ \Eprint
  {http://arxiv.org/abs/1306.0007} {arXiv:1306.0007 [astro-ph.SR]} \BibitemShut
  {NoStop}%
\bibitem [{\citenamefont {{Taylor}}\ and\ \citenamefont
  {{Gerosa}}(2018)}]{Taylor2018}%
  \BibitemOpen
  \bibfield  {author} {\bibinfo {author} {\bibfnamefont {S.~R.}\ \bibnamefont
  {{Taylor}}}\ and\ \bibinfo {author} {\bibfnamefont {D.}~\bibnamefont
  {{Gerosa}}},\ }\href {\doibase 10.1103/PhysRevD.98.083017} {\bibfield
  {journal} {\bibinfo  {journal} {\prd}\ }\textbf {\bibinfo {volume} {98}},\
  \bibinfo {eid} {083017} (\bibinfo {year} {2018})},\ \Eprint
  {http://arxiv.org/abs/1806.08365} {arXiv:1806.08365 [astro-ph.HE]}
  \BibitemShut {NoStop}%
\bibitem [{\citenamefont {{Fryer}}\ \emph {et~al.}(2012)\citenamefont
  {{Fryer}}, \citenamefont {{Belczynski}}, \citenamefont {{Wiktorowicz}},
  \citenamefont {{Dominik}}, \citenamefont {{Kalogera}},\ and\ \citenamefont
  {{Holz}}}]{Fryer2012}%
  \BibitemOpen
  \bibfield  {author} {\bibinfo {author} {\bibfnamefont {C.~L.}\ \bibnamefont
  {{Fryer}}}, \bibinfo {author} {\bibfnamefont {K.}~\bibnamefont
  {{Belczynski}}}, \bibinfo {author} {\bibfnamefont {G.}~\bibnamefont
  {{Wiktorowicz}}}, \bibinfo {author} {\bibfnamefont {M.}~\bibnamefont
  {{Dominik}}}, \bibinfo {author} {\bibfnamefont {V.}~\bibnamefont
  {{Kalogera}}}, \ and\ \bibinfo {author} {\bibfnamefont {D.~E.}\ \bibnamefont
  {{Holz}}},\ }\href {\doibase 10.1088/0004-637X/749/1/91} {\bibfield
  {journal} {\bibinfo  {journal} {\apj}\ }\textbf {\bibinfo {volume} {749}},\
  \bibinfo {eid} {91} (\bibinfo {year} {2012})},\ \Eprint
  {http://arxiv.org/abs/1110.1726} {arXiv:1110.1726 [astro-ph.SR]} \BibitemShut
  {NoStop}%
\bibitem [{\citenamefont {{Kidder}}(1995)}]{Kidder1995}%
  \BibitemOpen
  \bibfield  {author} {\bibinfo {author} {\bibfnamefont {L.~E.}\ \bibnamefont
  {{Kidder}}},\ }\href {\doibase 10.1103/PhysRevD.52.821} {\bibfield  {journal}
  {\bibinfo  {journal} {\prd}\ }\textbf {\bibinfo {volume} {52}},\ \bibinfo
  {pages} {821} (\bibinfo {year} {1995})},\ \Eprint
  {http://arxiv.org/abs/gr-qc/9506022} {arXiv:gr-qc/9506022 [gr-qc]}
  \BibitemShut {NoStop}%
\bibitem [{\citenamefont {{Lattimer}}\ and\ \citenamefont
  {{Yahil}}(1989)}]{Lattimer1989}%
  \BibitemOpen
  \bibfield  {author} {\bibinfo {author} {\bibfnamefont {J.~M.}\ \bibnamefont
  {{Lattimer}}}\ and\ \bibinfo {author} {\bibfnamefont {A.}~\bibnamefont
  {{Yahil}}},\ }\href {\doibase 10.1086/167404} {\bibfield  {journal} {\bibinfo
   {journal} {\apj}\ }\textbf {\bibinfo {volume} {340}},\ \bibinfo {pages}
  {426} (\bibinfo {year} {1989})}\BibitemShut {NoStop}%
\bibitem [{\citenamefont {{Timmes}}\ \emph {et~al.}(1996)\citenamefont
  {{Timmes}}, \citenamefont {{Woosley}},\ and\ \citenamefont
  {{Weaver}}}]{Timmes1996}%
  \BibitemOpen
  \bibfield  {author} {\bibinfo {author} {\bibfnamefont {F.~X.}\ \bibnamefont
  {{Timmes}}}, \bibinfo {author} {\bibfnamefont {S.~E.}\ \bibnamefont
  {{Woosley}}}, \ and\ \bibinfo {author} {\bibfnamefont {T.~A.}\ \bibnamefont
  {{Weaver}}},\ }\href {\doibase 10.1086/176778} {\bibfield  {journal}
  {\bibinfo  {journal} {\apj}\ }\textbf {\bibinfo {volume} {457}},\ \bibinfo
  {pages} {834} (\bibinfo {year} {1996})},\ \Eprint
  {http://arxiv.org/abs/astro-ph/9510136} {arXiv:astro-ph/9510136 [astro-ph]}
  \BibitemShut {NoStop}%
\bibitem [{\citenamefont {{Belczynski}}\ \emph
  {et~al.}(2016{\natexlab{b}})\citenamefont {{Belczynski}}, \citenamefont
  {{Heger}}, \citenamefont {{Gladysz}}, \citenamefont {{Ruiter}}, \citenamefont
  {{Woosley}}, \citenamefont {{Wiktorowicz}}, \citenamefont {{Chen}},
  \citenamefont {{Bulik}}, \citenamefont {{O'Shaughnessy}}, \citenamefont
  {{Holz}}, \citenamefont {{Fryer}},\ and\ \citenamefont {{Berti}}}]{BK2016}%
  \BibitemOpen
  \bibfield  {author} {\bibinfo {author} {\bibfnamefont {K.}~\bibnamefont
  {{Belczynski}}}, \bibinfo {author} {\bibfnamefont {A.}~\bibnamefont
  {{Heger}}}, \bibinfo {author} {\bibfnamefont {W.}~\bibnamefont {{Gladysz}}},
  \bibinfo {author} {\bibfnamefont {A.~J.}\ \bibnamefont {{Ruiter}}}, \bibinfo
  {author} {\bibfnamefont {S.}~\bibnamefont {{Woosley}}}, \bibinfo {author}
  {\bibfnamefont {G.}~\bibnamefont {{Wiktorowicz}}}, \bibinfo {author}
  {\bibfnamefont {H.~Y.}\ \bibnamefont {{Chen}}}, \bibinfo {author}
  {\bibfnamefont {T.}~\bibnamefont {{Bulik}}}, \bibinfo {author} {\bibfnamefont
  {R.}~\bibnamefont {{O'Shaughnessy}}}, \bibinfo {author} {\bibfnamefont
  {D.~E.}\ \bibnamefont {{Holz}}}, \bibinfo {author} {\bibfnamefont {C.~L.}\
  \bibnamefont {{Fryer}}}, \ and\ \bibinfo {author} {\bibfnamefont
  {E.}~\bibnamefont {{Berti}}},\ }\href {\doibase 10.1051/0004-6361/201628980}
  {\bibfield  {journal} {\bibinfo  {journal} {\aap}\ }\textbf {\bibinfo
  {volume} {594}},\ \bibinfo {eid} {A97} (\bibinfo {year}
  {2016}{\natexlab{b}})},\ \Eprint {http://arxiv.org/abs/1607.03116}
  {arXiv:1607.03116 [astro-ph.HE]} \BibitemShut {NoStop}%
\bibitem [{\citenamefont {{Janka}}(2017)}]{Janka2017}%
  \BibitemOpen
  \bibfield  {author} {\bibinfo {author} {\bibfnamefont {H.-T.}\ \bibnamefont
  {{Janka}}},\ }\enquote {\bibinfo {title} {{Neutrino Emission from
  Supernovae}},}\ in\ \href {\doibase 10.1007/978-3-319-21846-5_4} {\emph
  {\bibinfo {booktitle} {Handbook of Supernovae}}},\ \bibinfo {editor} {edited
  by\ \bibinfo {editor} {\bibfnamefont {A.~W.}\ \bibnamefont {{Alsabti}}}\ and\
  \bibinfo {editor} {\bibfnamefont {P.}~\bibnamefont {{Murdin}}}}\ (\bibinfo
  {year} {2017})\ p.\ \bibinfo {pages} {1575}\BibitemShut {NoStop}%
\bibitem [{\citenamefont {{O'Connor}}\ and\ \citenamefont
  {{Ott}}(2011)}]{OConnor2011}%
  \BibitemOpen
  \bibfield  {author} {\bibinfo {author} {\bibfnamefont {E.}~\bibnamefont
  {{O'Connor}}}\ and\ \bibinfo {author} {\bibfnamefont {C.~D.}\ \bibnamefont
  {{Ott}}},\ }\href {\doibase 10.1088/0004-637X/730/2/70} {\bibfield  {journal}
  {\bibinfo  {journal} {\apj}\ }\textbf {\bibinfo {volume} {730}},\ \bibinfo
  {eid} {70} (\bibinfo {year} {2011})},\ \Eprint
  {http://arxiv.org/abs/1010.5550} {arXiv:1010.5550 [astro-ph.HE]} \BibitemShut
  {NoStop}%
\bibitem [{\citenamefont {{Kesden}}\ \emph
  {et~al.}(2010{\natexlab{a}})\citenamefont {{Kesden}}, \citenamefont
  {{Sperhake}},\ and\ \citenamefont {{Berti}}}]{Kesden2010}%
  \BibitemOpen
  \bibfield  {author} {\bibinfo {author} {\bibfnamefont {M.}~\bibnamefont
  {{Kesden}}}, \bibinfo {author} {\bibfnamefont {U.}~\bibnamefont
  {{Sperhake}}}, \ and\ \bibinfo {author} {\bibfnamefont {E.}~\bibnamefont
  {{Berti}}},\ }\href {\doibase 10.1103/PhysRevD.81.084054} {\bibfield
  {journal} {\bibinfo  {journal} {\prd}\ }\textbf {\bibinfo {volume} {81}},\
  \bibinfo {eid} {084054} (\bibinfo {year} {2010}{\natexlab{a}})},\ \Eprint
  {http://arxiv.org/abs/1002.2643} {arXiv:1002.2643 [astro-ph.GA]} \BibitemShut
  {NoStop}%
\bibitem [{\citenamefont {{Kesden}}\ \emph
  {et~al.}(2010{\natexlab{b}})\citenamefont {{Kesden}}, \citenamefont
  {{Sperhake}},\ and\ \citenamefont {{Berti}}}]{Kesden2010b}%
  \BibitemOpen
  \bibfield  {author} {\bibinfo {author} {\bibfnamefont {M.}~\bibnamefont
  {{Kesden}}}, \bibinfo {author} {\bibfnamefont {U.}~\bibnamefont
  {{Sperhake}}}, \ and\ \bibinfo {author} {\bibfnamefont {E.}~\bibnamefont
  {{Berti}}},\ }\href {\doibase 10.1088/0004-637X/715/2/1006} {\bibfield
  {journal} {\bibinfo  {journal} {\apj}\ }\textbf {\bibinfo {volume} {715}},\
  \bibinfo {pages} {1006} (\bibinfo {year} {2010}{\natexlab{b}})},\ \Eprint
  {http://arxiv.org/abs/1003.4993} {arXiv:1003.4993 [astro-ph.CO]} \BibitemShut
  {NoStop}%
\bibitem [{\citenamefont {Maeder}\ and\ \citenamefont
  {Meynet}(2004)}]{Maeder2004}%
  \BibitemOpen
  \bibfield  {author} {\bibinfo {author} {\bibfnamefont {A.}~\bibnamefont
  {Maeder}}\ and\ \bibinfo {author} {\bibfnamefont {G.}~\bibnamefont
  {Meynet}},\ }\href {\doibase 10.1017/S0074180900196093} {\bibfield  {journal}
  {\bibinfo  {journal} {Symposium - International Astronomical Union}\ }\textbf
  {\bibinfo {volume} {215}},\ \bibinfo {pages} {500–509} (\bibinfo {year}
  {2004})}\BibitemShut {NoStop}%
\bibitem [{\citenamefont {{Maeder}}\ and\ \citenamefont
  {{Meynet}}(2008)}]{Maeder2008}%
  \BibitemOpen
  \bibfield  {author} {\bibinfo {author} {\bibfnamefont {A.}~\bibnamefont
  {{Maeder}}}\ and\ \bibinfo {author} {\bibfnamefont {G.}~\bibnamefont
  {{Meynet}}},\ }in\ \href@noop {} {\emph {\bibinfo {booktitle} {Mass Loss from
  Stars and the Evolution of Stellar Clusters}}},\ \bibinfo {series}
  {Astronomical Society of the Pacific Conference Series}, Vol.\ \bibinfo
  {volume} {388},\ \bibinfo {editor} {edited by\ \bibinfo {editor}
  {\bibfnamefont {A.}~\bibnamefont {{de Koter}}}, \bibinfo {editor}
  {\bibfnamefont {L.~J.}\ \bibnamefont {{Smith}}}, \ and\ \bibinfo {editor}
  {\bibfnamefont {L.~B.~F.~M.}\ \bibnamefont {{Waters}}}}\ (\bibinfo {year}
  {2008})\ p.~\bibinfo {pages} {3}\BibitemShut {NoStop}%
\bibitem [{\citenamefont {{Batta}}\ and\ \citenamefont
  {{Ramirez-Ruiz}}(2019)}]{Batta2019}%
  \BibitemOpen
  \bibfield  {author} {\bibinfo {author} {\bibfnamefont {A.}~\bibnamefont
  {{Batta}}}\ and\ \bibinfo {author} {\bibfnamefont {E.}~\bibnamefont
  {{Ramirez-Ruiz}}},\ }\href@noop {} {\bibfield  {journal} {\bibinfo  {journal}
  {arXiv e-prints}\ ,\ \bibinfo {eid} {arXiv:1904.04835}} (\bibinfo {year}
  {2019})},\ \Eprint {http://arxiv.org/abs/1904.04835} {arXiv:1904.04835
  [astro-ph.HE]} \BibitemShut {NoStop}%
\bibitem [{\citenamefont {{de Mink}}\ \emph {et~al.}(2010)\citenamefont {{de
  Mink}}, \citenamefont {{Cantiello}}, \citenamefont {{Langer}},\ and\
  \citenamefont {{Pols}}}]{deMink2010}%
  \BibitemOpen
  \bibfield  {author} {\bibinfo {author} {\bibfnamefont {S.~E.}\ \bibnamefont
  {{de Mink}}}, \bibinfo {author} {\bibfnamefont {M.}~\bibnamefont
  {{Cantiello}}}, \bibinfo {author} {\bibfnamefont {N.}~\bibnamefont
  {{Langer}}}, \ and\ \bibinfo {author} {\bibfnamefont {O.~R.}\ \bibnamefont
  {{Pols}}},\ }in\ \href {\doibase 10.1063/1.3536387} {\emph {\bibinfo
  {booktitle} {American Institute of Physics Conference Series}}},\ \bibinfo
  {series} {American Institute of Physics Conference Series}, Vol.\ \bibinfo
  {volume} {1314},\ \bibinfo {editor} {edited by\ \bibinfo {editor}
  {\bibfnamefont {V.}~\bibnamefont {{Kalogera}}}\ and\ \bibinfo {editor}
  {\bibfnamefont {M.}~\bibnamefont {{van der Sluys}}}}\ (\bibinfo {year}
  {2010})\ pp.\ \bibinfo {pages} {291--296}\BibitemShut {NoStop}%
\bibitem [{\citenamefont {{Marchant}}\ \emph {et~al.}(2016)\citenamefont
  {{Marchant}}, \citenamefont {{Langer}}, \citenamefont {{Podsiadlowski}},
  \citenamefont {{Tauris}},\ and\ \citenamefont {{Moriya}}}]{Marchant2016}%
  \BibitemOpen
  \bibfield  {author} {\bibinfo {author} {\bibfnamefont {P.}~\bibnamefont
  {{Marchant}}}, \bibinfo {author} {\bibfnamefont {N.}~\bibnamefont
  {{Langer}}}, \bibinfo {author} {\bibfnamefont {P.}~\bibnamefont
  {{Podsiadlowski}}}, \bibinfo {author} {\bibfnamefont {T.~M.}\ \bibnamefont
  {{Tauris}}}, \ and\ \bibinfo {author} {\bibfnamefont {T.~J.}\ \bibnamefont
  {{Moriya}}},\ }\href {\doibase 10.1051/0004-6361/201628133} {\bibfield
  {journal} {\bibinfo  {journal} {\aap}\ }\textbf {\bibinfo {volume} {588}},\
  \bibinfo {eid} {A50} (\bibinfo {year} {2016})},\ \Eprint
  {http://arxiv.org/abs/1601.03718} {arXiv:1601.03718 [astro-ph.SR]}
  \BibitemShut {NoStop}%
\bibitem [{\citenamefont {{Song}}\ \emph {et~al.}(2016)\citenamefont {{Song}},
  \citenamefont {{Meynet}}, \citenamefont {{Maeder}}, \citenamefont
  {{Ekstr{\"o}m}},\ and\ \citenamefont {{Eggenberger}}}]{Song2016}%
  \BibitemOpen
  \bibfield  {author} {\bibinfo {author} {\bibfnamefont {H.~F.}\ \bibnamefont
  {{Song}}}, \bibinfo {author} {\bibfnamefont {G.}~\bibnamefont {{Meynet}}},
  \bibinfo {author} {\bibfnamefont {A.}~\bibnamefont {{Maeder}}}, \bibinfo
  {author} {\bibfnamefont {S.}~\bibnamefont {{Ekstr{\"o}m}}}, \ and\ \bibinfo
  {author} {\bibfnamefont {P.}~\bibnamefont {{Eggenberger}}},\ }\href {\doibase
  10.1051/0004-6361/201526074} {\bibfield  {journal} {\bibinfo  {journal}
  {\aap}\ }\textbf {\bibinfo {volume} {585}},\ \bibinfo {eid} {A120} (\bibinfo
  {year} {2016})},\ \Eprint {http://arxiv.org/abs/1508.06094} {arXiv:1508.06094
  [astro-ph.SR]} \BibitemShut {NoStop}%
\bibitem [{\citenamefont {{Cui}}\ \emph {et~al.}(2018)\citenamefont {{Cui}},
  \citenamefont {{Wang}}, \citenamefont {{Zhu}}, \citenamefont {{L{\"u}}},
  \citenamefont {{Chen}},\ and\ \citenamefont {{Han}}}]{Cui2018}%
  \BibitemOpen
  \bibfield  {author} {\bibinfo {author} {\bibfnamefont {Z.}~\bibnamefont
  {{Cui}}}, \bibinfo {author} {\bibfnamefont {Z.}~\bibnamefont {{Wang}}},
  \bibinfo {author} {\bibfnamefont {C.}~\bibnamefont {{Zhu}}}, \bibinfo
  {author} {\bibfnamefont {G.}~\bibnamefont {{L{\"u}}}}, \bibinfo {author}
  {\bibfnamefont {H.}~\bibnamefont {{Chen}}}, \ and\ \bibinfo {author}
  {\bibfnamefont {Z.}~\bibnamefont {{Han}}},\ }\href {\doibase
  10.1088/1538-3873/aac55e} {\bibfield  {journal} {\bibinfo  {journal} {\pasp}\
  }\textbf {\bibinfo {volume} {130}},\ \bibinfo {pages} {084202} (\bibinfo
  {year} {2018})},\ \Eprint {http://arxiv.org/abs/1805.08397} {arXiv:1805.08397
  [astro-ph.SR]} \BibitemShut {NoStop}%
\bibitem [{\citenamefont {{Schmidt}}\ \emph {et~al.}(2015)\citenamefont
  {{Schmidt}}, \citenamefont {{Ohme}},\ and\ \citenamefont
  {{Hannam}}}]{Schmidt2015}%
  \BibitemOpen
  \bibfield  {author} {\bibinfo {author} {\bibfnamefont {P.}~\bibnamefont
  {{Schmidt}}}, \bibinfo {author} {\bibfnamefont {F.}~\bibnamefont {{Ohme}}}, \
  and\ \bibinfo {author} {\bibfnamefont {M.}~\bibnamefont {{Hannam}}},\ }\href
  {\doibase 10.1103/PhysRevD.91.024043} {\bibfield  {journal} {\bibinfo
  {journal} {\prd}\ }\textbf {\bibinfo {volume} {91}},\ \bibinfo {eid} {024043}
  (\bibinfo {year} {2015})},\ \Eprint {http://arxiv.org/abs/1408.1810}
  {arXiv:1408.1810 [gr-qc]} \BibitemShut {NoStop}%
\bibitem [{\citenamefont {{Steinle}}\ \emph {et~al.}()\citenamefont
  {{Steinle}}, \citenamefont {{Gangardt}}, \citenamefont {{Gerosa}},
  \citenamefont {{Kesden}},\ and\ \citenamefont
  {{Stoikos}}}]{SteinleGangardt2020}%
  \BibitemOpen
  \bibfield  {author} {\bibinfo {author} {\bibfnamefont {N.}~\bibnamefont
  {{Steinle}}}, \bibinfo {author} {\bibfnamefont {D.}~\bibnamefont
  {{Gangardt}}}, \bibinfo {author} {\bibfnamefont {D.}~\bibnamefont
  {{Gerosa}}}, \bibinfo {author} {\bibfnamefont {M.}~\bibnamefont {{Kesden}}},
  \ and\ \bibinfo {author} {\bibfnamefont {E.}~\bibnamefont {{Stoikos}}},\
  }\href@noop {} {\ }\bibinfo {note} {(in preparation)}\BibitemShut {NoStop}%
\bibitem [{\citenamefont {{Vink}}(2015)}]{Vink2015}%
  \BibitemOpen
  \bibfield  {author} {\bibinfo {author} {\bibfnamefont {J.~S.}\ \bibnamefont
  {{Vink}}},\ }in\ \href@noop {} {\emph {\bibinfo {booktitle} {Wolf-Rayet
  Stars}}},\ \bibinfo {editor} {edited by\ \bibinfo {editor} {\bibfnamefont
  {W.-R.}\ \bibnamefont {{Hamann}}}, \bibinfo {editor} {\bibfnamefont
  {A.}~\bibnamefont {{Sander}}}, \ and\ \bibinfo {editor} {\bibfnamefont
  {H.}~\bibnamefont {{Todt}}}}\ (\bibinfo {year} {2015})\ pp.\ \bibinfo {pages}
  {133--138},\ \Eprint {http://arxiv.org/abs/1510.00227} {arXiv:1510.00227
  [astro-ph.SR]} \BibitemShut {NoStop}%
\bibitem [{\citenamefont {{Yoon}}(2017)}]{Yoon2017}%
  \BibitemOpen
  \bibfield  {author} {\bibinfo {author} {\bibfnamefont {S.-C.}\ \bibnamefont
  {{Yoon}}},\ }\href {\doibase 10.1093/mnras/stx1496} {\bibfield  {journal}
  {\bibinfo  {journal} {\mnras}\ }\textbf {\bibinfo {volume} {470}},\ \bibinfo
  {pages} {3970} (\bibinfo {year} {2017})},\ \Eprint
  {http://arxiv.org/abs/1706.04716} {arXiv:1706.04716 [astro-ph.SR]}
  \BibitemShut {NoStop}%
\bibitem [{\citenamefont {{Belczynski}}\ \emph {et~al.}(2010)\citenamefont
  {{Belczynski}}, \citenamefont {{Bulik}}, \citenamefont {{Fryer}},
  \citenamefont {{Ruiter}}, \citenamefont {{Valsecchi}}, \citenamefont
  {{Vink}},\ and\ \citenamefont {{Hurley}}}]{Belczynski2010}%
  \BibitemOpen
  \bibfield  {author} {\bibinfo {author} {\bibfnamefont {K.}~\bibnamefont
  {{Belczynski}}}, \bibinfo {author} {\bibfnamefont {T.}~\bibnamefont
  {{Bulik}}}, \bibinfo {author} {\bibfnamefont {C.~L.}\ \bibnamefont
  {{Fryer}}}, \bibinfo {author} {\bibfnamefont {A.}~\bibnamefont {{Ruiter}}},
  \bibinfo {author} {\bibfnamefont {F.}~\bibnamefont {{Valsecchi}}}, \bibinfo
  {author} {\bibfnamefont {J.~S.}\ \bibnamefont {{Vink}}}, \ and\ \bibinfo
  {author} {\bibfnamefont {J.~R.}\ \bibnamefont {{Hurley}}},\ }\href {\doibase
  10.1088/0004-637X/714/2/1217} {\bibfield  {journal} {\bibinfo  {journal}
  {\apj}\ }\textbf {\bibinfo {volume} {714}},\ \bibinfo {pages} {1217}
  (\bibinfo {year} {2010})},\ \Eprint {http://arxiv.org/abs/0904.2784}
  {arXiv:0904.2784 [astro-ph.SR]} \BibitemShut {NoStop}%
\bibitem [{\citenamefont {{Crowther}}\ \emph {et~al.}(2010)\citenamefont
  {{Crowther}}, \citenamefont {{Schnurr}}, \citenamefont {{Hirschi}},
  \citenamefont {{Yusof}}, \citenamefont {{Parker}}, \citenamefont
  {{Goodwin}},\ and\ \citenamefont {{Kassim}}}]{Crowther2010}%
  \BibitemOpen
  \bibfield  {author} {\bibinfo {author} {\bibfnamefont {P.~A.}\ \bibnamefont
  {{Crowther}}}, \bibinfo {author} {\bibfnamefont {O.}~\bibnamefont
  {{Schnurr}}}, \bibinfo {author} {\bibfnamefont {R.}~\bibnamefont
  {{Hirschi}}}, \bibinfo {author} {\bibfnamefont {N.}~\bibnamefont {{Yusof}}},
  \bibinfo {author} {\bibfnamefont {R.~J.}\ \bibnamefont {{Parker}}}, \bibinfo
  {author} {\bibfnamefont {S.~P.}\ \bibnamefont {{Goodwin}}}, \ and\ \bibinfo
  {author} {\bibfnamefont {H.~A.}\ \bibnamefont {{Kassim}}},\ }\href {\doibase
  10.1111/j.1365-2966.2010.17167.x} {\bibfield  {journal} {\bibinfo  {journal}
  {\mnras}\ }\textbf {\bibinfo {volume} {408}},\ \bibinfo {pages} {731}
  (\bibinfo {year} {2010})},\ \Eprint {http://arxiv.org/abs/1007.3284}
  {arXiv:1007.3284 [astro-ph.SR]} \BibitemShut {NoStop}%
\bibitem [{\citenamefont {{Bardeen}}\ and\ \citenamefont
  {{Petterson}}(1975)}]{Bardeen1975}%
  \BibitemOpen
  \bibfield  {author} {\bibinfo {author} {\bibfnamefont {J.~M.}\ \bibnamefont
  {{Bardeen}}}\ and\ \bibinfo {author} {\bibfnamefont {J.~A.}\ \bibnamefont
  {{Petterson}}},\ }\href {\doibase 10.1086/181711} {\bibfield  {journal}
  {\bibinfo  {journal} {\apjl}\ }\textbf {\bibinfo {volume} {195}},\ \bibinfo
  {pages} {L65} (\bibinfo {year} {1975})}\BibitemShut {NoStop}%
\bibitem [{\citenamefont {{Packet}}(1981)}]{Packet1981}%
  \BibitemOpen
  \bibfield  {author} {\bibinfo {author} {\bibfnamefont {W.}~\bibnamefont
  {{Packet}}},\ }\href@noop {} {\bibfield  {journal} {\bibinfo  {journal}
  {\aap}\ }\textbf {\bibinfo {volume} {102}},\ \bibinfo {pages} {17} (\bibinfo
  {year} {1981})}\BibitemShut {NoStop}%
\bibitem [{\citenamefont {{de Mink}}\ \emph {et~al.}(2013)\citenamefont {{de
  Mink}}, \citenamefont {{Langer}}, \citenamefont {{Izzard}}, \citenamefont
  {{Sana}},\ and\ \citenamefont {{de Koter}}}]{deMink2013}%
  \BibitemOpen
  \bibfield  {author} {\bibinfo {author} {\bibfnamefont {S.~E.}\ \bibnamefont
  {{de Mink}}}, \bibinfo {author} {\bibfnamefont {N.}~\bibnamefont {{Langer}}},
  \bibinfo {author} {\bibfnamefont {R.~G.}\ \bibnamefont {{Izzard}}}, \bibinfo
  {author} {\bibfnamefont {H.}~\bibnamefont {{Sana}}}, \ and\ \bibinfo {author}
  {\bibfnamefont {A.}~\bibnamefont {{de Koter}}},\ }\href {\doibase
  10.1088/0004-637X/764/2/166} {\bibfield  {journal} {\bibinfo  {journal}
  {\apj}\ }\textbf {\bibinfo {volume} {764}},\ \bibinfo {eid} {166} (\bibinfo
  {year} {2013})},\ \Eprint {http://arxiv.org/abs/1211.3742} {arXiv:1211.3742
  [astro-ph.SR]} \BibitemShut {NoStop}%
\bibitem [{\citenamefont {{Bardeen}}\ \emph {et~al.}(1972)\citenamefont
  {{Bardeen}}, \citenamefont {{Press}},\ and\ \citenamefont
  {{Teukolsky}}}]{Bardeen1972}%
  \BibitemOpen
  \bibfield  {author} {\bibinfo {author} {\bibfnamefont {J.~M.}\ \bibnamefont
  {{Bardeen}}}, \bibinfo {author} {\bibfnamefont {W.~H.}\ \bibnamefont
  {{Press}}}, \ and\ \bibinfo {author} {\bibfnamefont {S.~A.}\ \bibnamefont
  {{Teukolsky}}},\ }\href {\doibase 10.1086/151796} {\bibfield  {journal}
  {\bibinfo  {journal} {\apj}\ }\textbf {\bibinfo {volume} {178}},\ \bibinfo
  {pages} {347} (\bibinfo {year} {1972})}\BibitemShut {NoStop}%
\bibitem [{\citenamefont {{Salpeter}}(1964)}]{Salpeter1964}%
  \BibitemOpen
  \bibfield  {author} {\bibinfo {author} {\bibfnamefont {E.~E.}\ \bibnamefont
  {{Salpeter}}},\ }\href {\doibase 10.1086/147973} {\bibfield  {journal}
  {\bibinfo  {journal} {\apj}\ }\textbf {\bibinfo {volume} {140}},\ \bibinfo
  {pages} {796} (\bibinfo {year} {1964})}\BibitemShut {NoStop}%
\end{thebibliography}%

\begin{acknowledgments}
The authors would like to thank Davide Gerosa and Nicola Giacobbo for insigthful comments and the National Science Foundation (NSF) for support through NSF Grant No. PHY-1607031.
\end{acknowledgments}

\section*{Appendix}
\label{sec:app}

This Appendix reviews the prescriptions for binary stellar evolution adopted in this paper.  Our goal has not been to advance the state of the art in this extraordinarily rich field, but rather to identify scenarios that might produce BBHs with large misaligned spins whose GWs exhibit detectable signatures of spin precession.

\appendix

\section{Single Stellar Evolution}
\label{app:SSE}
\begin{figure}[t!]
  \centering
  \includegraphics[width=0.48\textwidth]{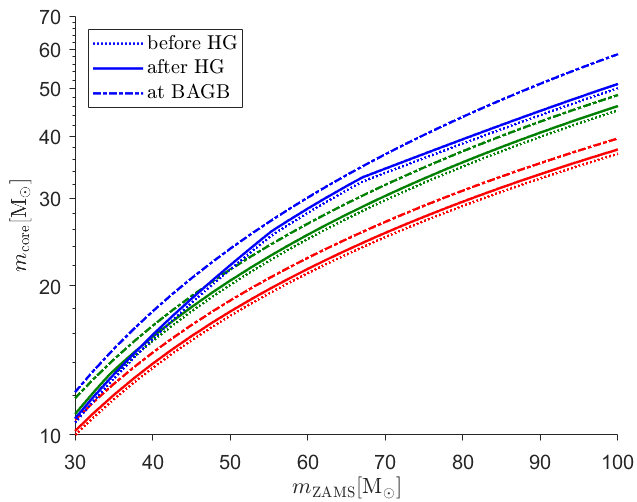}
   \caption{The core mass as a function of ZAMS mass at the end of the three main stages of stellar evolution. The red, green, and blue lines correspond to metallicities of $Z$ = 0.0002, 0.002, and 0.02, while the dotted, solid, and dot-dashed lines correspond to core masses at the ends of the MS, HG, and CHeB stages.}
   \label{F:CoreMVSzamsM}
\end{figure}
We use the formulae of \citeauthor{Hurley2000}~\cite{Hurley2000} to model single stellar evolution as a function of the zero-age main sequence (ZAMS) mass and metallicity.  Our BH progenitors are O-type stars ($m_{\rm ZAMS} \geq 30$~M$_{\odot}$) at sub-solar metallicity ($Z \leq Z_{\odot}$) that burn hydrogen in their cores over the main sequence (MS). After exhausting the hydrogen in their cores, the stars enter the short-lived Hertzsprung gap (HG) in which they are powered by hydrogen-shell burning. Helium burning, which commenced on the HG, comes to dominate the stellar luminosity in the core helium-burning (CHeB) stage; this stage can be further divided into the blue loop (BL) and red giant (RG) phases. After the stellar radius expands greatly during the HG and CHeB, the star eventually loses its envelope since the late-CHeB radial expansion drives extreme wind mass loss.  The stellar core emerges as a Wolf-Rayet (WR) star with an inner iron core surrounded by shells of lighter elements. The WR star experiences strong winds for its small size and inevitably collapses into a black hole (BH) for our choice of ZAMS masses and metallicities \cite{Heger2003}.

\begin{figure}[t!] 
  \centering
  \includegraphics[width=0.48\textwidth]{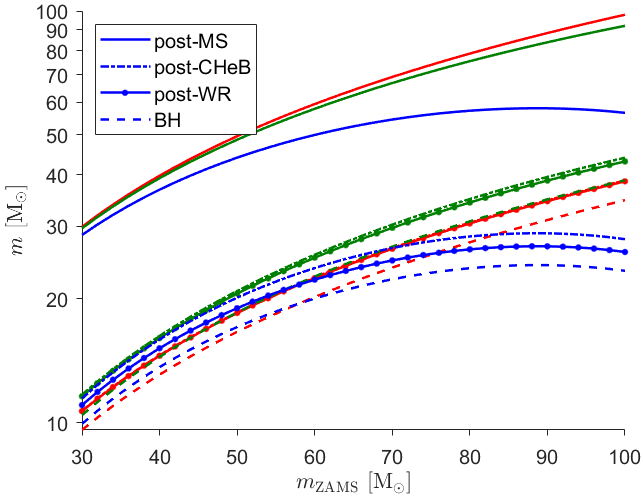}
   \caption{Mass evolution of an isolated star as a function of ZAMS mass. The red, green, and blue lines correspond to metallicities of $Z$ = 0.0002, 0.002, and 0.02.  The solid, dot-dashed, and dotted lines correspond to the ends of the MS, CHeB, and WR stages, while the dashed lines gives the BH mass after additional mass loss during core collapse.} \label{F:SSEmass}
\end{figure} 

\begin{figure*}[t!] 
  \centering
  \includegraphics[width=1.0\linewidth]{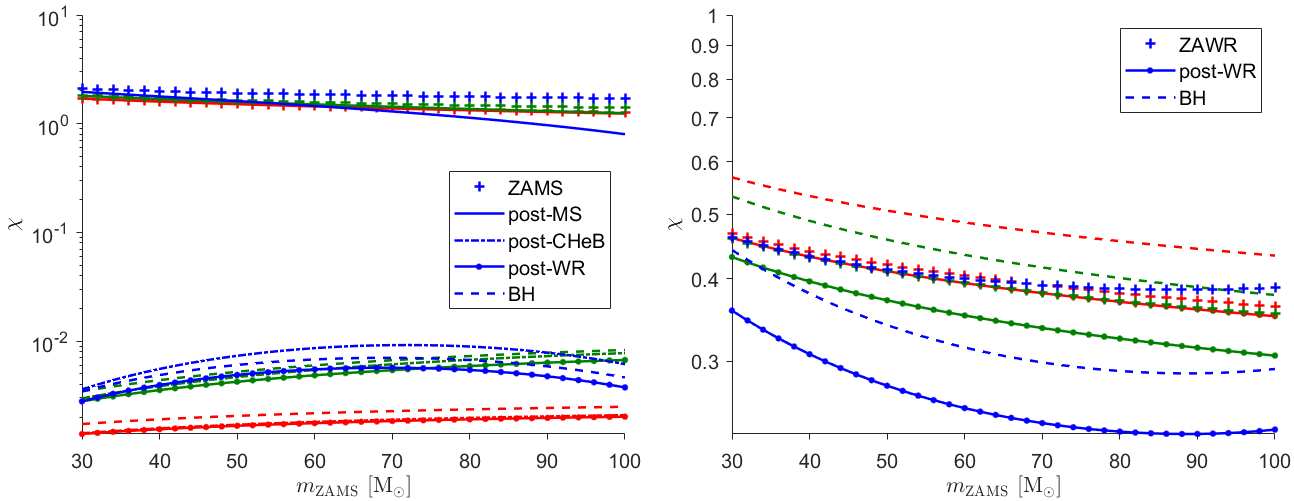}
   \caption{Evolution of the dimensionless spin as a function of ZAMS mass.  The left (right) panels correspond to maximal (minimal) core-envelope coupling in which the spin is initialized at a fraction $f_{\rm B} = 0.03$ of the breakup value at ZAMS (ZAWR).  The lines marked by crosses show these initial spins, while the other line styles and the line colors are the same as in Fig.~\ref{F:SSEmass}.} \label{F:SSEspin}
\end{figure*}

Fig.~\ref{F:CoreMVSzamsM} displays the core mass as a function of ZAMS mass at the end of the MS, HG, and CHeB stages for three different metallicites in the absence of winds. These core masses monotonically increase with ZAMS mass and metallicity, but this monotonicity does not hold when the effects of wind mass loss are included. The WR radius and lifetime as functions of WR mass $m$ given by
\begin{align}
\frac{R_{\rm WR}}{{\rm R}_{\odot}} &= \frac{0.2391\left(m/{\rm M_{\odot}}\right)^{4.6}}{\left(m/{\rm M_{\odot}}\right)^4 + 0.162\left(m/{\rm M_{\odot}}\right)^3 + 0.0065}
\label{E:WRradius} \\
\frac{t_{\rm WR}}{{\rm Myr}} &= \frac{0.4129 + 18.81\left(m/{\rm M_{\odot}}\right)^4 + 1.853\left(m/{\rm M_{\odot}}\right)^6}{\left(m/{\rm M_{\odot}}\right)^{6.5}}~.
\label{E:WRlifetime}
\end{align}
We adopt these formulae because their simplicity, but WR stars are an active area of research \cite{Crowther2007,Vink2015} and their evolution has important consequences for compact-object formation \cite{Yoon2017}.

Fig.~\ref{F:SSEmass} shows isolated stellar mass evolution from birth to core collapse.  Winds drive metallicity-dependent mass loss during the MS stage (solid lines), then the core emerges as a WR star at the end of the CHeB stage (dot-dashed lines).  Winds drive further mass loss during the WR stage (dotted lines), until core collapse produces a BH whose mass has been reduced by $10\%$ due to neutrino emission during the collapse.  At lower ZAMS masses, winds are negligible and the larger WR masses at high metallicity lead to larger BH masses.  At higher ZAMS masses, greater wind-driven mass loss at high metallicity leads to smaller BH masses.  

The left panel of Fig.~\ref{F:SSEspin} shows the evolution of the dimensionless spin $\rchi$ as a function of ZAMS mass $m_{\rm ZAMS}$ for maximal core-envelope coupling. In the case of maximal coupling, an initial spin that is a fraction $f_{\rm B} = 0.03$ of the breakup value is assigned at ZAMS implying a dimensionless spin $\rchi \approx 2$ (lines marked by crosses).  Mass loss on the MS (solid lines) leads to a modest decrease in the spin, particular for high $m_{\rm ZAMS}$ and $Z$. The larger stellar radius and greater mass loss during the CHeB stage (dot-dashed lines) yield $\rchi \lesssim 10^{-2}$ for all masses and metallicities, which we assume removes the envelope.  Winds during the WR stage (dotted lines) further reduce $\rchi$, most significantly for $Z = 0.02$. Finally, the change $\delta m/m = -0.1$ in the mass due to neutrino emission during core collapse fractionally increases the dimensionless spin by an amount $\delta\rchi/\rchi = -2\delta m/m = 0.2$ (dashed lines).

The right panel of Fig.~\ref{F:SSEspin} shows the evolution of the dimensionless spin $\rchi$ as a function of ZAMS mass $m_{\rm ZAMS}$ for minimal core-envelope coupling.  As the spin of the stellar core is independent of that of the envelope under this assumption, it can be initialized to a fraction $f_{\rm B} = 0.03$ of the breakup value at the start of the WR stage, implying a dimensionless spin $\rchi \approx 0.45$ (lines marked by crosses).  As $f_{\rm B}$ is a free parameter in our model, it can be chosen to yield near maximal spins. As in the case of maximal coupling, WR winds reduce the dimensionless spins by an amount that increases from a few percent at $Z = 0.0002$ to $\lesssim 50\%$ at $Z = 0.02$ (dotted lines).  Neutrino emission during core collapse again increases the dimensionless spin of the BH by $20\%$ compared to its value at the end of the WR stage (dashed lines).

Our simplified model of single stellar evolution described above reproduces the essential features of more sophisticated models. For the ranges of ZAMS mass and metallicity that we consider, the mass evolution depicted in Fig.~\ref{F:SSEmass} is broadly consistent with previous studies \cite{Hurley2000,BK2008,Qin2018,Giacobbo2018,Sedda2018}, as is the range of BBH masses \cite{Fryer1999,Belczynski2010,Crowther2010}.  
Previous work has also explored the spin evolution of BBH progenitors under the assumption of maximal \cite{Belczynski2020,Gerosa2018,Fuller2019} and minimal \cite{Qin2018,Postnov2018} core-envelope coupling and arrived at qualitatively similar results for the final BBH spins depicted in Fig.~\ref{F:SSEspin}.  We do not consider the possibility that large spins could lead to enhanced rotational mixing and chemically homogeneous evolution \cite{deMink2010,Marchant2016,Mandel2016,Song2016,Cui2018}.

\section{Binary Stellar Evolution}
\label{app:BinEvo}

\subsection{Roche lobe overflow}
\label{app:RLOF}
\begin{figure}[t!] 
  \centering
   \includegraphics[width=0.48\textwidth]{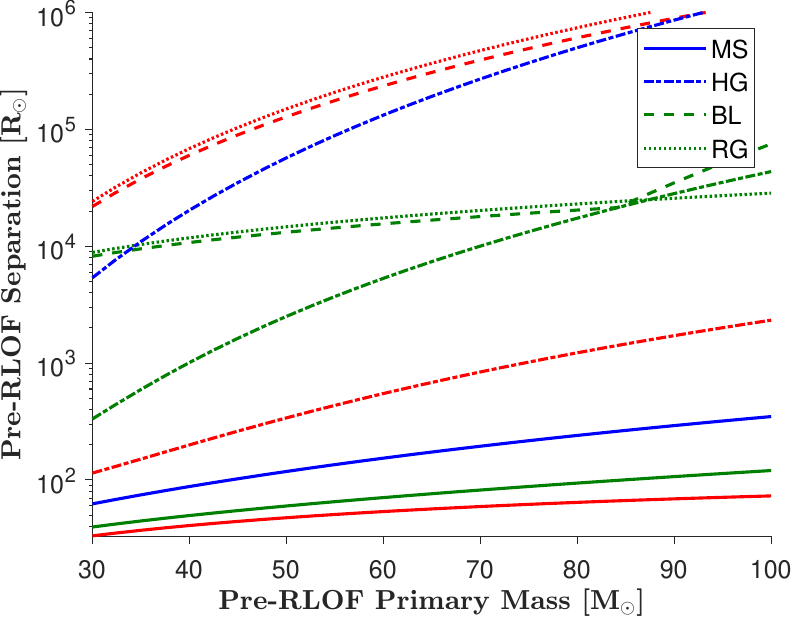}
   \caption{The maximum binary separation at which Roche-lobe overflow (RLOF) occurs during a given stage of stellar evolution as a function of the donor mass. The blue, green, and red lines correspond to metallicities $Z = 0.02$, 0.002, and 0.0002, while the solid, dot-dashed, dashed, and dotted lines correspond to the main sequence (MS), Hertzsprung gap (HG), blue loop (BL), and red giant (RG) stages of stellar evolution. The companion mass is chosen to be $10\%$ less than that of the donor star.
   } \label{F:RLOF}
\end{figure}

A binary star experiences Roche-lobe overflow (RLOF) when its radius $R_\ast$ is equal to its Roche-lobe radius $R_{\rm RL}$ given by Eq.~(\ref{E:RL}). For our high-mass BH progenitors ($m_{\rm ZAMS} \geq 30~{\rm M}_\odot$), $R_\ast$ increases monotonically during each stage of stellar evolution. This implies that, for circular orbits, there is a maximum binary separation at which RLOF occurs during a given stage for each pair of stellar masses $m_i$ and metallicity $Z$.

Fig.~\ref{F:RLOF} shows these maximum separations as a function of the donor star's mass assuming a companion that $10\%$ less massive. They are linearly proportional to the stellar radius and therefore increase with metallicity during the MS and HG stages. The stellar radius is maximized at the end of the HG for $Z = 0.02$, so a star that does not fill its Roche lobe at that time will never experience RLOF. We therefore only show the maximum separations for the BL and RG stages for stars with $Z = 0.0002$ and 0.002 that continue to expand during these stages.

Binaries with ZAMS separations $a_{\rm ZAMS}$ below the maximum separation for RLOF on the MS will not form BBHs because the density gradient in the stellar interior during this stage is insufficient to define a core-envelope boundary \cite{Ivanova2004,BK2007}. This implies that RLOF will completely disrupt the star rather than leaving behind a WR star. Binaries with $a_{\rm ZAMS}$ above the maximum separation for RLOF (on the HG for $Z = 0.02$ and during the RG stage for lower metallicity) will also fail to form GW sources because their merger times will be greater than the age of the Universe in the absence of CEE. RLOF will also destroy binaries if the change in separation during primary or secondary RLOF causes either star to fill its Roche lobe on the MS or during the WR stage.  

Once we have identified the stage of stellar evolution during which RLOF occurs, we find the time of RLOF by equating the time-dependent stellar radius during this stage to the Roche-lobe radius given by Eq.~(\ref{E:RL}). We then determine the core and envelope masses at this time, and use them to calculate the WR mass and the new binary separation according to Eqs.~(\ref{E:CE}) or (\ref{E:SMT}) depending on whether the RLOF leads to CEE or SMT in the formation scenario under consideration.

\subsection{Accretion during Stable Mass Transfer}
\label{app:AccSMT}

In our model, the companion accretes a fraction $f_{\rm a}$ of the envelope of the donor star during stable mass transfer (SMT). The value of $f_{\rm a}$ can range from 0 to 1 \cite{Meurs1989}, but we only consider $f_{\rm a} = 0.2$ and $f_{\rm a} = 0.5$ \cite{Belczynski2020,Stevenson2017}. We assume that accretion occurs through a thin disk and that the Bardeen-Petterson effect \cite{Bardeen1975} drives the inner edge of this disk into the equatorial plane of the accretor.

In Pathways A1, A2, and B2, the accretor is a star as shown in Fig.~\ref{F:Diagram}. The increase in its dimensionless spin per unit of accreted mass is given by Eq.~(\ref{E:accB2}) in which the stellar radius $R$ ss a function of mass given by Eq.~(2) of Tout~{\it et al.} (1996) \cite{Tout1996} for a MS star (with $r_{\rm g}^2 = 0.2$) or Eq.~(\ref{E:WRradius}) for a WR star (with $r_{\rm g}^2 = 0.075$). As $R \gg Gm/c^2$ for both MS and WR stars, even a modest amount of mass may significantly increase the accreting star's spin \cite{Packet1981,deMink2013}. This increase in spin has little effect on the spin of the secondary BH in Scenario A, due to the secondary being subsequently spun down during CEE as shown in the bottom panels of Fig.~\ref{F:ScenAChi}. However, accretion by the primary WR star is responsible for the maximal spin of the primary BH in Pathway B2, as shown in Table~\ref{T:table1} and the top right panel of Fig.~\ref{F:ScenBChi}. 

\begin{figure*}[!t]
\centering
\includegraphics[width=1.0\textwidth]{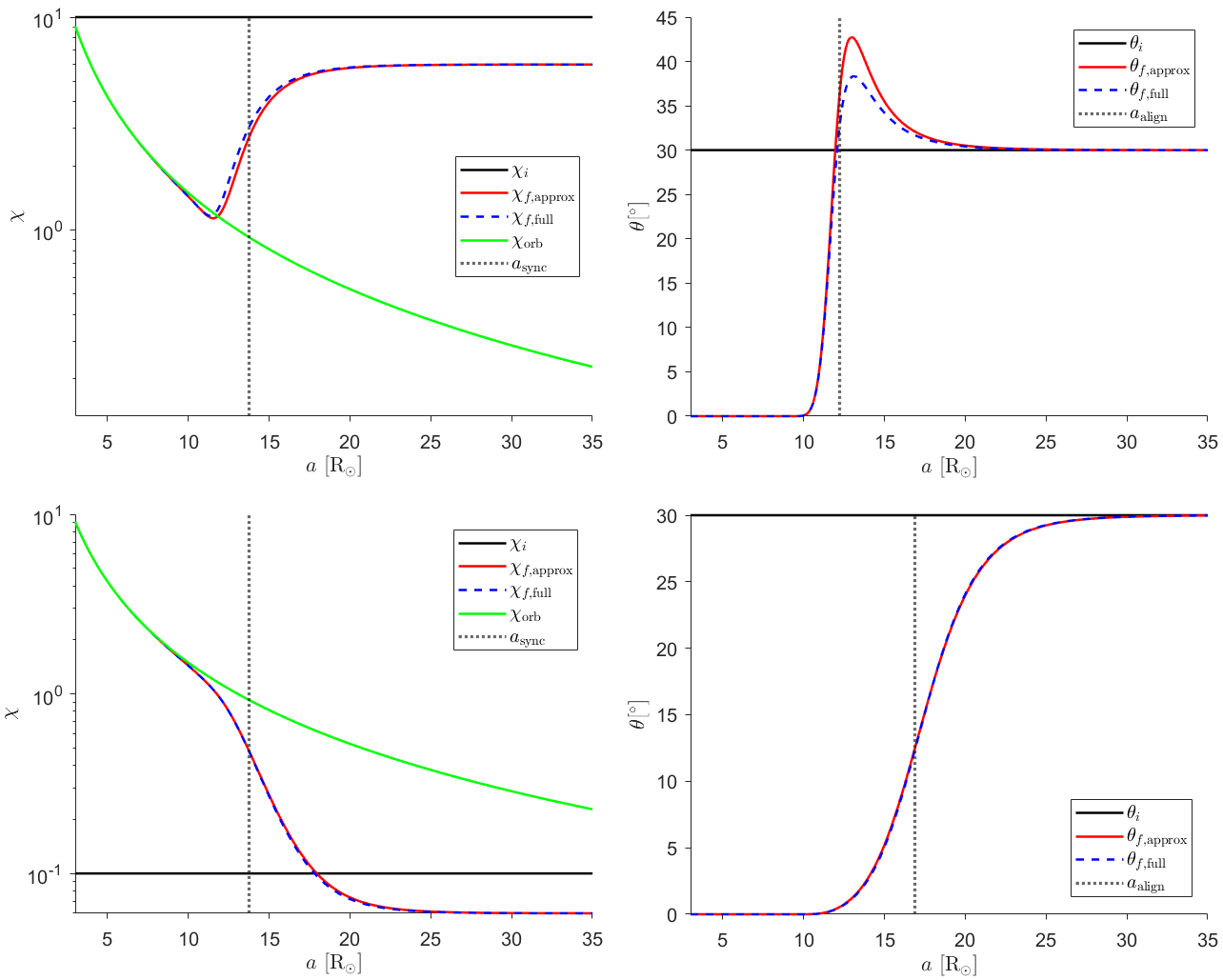}%
\caption{Dimensionless spin $\rchi$ and spin-orbit misalignment angle $\theta$ as a function of binary separation $a$ at the end of the lifetime of a $Z = 0.02$, $30~{\rm M}_\odot$ WR star with an equal-mass companion.  The top (bottom) panels show an initial spin $\rchi_i = 10$ (0.1); both cases have an initial misalignment $\theta_i = 30^\circ$.  The solid red (dashed blue) lines show the linearized (full nonlinear) spin evolution.  The solid green lines show the tidally synchronized spin $\rchi_{\rm orb}$.  The dotted vertical lines $a_{\rm sync}$ ($a_{\rm align}$) in the left (right) panels shows the separations at which the synchronization (alignment) timescales $t_{\rm sync}$ ($t_{\rm align}$) equal the WR lifetime.} \label{F:TidesSep}%
\end{figure*}

In Pathway B1, the accretor is a BH and the increase in its mass $m_{\rm BH}$ and dimensionless spin $\rchi$ per unit of accreted rest mass are given by
\begin{subequations} \label{E:BHAcc}
\begin{align}
\frac{dm_{\rm BH}}{dm} &= E(\rchi)~,  \\ 
\frac{d\rchi}{dm} &= \frac{L(\rchi)}{m_{\rm BH}^2} - \frac{2\rchi E(\rchi)}{m_{\rm BH}}~,
\end{align}
\end{subequations}
where $E(\rchi)$ and $L(\rchi)$ are the specific energy and orbital angular momentum of massive particles at the prograde innermost stable circular orbit (ISCO) of a BH with dimensionless spin $\rchi$ \cite{Bardeen1972}. According to these equations, a BH accreting at the Eddington rate $\dot{m}_{\rm Edd} = L_{\rm Edd}/\epsilon c^2$ will change its mass and spin on the Salpeter timescale \cite{Salpeter1964}
\begin{equation} \label{E:T_Sal}
t_{\rm Edd} = \frac{m_{\rm BH}}{\dot{m}_{\rm Edd}} = \frac{\kappa_T\epsilon c}{4\pi G} = 45.1~{\rm Myr} \left( \frac{\epsilon}{0.1} \right)~,
\end{equation}
where $\kappa_T$ is the Thomson opacity and $\epsilon$ is the radiative efficiency. Since this timescale is much longer than the duration of the HG and CHeB stages on which SMT occurs ($\lesssim 0.01$~Myr and $\lesssim 0.4$~Myr respectively), the BH mass and spin will change by a negligible amount if accretion is Eddington limited. We allow super-Eddington accretion in the results presented in Sec.~\ref{sec:Res}; imposing the restriction $\dot{m} < \dot{m}_{\rm Edd}$ would reduce the average primary mass $\overline{m}_1$ and dimensionless spin $\overline{\rchi}_1$ to their post-BH values in the upper left panels of Figs.~\ref{F:ScenBMasses} and \ref{F:ScenBChi}. It would also decrease the average aligned effective spin $\overline{\rchi}_{\rm eff}$ in the bottom left panel of Fig.~\ref{F:ScenBChiEffTilts} and the entries corresponding to Pathway B1 in Table~\ref{T:table1}.

\subsection{Tidal Evolution}
\label{app:Tides}

Tides and stellar winds determine the dimensionless WR spin $\rchi$ and spin-orbit misalignment angle $\theta$ according to the coupled Eqs.~(\ref{E:TWspinODE}) and (\ref{E:TWalignODE}).  In Fig.~\ref{F:TidesSep}, we show the solutions to these equations at the end of the WR lifetime as a function of binary separation $a$ for the initial conditions $\rchi_i = 10$ or 0.1 and $\theta_i = 30^\circ$.  As the tidal synchronization timescale $t_{\rm sync} \propto a^{17/2}$ according to Eq.~(\ref{E:Tsync}), tides are irrelevant for $a \gtrsim 20~{\rm R}_\odot$ and $\rchi$ decreases by $\sim 40\%$ due to winds while $\theta_f = \theta_i$.  For intermediate separations $10~{\rm R}_\odot \lesssim a \lesssim 20~{\rm R}_\odot$, both tides and winds play a role and $\rchi$ approaches its synchronized value $\rchi_{\rm orb}$.  For $\rchi > 2\rchi_{\rm orb}/(1 - \eta)$ and $a \gtrsim a_{\rm align}$ as in the upper right panel of Fig.~\ref{F:TidesSep}, $\dot{\theta} > 0$ according to Eq.~(\ref{E:TWalignODE}) implying that $\theta$ can increase by as much as $\sim 25\%$.  This mechanism for increasing the spin-orbit misalignment is interesting but unlikely to affect many BBHs, given the narrow range of binary separations $a_{\rm align} \lesssim a \simeq a_{\rm sync}$ at which it operates.  At small separations, $a \lesssim 10~{\rm R}_\odot$, tides dominate the spin evolution and $\rchi \to \rchi_{\rm orb}, \theta \to 0^\circ$.

In Fig.~\ref{F:TidesSep}, the solid red and dashed blue lines show the linearized spin evolution given by Eqs.~(\ref{E:TWspinODE}) and (\ref{E:TWalignODE}) and the full nonlinear evolution given by \cite{Hut1981}
\begin{subequations} \label{E:NLtides}
\begin{align}
\left( \frac{d\rchi}{dm} \right)_{\rm tid, NL} &= \frac{1}{\dot{m} t_{\rm sync}} \left[ \rchi_{\rm orb}\cos\theta - \rchi \left( 1 - \frac{1}{2} \sin^2\theta \right) \right]~, \\
\left( \frac{d\theta}{dm} \right)_{\rm tid, NL} &= - \frac{\sin\theta}{\dot{m} t_{\rm sync}} \left[ \frac{\rchi_{\rm orb}}{\rchi} - \frac{1}{2} \left( \cos\theta - \eta \right) \right]~,
\end{align}
\end{subequations}
The linear approximation is excellent except that it can overestimate the increase in the spin-orbit misalignment by as much as $\sim 70\%$ as shown in the top right panel.

\end{document}